\documentclass[twocolumn]{aastex62}

\usepackage{graphicx}
\usepackage{amsmath}
\usepackage{amssymb}
\usepackage{enumitem}

\defcitealias{lehner15}{LHW15}
\defcitealias{howk17}{Paper~I}

\newcommand{\dvrange}{$-700 \le \vlsr\ \le -150$ \km}
\newcommand{\dvobs}{$-510 \le \vlsr\ \le -150$ \km}

\def\nodata{ ~$\cdots$~ }

\newcommand{\zem}{\ensuremath{z_{\rm em}}}

\newcommand{\cmm}{cm$^{-2}$}

\newcommand{\lya}{Ly$\alpha$}

\newcommand{\lyb}{Ly$\beta$}
\newcommand{\lyg}{Ly$\gamma$}
\newcommand{\nhi}{$N_{\rm H\,I}$}

\newcommand{\novi}{$N_{\rm O\,VI}$}

\newcommand{\mnh}{N_{\rm H}}
\newcommand{\mnhi}{N_{\rm H\,I}}

\newcommand{\mlnovi}{\log N_{\rm O\,VI}}

\newcommand{\mlnhi}{\log N_{\rm H\,I}}
\newcommand{\mlnh}{\log N_{\rm H}}
\newcommand{\mlnnh}{\log n_{\rm H}}

\newcommand{\km}{${\rm km\,s}^{-1}$}

\newcommand{\hst}{{\em HST}}
\newcommand{\fuse}{{\em FUSE}}
\newcommand{\vlsr}{\ensuremath{v_{\rm LSR}}}

\newcommand{\lms}{\ensuremath{l_{\rm MS}}}
\newcommand{\bms}{\ensuremath{b_{\rm MS}}}

\newcommand{\ca}{\ensuremath{{\rm [C/\alpha]}}}
\newcommand{\rvir}{\ensuremath{R_{\rm vir}}}
\newcommand{\Mvir}{\ensuremath{M_{\rm vir}}}
\newcommand{\fc}{\ensuremath{f_{\rm c}}}

\newcommand{\hi}{\ion{H}{1}}
\newcommand{\hii}{\ion{H}{2}}

\newcommand{\heii}{\ion{He}{2}}
\newcommand{\alii}{\ion{Al}{2}}

\newcommand{\cii}{\ion{C}{2}}
\newcommand{\ciii}{\ion{C}{3}}
\newcommand{\civ}{\ion{C}{4}}
\newcommand{\nni}{\ion{N}{1}}

\newcommand{\nv}{\ion{N}{5}}
\newcommand{\oi}{\ion{O}{1}}

\newcommand{\ovi}{\ion{O}{6}}

\newcommand{\pii}{\ion{P}{2}}

\newcommand{\sii}{\ion{S}{2}}
\newcommand{\Siii}{\ion{S}{3}}
\newcommand{\siv}{\ion{S}{4}}

\newcommand{\siii}{\ion{Si}{2}}
\newcommand{\siiii}{\ion{Si}{3}}
\newcommand{\siiv}{\ion{Si}{4}}

\newcommand{\feii}{\ion{Fe}{2}}

\shortauthors{Lehner et al.}
\shorttitle{Project AMIGA: The Circumgalactic Medium of Andromeda}
\begin{document}

\title{Project AMIGA: The Circumgalactic Medium of Andromeda \footnote{Based on observations made with the NASA/ESA Hubble Space Telescope, obtained from the data archive at the Space Telescope Science Institute. STScI is operated by the Association of Universities for Research in Astronomy, Inc. under NASA contract NAS 5-26555.}}

\author[0000-0001-9158-0829]{Nicolas Lehner}
\affiliation{Department of Physics, University of Notre Dame, Notre Dame, IN 46556}

\author{Samantha C. Berek}
\altaffiliation{NSF REU student.}
\affiliation{Department of Physics, University of Notre Dame, Notre Dame, IN 46556}
\affiliation{Department of Astronomy, Yale University, New Haven, CT 06511 USA}

\author[0000-0002-2591-3792]{J. Christopher Howk}
\affiliation{Department of Physics, University of Notre Dame, Notre Dame, IN 46556}

\author[0000-0002-0507-7096]{Bart P. Wakker}
\affiliation{Department of Astronomy, University of Wisconsin Madison, WI 53706}

\author[0000-0002-7982-412X]{Jason Tumlinson}
\affiliation{Space Telescope Science Institute, 3700 San Martin Drive, Baltimore, MD, 21218}
\affiliation{Department of Physics \& Astronomy, Johns Hopkins University, 3400 N.\ Charles Street, Baltimore, MD 21218}

\author[0000-0003-1892-4423]{Edward B. Jenkins}
\affiliation{Department of Astrophysical Sciences, Princeton University,  Princeton, NJ 08544}

\author[0000-0002-7738-6875]{J. Xavier Prochaska}
\affiliation{UCO/Lick Observatory, Department of Astronomy \& Astrophysics, University of Califorinia Santa Cruz, 1156 High Street, Santa Cruz, CA 95064}

\author[0000-0001-7472-3824]{Ramona Augustin}
\affiliation{Space Telescope Science Institute, 3700 San Martin Drive, Baltimore, MD, 21218}

\author[0000-0001-9658-0588]{Suoqing Ji}
\affiliation{TAPIR, Walter Burke Institute for Theoretical Physics, California Institute of Technology, Pasadena, CA, 91125}

\author[0000-0002-4900-6628]{Claude-Andr\'e Faucher-Gigu\`ere}
\affiliation{CIERA and Department of Physics and Astronomy, Northwestern University, 2145 Sheridan Road, Evanston, IL 60208}

\author[0000-0001-7326-1736]{Zachary Hafen}
\affiliation{CIERA and Department of Physics and Astronomy, Northwestern University, 2145 Sheridan Road, Evanston, IL 60208}

\author[0000-0003-1455-8788]{Molly S.\ Peeples}
\affiliation{Space Telescope Science Institute, 3700 San Martin Drive, Baltimore, MD, 21218}
\affiliation{Department of Physics \& Astronomy, Johns Hopkins University, 3400 N.\ Charles Street, Baltimore, MD 21218}

\author[0000-0001-5817-0932]{Kat A. Barger}
\affiliation{Department of Physics \& Astronomy, Texas Christian University, Fort Worth, TX 76129}

\author[0000-0002-8518-6638]{Michelle A. Berg}
\affiliation{Department of Physics, University of Notre Dame, Notre Dame, IN 46556}

\author[0000-0002-3120-7173]{Rongmon Bordoloi}
\affil{North Carolina State University, Department of Physics, Raleigh, NC 27695-8202}

\author[0000-0002-1793-9968]{Thomas M. Brown}
\affiliation{Space Telescope Science Institute, 3700 San Martin Drive, Baltimore, MD, 21218}

\author[0000-0003-0724-4115]{Andrew J. Fox}
\affil{AURA for ESA, Space Telescope Science Institute, 3700 San Martin Drive, Baltimore, MD 21218}

\author[0000-0003-0394-8377]{Karoline M. Gilbert}
\affiliation{Space Telescope Science Institute, 3700 San Martin Drive, Baltimore, MD, 21218}
\affiliation{Department of Physics \& Astronomy, Johns Hopkins University, 3400 N.\ Charles Street, Baltimore, MD 21218}

\author[0000-0001-8867-4234]{Puragra Guhathakurta}
\affiliation{UCO/Lick Observatory, Department of Astronomy \& Astrophysics, University of Califorinia Santa Cruz, 1156 High Street, Santa Cruz, CA 95064}

\author[0000-0001-9690-4159]{Jason S. Kalirai}
\affiliation{Johns Hopkins Applied Physics Laboratory, 11100 Johns Hopkins Road, Laurel, MD 20723}

\author[0000-0002-6050-2008]{Felix J.\ Lockman}
\affiliation{Green Bank Observatory, Green Bank, WV 24944}

\author[0000-0002-7893-1054]{John M. O'Meara}
\affiliation{W.M. Keck Observatory 65-1120 Mamalahoa Highway Kamuela, HI 96743}

\author{D.J. Pisano}
\altaffiliation{Adjunct Astronomer, Green Bank Observatory.}
\affiliation{Department of Physics \& Astronomy, West Virginia University, P.O. Box 6315, Morgantown, WV 26506}
\affiliation{Center for Gravitational Waves and Cosmology, West Virginia University, Chestnut Ridge Research Building, Morgantown, WV 26505}

\author[0000-0003-3381-9795]{Joseph Ribaudo}
\affiliation{Department of Engineering and Physics, Providence College, Providence, RI, 02918}

\author{Jessica K. Werk}
\affiliation{Department of Astronomy, University of Washington, Seattle, WA 98195}

\begin{abstract}
Project AMIGA (Absorption Maps In the Gas of Andromeda)  is a large ultraviolet {\it Hubble Space Telescope}\ program, which has assembled a sample of 43 QSOs that pierce  the circumgalactic medium (CGM) of Andromeda (M31) from $R=25$ to 569 kpc (25 of them probing gas from 25 kpc to about the virial radius--$\rvir =300$ kpc--of M31). Our large sample provides an unparalleled look at the physical conditions and distribution of metals in the CGM of a single galaxy using ions that probe a wide range of gas phases (\siii, \siiii, \siiv, \cii, \civ, and \ovi, the latter being from the {\it Far Ultraviolet Spectroscopic Explorer}). We find that \siiii\ and \ovi\  have near unity covering factor maintained all the way out to  $1.2\rvir$ and $1.9\rvir$, respectively. We show that \siiii\ is the dominant ion over \siii\ and \siiv\ at any $R$. While we do not find that the properties of the CGM of M31 depend strongly on the azimuth, we show that they change remarkably around $0.3$--$0.5\rvir$, conveying that the inner regions of the CGM of M31 are more dynamic and have more complicated multi-phase gas-structures than at $R\ga 0.5\rvir$. We estimate the metal mass of the CGM within \rvir\ as probed by \siii, \siiii, and \siiv\ is $2\times 10^7$ M$_\sun$ and by \ovi\ is $>8 \times 10^7$ M$_\sun$, while the baryon mass of the $\sim 10^4$--$10^{5.5}$ K gas is $\ga 4 \times 10^{10}\,(Z/0.3\,Z_\sun)^{-1}$  M$_\sun$ within \rvir. We show that different zoom-in cosmological simulations of $L^*$ galaxies better reproduce the column density profile of \ovi\ with $R$ than \siiii\ or the other studied ions. We find that observations of the M31 CGM and zoom-in simulations of $L^*$ galaxies have both lower ions showing higher column density dispersion and dependence on $R$ than higher ions, indicating that the higher ionization structures are larger and/or more broadly distributed.
\end{abstract}


\section{Introduction}\label{s-intro}

Over the last 10 years, in particular since the installation of the Cosmic Origins Spectrograph (COS) on the {\it Hubble Space Telescope} (\hst), we have made significant leaps in empirically characterizing the circumgalactic medium (CGM) of galaxies at low redshift where a wide range of galaxy masses can be studied (see recent review by \citealt{tumlinson17}). We appreciate now that the CGM of typical star-forming or quiescent galaxies have a large share of galactic baryons and metals in relatively cool gas-phases ($10^4$--$10^{5.5}$ K) \citep[e.g.,][]{stocke13,bordoloi14,liang14,peeples14,werk14,johnson15,burchett16,prochaska17,chen19,poisson19}. We have come to understand that the CGM of galaxies at $z \la 1$ is not just filled with metal-enriched gas ejected by successive galaxy outflows, but has also a large amount of metal poor gas ($<1$--2\% solar) in which little net chemical enrichment has occurred over several billions of years \citep[e.g.,][]{ribaudo11,thom11,lehner13,lehner18,lehner19,wotta16,wotta19,prochaska17,kacprzak19,poisson19,zahedy19}. The photoionized gas around $z\la 1$ galaxies is very chemically inhomogeneous, as shown by large metallicity ranges and the large metallicity variations among kinematically distinct components in a single halo (\citealt{wotta19,lehner19}, and see also \citealt{crighton13a,muzahid15,muzahid16, rosenwasser18}). Such a large metallicity variation is not only observed in the CGM of star-forming galaxies, but also in the CGM of passive and massive galaxies where there appears to be as much cold, bound \hi\ gas as in their star-forming counterparts \citep[e.g.,][]{thom12,tumlinson13,berg19,zahedy19}.

These empirical results have revealed both expected and unexpected properties of the CGM of galaxies and they all provide new means to understand the complex relationship between galaxies and their CGM. Prior to these empirical results, the theory of galaxy formation and evolution was mostly left constraining the CGM properties indirectly by their outcomes, such as galaxy stellar mass and ISM properties. Thus the balance between outflows, inflows, recycling, and ambient gas--and the free parameters controlling them--were tuned to match the optical properties of galaxies rather than implemented directly as physically-rigorous and self-consistent models. These indirect constraints suffer from problems of model uniqueness: it is possible to match stellar masses and metallicities with very different treatments of feedback physics \citep[e.g.,][]{hummels13,liang16}.  Recent empirical and theoretical advances offer a way out of this model degeneracy. New high-resolution, zoom-in simulations employ explicit treatments of the multiple gas-phase nature and feedback from stellar population models  \citep[e.g.,][]{hopkins14,hopkins18}. It is also becoming clear that not only high resolution inside the galaxies but also in their CGM is required to capture more accurately the complex processes in the cool CGM, such as metal mixing \citep{hummels19,peeples19,suresh19,vandevoort18,corlies19}.

A significant limitation in interpreting the new empirical results in the context of new high-resolution zoom simulations is that only average properties of the CGM are robustly derived from traditional QSO absorption-line techniques for examining halo gas. In the rare cases where there is a UV-bright QSO behind a given galaxy, the CGM is typically probed along a single ``core sample" through the halo of each galaxy. These measurements are then aggregated into a statistical map, where galaxies with different inclinations, sizes, and environments are blended together and the radial-azimuthal dependence of the CGM is essentially lost.  All sorts of biases can result: phenomena that occur strongly in only a subset of galaxies can be misinterpreted as being weaker but more common, and genuine trends with mass or star formation rate can be misinterpreted as simply scatter with no real physical meaning (see also \citealt{bowen16}). Simulations also suggest that time-variable winds, accretion flows, and satellite halos can induce strong halo-to-halo variability, further complicating interpretation \citep[e.g.][]{hafen17,oppenheimer18a}. Observational studies of single galaxy CGM with multiple sightlines are therefore required to gain spatial information on the properties of the CGM. 

Multi-sightline information on the CGM of single galaxies has been obtained in a few cases from binary or multiple (2--4) grouped QSOs behind foreground galaxies \citep[e.g.,][]{bechtold94,martin10,keeney13,bowen16}, gravitationally-lensed quasars \citep[e.g.,][]{smette92,rauch01,ellison04,lopez05,zahedy16,rubin18,kulkarni19}, giant gravitational arcs \citep[e.g.,][]{lopez19}, or extended bright background objects observed with integral field units \citep[e.g.,][]{peroux18}. These observations provide better constraints on the kinematic relationship between the CGM gas and the galaxy and on the size of CGM structures. However, they yield limited information on the gas-phase structure owing to a narrow range of ionization diagnostics or poor quality spectral data. Thus, it remains unclear how tracers of different gas phases vary with projected distance $R$ or azimuth $\Phi$ around the galaxy. 

The CGM that has been pierced the most is that of the Milky Way (MW), with several hundred QSO sightlines \citep{wakker03,shull09,lehner12,putman12,richter17} through the Galactic halo. However, our position as observers within the MW disk severely limits the interpretation of these data (especially for the extended CGM, see \citealt{zheng15, zheng20}) and makes it difficult to compare with observations of other galaxies. 

With a virial radius that spans over $30\degr$ on the sky, M31 is the only $L^*$ galaxy where we can access more than 5 sightlines without awaiting the next generation of UV space-based telescope (e.g., \citealt{luvoir19}). With current UV capabilities, it is the only single galaxy where we can study the global distribution and properties of metals and baryons in some detail.

In our pilot study (\citealt{lehner15}, hereafter \citetalias{lehner15}), we mined the \hst/COS G130M/G160M archive available at the \textit{Barbara A. Mikulski Archive for Space Telescopes} (MAST) for sightlines piercing the M31 halo within a projected distance of $\sim 2 \rvir$ (where $R_{\rm vir}=300$ kpc for M31, see below). There were 18 sightlines, but only 7 at projected distance $R \la \rvir$. Despite the small sample, the results of this study were  quite revealing, demonstrating a high covering factor (6/7) of M31 CGM absorption by \siiii\ (and other ions including, e.g., \civ, \siii) within $1.1\rvir$ and a covering factor near zero (1/11) between $1.1 \rvir$ and $2 \rvir$. We found also a drastic change in the ionization properties, as the gas is more highly ionized at $R \sim \rvir $ than at $R<0.2\rvir$. The \citetalias{lehner15} results strongly suggest that the CGM of M31 as seen in absorption of low ions (\cii, \siii) through intermediate (\siiii, \siiv) and high ions (\civ, \ovi) is very extended out to at least the virial radius. However, owing to the small sample within \rvir, the variation of the column densities ($N$) and covering factors (\fc) with projected distances and azimuthal angle remain poorly constrained.

Our Project AMIGA (Absorption Maps In the Gas of Andromeda) is a large \hst\ program (PID: 14268, PI: Lehner) that aims to fill the CGM with 18 additional sightlines at various $R$ and $\Phi$ within $1.1 \rvir$ of M31 using high-quality COS G130M and G160M observations, yielding a sample of 25 background QSOs probing the CGM of M31. We have also searched MAST for additional QSOs beyond $1.1 \rvir$ up to $R=569$ kpc from M31 ($\sim 1.9 \rvir$) to characterize the extended gas around M31 beyond its virial radius. This archival search yielded 18 suitable QSOs. Our total sample of 43 QSOs probing the CGM of a single galaxy from 25 to 569 kpc is the first to explore simultaneously the azimuthal and radial dependence of the kinematics, ionization level, surface-densities, and mass of the CGM of a galaxy over its entire virial radius and beyond. With these observations, we can also test how the CGM properties derived from one galaxy using multiple sightlines  compares with a sample of galaxies with single sightline information and we can directly compare the results with cosmological zoom-in simulations.

With the COS G130M and G160M wavelength coverage, the key ions in our study are \cii, \civ, \siii, \siiii, \siiv\ (other ions and atoms include \feii, \sii, \oi, \nni, \nv, but are typically not detected, although the limit on \oi\ constrains the level of ionization). These species span ionization potentials from $<1$ to $\sim$4 Rydberg and thus trace neutral to highly ionized gas at a wide range of temperatures and densities. We have also searched the {\it Far Ultraviolet Spectroscopic Explorer}\ (\fuse) to have coverage of \ovi, which resulted in 11 QSOs in our sample having both COS and \fuse\ observations. The \hi\ \lya\ absorption can unfortunately not be used because the MW dominates the entire \lya\ absorption. Instead we have obtained deep \hi\ 21-cm observations with the Robert C. Byrd Green Bank Telescope (GBT) toward all the targets in our sample and several additional ones (\citealt{howk17}, hereafter \citetalias{howk17}), showing no detection of any \hi\ down to a level $\mnhi \simeq 4\times 10^{17}$ \cmm\  ($5\sigma$; averaged over an area that is 2 kpc at the distance of M31). Our non-detections place a limit on the covering factor of such optically thick \hi\ gas around M31 to $\fc\ < 0.051$ (at 90\% confidence level) for $R\la \rvir$.

This paper is organized as follows. In \S\ref{s-data}, we provide more information about the criteria used to assemble our sample of QSOs and explain the various steps to derive the properties (velocities and column densities) of the absorption. In that section, we also present the line identification for each QSO spectrum, which resulted in the identification of 5,642 lines. In \S\ref{s-ms}, we explain in detail how we remove the foreground contamination from the Magellanic Stream (MS, e.g., \citealt{putman03,nidever08,fox14}), which extends to the M31 CGM region of the sky with radial velocities that overlap with those expected from the CGM of M31. For this work, we have developed a more systematic and automated methodology than in \citetalias{lehner15} to deal with this contamination. In \S\ref{s-dwarfs}, we present the sample of the M31 dwarf satellite galaxies to which we compare the halo gas measurements. In \S\ref{s-properties}, we derive the empirical properties of the CGM of M31 including how the column densities and velocities vary with $R$ and $\Phi$, the covering factors of the ions and how they change with $R$, and the metal and baryon masses of the CGM of M31. In \S\ref{s-disc}, we discuss the results derived in \S\ref{s-properties} and compare them to observations from the COS-Halos survey \citep{tumlinson13,werk14} and to state-of-the-art cosmological zoom-ins from in particular the Feedback in Realistic Environments (FIRE, \citealt{hopkins19}) and and Figuring Out Gas \& Galaxies In Enzo (FOGGIE, \citealt{peeples19})  simulations projects. In \S\ref{s-sum}, we summarize our main conclusions.

To properly compare to other work, and to simulations, we must estimate a characteristic radius for M31. We use the radius $R_{200}$ enclosing a mean overdensity of $200$ times the critical density: $R_{200} = (3 M_{\rm 200} / 4 \pi \, \Delta\; \rho_{\rm crit})^{1/3}$, where $\Delta = 200$ and $\rho_{\rm crit}$ is the critical density. For M31, we adopt $M_{200} = 1.26 \times 10^{12}$  M$_\sun$ (e.g., \citealt{watkins10,vandermarel12}), implying $R_{200} \simeq 230$ kpc. For the virial mass and radius (\Mvir\ and \rvir), we use the definition that follows from the top-hat model in an expanding universe with a cosmological constant where $\Mvir = 4\pi/3 \; \rho_{\rm vir} \rvir^3$ where the virial density $ \rho_{\rm vir} = \Delta_{\rm vir} \Omega_{\rm m} \rho_{\rm crit}$ \citep{klypin11,vandermarel12}. The average virial overdensity is $\Delta_{\rm vir} = 360$ assuming a cosmology with $h=0.7$ and $\Omega_{\rm m} = 0.27$ \citep{klypin11}. Following, e.g, \citet{vandermarel12}, $\Mvir \simeq 1.2 M_{200} \simeq 1.5 \times 10^{12}$  M$_\sun$ and $\rvir \simeq 1.3 R_{200} \simeq 300$ kpc. The escape velocity at $R_{200}$ for M31 is then $v_{200}\simeq 212$ \km. A distance of M31 of $d_{\rm M31} = 752 \pm 27$ kpc based on the measurements of Cepheid variables \citep{riess12} is assumed throughout. We note that this distance is somewhat smaller than the other often adopted distance of M31 of 783 kpc \citep[e.g.,][]{brown04,mcconnachie05}, but for consistency with our previous survey as well as the original design of Project AMIGA, we have adopted $d_{\rm M31} = 752$ kpc. All the projected distances were computed using the three dimensional separation (coordinates of the target and distance of M31).

\begin{figure*}[tbp]
\epsscale{0.9}
\plotone{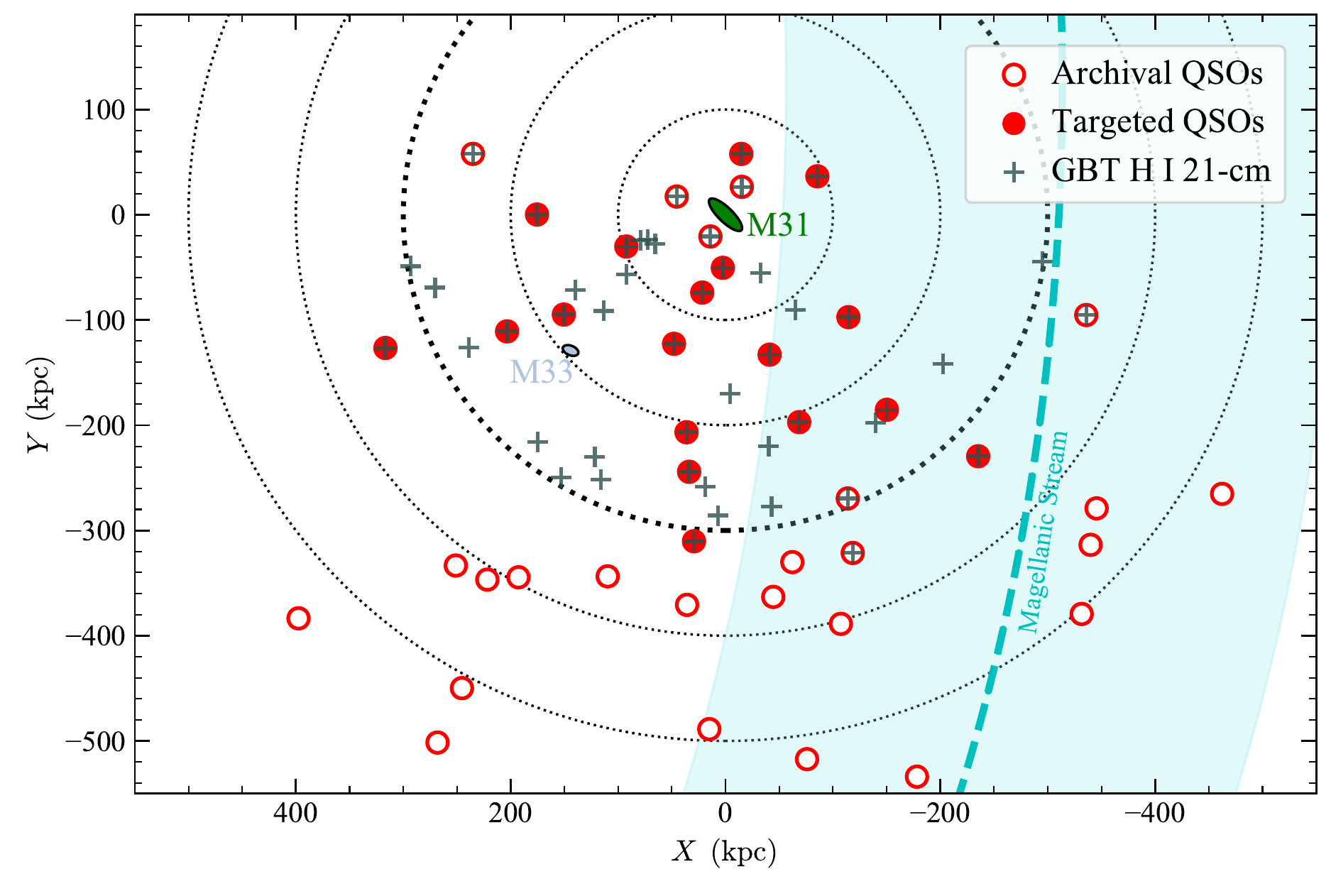}
\caption{Locations of the Project AMIGA pointings relative to the M31--M33 system. The axes show the angular separations converted into physical coordinates relative to the center of M31. North is up and east to the left. The 18 sightlines from our large \hst\ program are in red filled circles; the 25 archival COS targets are in open red circles. Crosses show the GBT \hi\ 21-cm observations described in  \citetalias{howk17}. Dotted circles show impact parameters $R = 100$, 200, 300, 400, 500 kpc. $\rvir = 300$ kpc is marked with a heavy dashed line. The sizes and orientations of the two galaxies are taken from RC3 \citep{devaucouleurs91} and correspond to the optical $R_{25}$ values. The light blue dashed line shows the plane of the Magellanic Stream ($b_{\rm MS} =0\degr$) as defined by \citet{nidever08}. The shaded region within $b_{\rm MS} \pm 20\degr$ of the MS midplane is the approximate region where we identify most of the MS absorption components  contaminating the M31 CGM absorption (see \S\ref{s-ms}).\label{f-map}}
\end{figure*}

\section{Data and Analysis}\label{s-data}
\subsection{The Sample}\label{s-sample}

The science goals of our \hst\ large program require estimating the spatial distributions of the kinematics and metal column densities of the M31 CGM gas within about $1.1\rvir$ as a function of azimuthal angle and impact parameter. The search radius was selected based on our pilot study where we detected M31 CGM gas up to $\sim 1.1\rvir$, but essentially not beyond \citepalias{lehner15} (a finding that we revisit in this paper with a larger archival sample, see below). With our \hst\ program, we observed 18 QSOs at $R\la 1.1\rvir$ that were selected to span the M31 projected major axis, minor axis, and intermediate orientations. The sightlines do not sample the impact parameter space or azimuthal distribution at random. Instead, the sightlines were selected to probe the azimuthal variations systematically. The sample was also limited by a general lack of identified UV-bright AGNs behind the northern half of M31ʼs CGM owing to higher foreground MW dust extinction near the plane of the Milky Way disk. Combined with 7 archival QSOs, these sightlines probe the CGM of M31 in azimuthal sectors spanning the major and minor axes with a radial sample of 7--10 QSOs in each $\sim$100 kpc bin in $R$.

In addition to target locations, the 18 QSOs were optimized to be the brightest available QSOs (to minimize exposure time) and to have the lowest available redshifts (in order to minimize the contamination from unrelated absorption from high redshift absorbers). For targets with no existing UV spectra prior to our observations, we also required that the GALEX NUV and FUV flux magnitudes are about the same to minimize the likelihood of an intervening Lyman limit system (LLS) with optical depth at the Lyman limit $\tau_{\rm LL}>2$. An intervening LLS could absorb more or all of the QSO flux we would need to measure foreground absorption in M31. This strategy successfully kept QSOs with intervening LLS out of the sample. 

As we discuss below and as detailed by \citetalias{lehner15}, the MS crosses through the M31 region of the sky at radial velocities that can overlap with those of M31 (see also \citealt{nidever08,fox14}). To understand the extent of MS contamination and the extended gas around M31 beyond the virial radius, we also searched for targets beyond $1.1\rvir$  with COS G130M and/or G160M data. This search identified another 18 QSOs at $1.1\la R/\rvir <1.9$ that met the data quality criteria for inclusion in the sample\footnote{This search found eight additional targets at $R>1.6 \rvir$ that are not included in our sample. SDSSJ021348.53+125951.4, 4C10.08, LBQS0052-0038 were excluded because of low S/N in the COS data. NGC7714 has smeared absorption lines.  LBQS0107-0232/3/5 lie at $z_{\rm em}\simeq 0.7$--1 and have extremely complex spectra. HS2154+2228 at $z_{\rm em} = 1.29$ has no G130M wavelength coverage making the line identification highly uncertain. \label{foot-reason}}. Our final sample consists of 43 sightlines probing the CGM of M31 from 25 to 569 kpc; 25 of these probe the M31 CGM from 25 to 342 kpc, corresponding to $0.08 - 1.1 \rvir$. Fig.~\ref{f-map} shows the locations of each QSO in the M31--M33 system (the filled circles being targets obtained as part of our \hst\ program PID: 14268 and the open circles being QSOs with archival \hst\ COS G130M/G160M data), and  Table~\ref{t-sum} lists the properties of our sample QSOs ordered by increasing projected distances from M31. In this table, we list the redshift of the QSOs (\zem), the J2000 right ascension (RA) and declination (Dec.), the MS coordinate (\lms, \bms, see \citealt{nidever08} for the definition of this coordinate system), the radially ($R$) and cartesian ($X,Y$) projected distances, the program identification of the \hst\ program (PID), the COS grating used for the observations of the targets, and the signal-to-noise ratio (SNR) per COS resolution element of the COS spectra near the \siiii\ transition (except otherwise stated in the footnote of this table).  

\subsection{UV Spectroscopic Calibration}\label{s-calib}

To search for M31 CGM absorption and to determine the properties of the CGM gas, we use ions and atoms that have their wavelengths in the UV (see \S\ref{s-prop}). Any transitions with $\lambda>1144$ \AA\ are in the \hst\ COS bandpass. All the targets in our sample were observed with \hst\ using the COS G130M grating ($R_\lambda \approx 17,000$). All the targets observed as part of our new \hst\ program were also observed with COS G160M, and all the targets but one within $R<1.1 \rvir$ have both G130M and G160M wavelength coverage. 

We also searched for additional archival UV spectra in MAST, including the \fuse\ ($R_\lambda \approx 15,000$) archive to complement the gas-phase diagnostics from the COS spectra with information from the \ovi\ absorption. We use the \fuse\ observations for 11 targets with adequate SNR near \ovi\ (i.e., $\ga 5$): RX\_J0048.3+3941, IRAS\_F00040+4325, MRK352, PG0052+251, MRK335, UGC12163, PG0026+129, MRK1502, NGC7469, MRK304, PG2349-014 (only the first 6 targets in this list are at $R\la 1.1 \rvir$). We did not consider \fuse\ data for quasars without COS observations because the available UV transitions in the far-UV spectrum (\ovi, \cii, \ciii, \siii, \feii) are either too weak or too contaminated to allow for a reliable identification of the individual velocity components in their absorption profiles. 

There are also 3 targets (MRK335, UGC12163, and NGC7469) with \hst\ STIS E140M ($R_\lambda \simeq 46,500$) observations that provide higher resolution information.\footnote{For 2 targets, we also use COS G225M (3C454.3) and FOS NUV (3C454.3, PG0044+030) observations to help with the line identification (see \S\ref{s-lineid}). The data processing follows the same procedure as the other data.}

Information on the design and performance of COS, STIS, \fuse\ can be found in \citet{green12}, \citet{woodgate98}, and \citet{moos00}, respectively. For the \hst\ data, we use the pipeline-calibrated final data products available in MAST. The \hst\ STIS E140M data have an accurate wavelength calibration and the various exposure and echelle orders are combined into a single spectrum by interpolating the photon counts and errors onto a common grid, adding the photon counts and converting back to a flux.

The processing of the \fuse\ data is described in detail by \citet{wakker03} and \citet{wakker06}. In short, the spectra are calibrated using version 2.1 or version 2.4 of the \fuse\ calibration pipeline. The wavelength calibration of \fuse\ can suffer from stretches and misalignments. To correct for residual wavelength shifts, the central velocities of the MW interstellar lines are determined for each detector segment of each individual observation. The \fuse\ segments are then aligned with the interstellar velocities implied by the STIS E140M spectra or with the velocity of the strongest component seen in the 21-cm \hi\ spectrum. Since the \ovi\ absorption can be contaminated by H$_2$ absorption, we remove this contamination following the method described in \citet{wakker06}. This contamination can be removed fairly accurately with an uncertainty of about $\pm 0.1$ dex on the \ovi\ column density \citep{wakker03}.

For the COS G130M and G160M spectra, the spectral lines in separate observations of the same target are not always aligned, with misalignments of up to $\pm 40$ \km\ that varying as function of wavelength. This is a known issue that has been reported previously \citep[e.g.,][]{savage14,wakker15}. While the COS team has improved the wavelength solution, we find that this problem can still be present sometimes. 
Since accurate alignment is critical for studying multiple gas-phases probed by different ions and since there is no way to determine {\it a priori} which targets are affected, we uniformly apply the \cite{wakker15} methodology to coadd the different exposures of the COS data to ensure proper alignment of the absorption lines. In short, we identify the various strong ISM and IGM weak lines and record the component structures and identify possible contamination of the ISM lines by IGM lines. We cross-correlate each line in each exposure, using a $\sim$3 \AA\ wide region, and apply a shift as a function of wavelength to each spectrum. To determine the absolute wavelength calibration, we compare the velocity centroids of the Gaussian fits to the interstellar UV absorption lines (higher velocity absorption features being Gaussian fitted separately) and the \hi\ emission observed from our 9$\arcmin$ GBT \hi\ survey \citepalias{howk17} or otherwise from 21-cm data from the Leiden/Argentine/Bonn (LAB) survey \citep{kalberla05} or the Parkes Galactic All Sky Survey (GASS) \citep{kalberla10}. The alignment is coupled with the line identification into an iterative process to simultaneously determine the most accurate alignment and line identification (see \S\ref{s-lineid}). To combine the aligned spectra, we add the total counts in each pixel and then convert back to flux, using the average flux/count ratio at each wavelength (see also \citealt{tumlinson11a,tripp11}); the flux error is estimated from the Poisson noise implied by the total count rate.

\subsection{Line Identification}\label{s-lineid}
We are interested in the velocity range  $-700 \le \vlsr\ \le -150$ \km\ where absorption from the M31 CGM may occur (see \S\ref{s-prop} for the motivation of this velocity range). It is straightforward to identify M31 absorption or its absence in this pre-defined velocity range, but we must ensure that there is either no contamination from higher redshift absorbers, or if there is, that we can correct for it.

For ions with multiple transitions, it is relatively simple to determine whether contamination is at play by comparing the column densities and the shapes of the velocity profiles of the available transitions. The profiles of atoms or ions with a single transition can be compared to other detected ions to check if there is some obvious contamination in the single transition absorption. However, some contamination may still remain undetected if it directly coincides with the absorption under consideration. Furthermore, when only a single ion with a single transition is detected (\siiii\ $\lambda$1206 being the prime example), the only method that determines if it is contaminated or not is to undertake a complete line identification of all absorption features in each QSO spectrum.

For the 18 targets in our large \hst\ program, our instrument setup ensures that we have the complete wavelength coverage with no gap between 1140 and 1800 \AA. As part of our target selection, we also favor QSOs at low redshift (44\% are at $\zem \le 0.1$, 89\% at $\zem \le 0.3$). This assures that \lya\ remains in the observed wavelength range out to the redshift of the QSO (\lya\ redshifts out the long end of the COS band at $z = 0.48$) and greatly reduces the contamination from EUV transitions in the COS bandpass. The combination of wavelength coverage and low QSO redshift ensures the most accurate line identification.  At $R<351$ kpc (i.e., $\la 1.2 \rvir$), 93\% have  \lya\ coverage down to $z = \zem$  that remains in the observed wavelength range (one target has only observation of G130M and another QSO is at $z=0.5$, see Table~\ref{t-sum}). On the other hand, for the targets at $R>351$ kpc, the wavelength coverage is not as complete over 1140--1800 \AA\ (55\% of the QSOs have only 1 COS grating---all but one have G130M, and 4 QSOs  have $\zem \ga 0.48$). We note that the QSOs of 6/10 G130M observations have $\zem <0.17$, setting all the \lya\ transitions within the COS G130M bandpass. 

The overall line identification process is as follows. First, we mark all the ISM absorption features (i.e., any absorption that could arise from the MW or M31) and the velocity components (which is done as part of the overall alignment of the spectra, see \S\ref{s-calib}). Local (approximate) continua are fitted near the absorption lines to estimate the equivalent widths ($W_\lambda$) and their ratios for ions with several transitions are checked to determine if any are potentially contaminated. We then search for any absorption features at $z = \zem$, again identifying any velocity component structures in the absorption. We then identify possible \lya\ absorption and any other associated lines (other \hi\ transitions and metal transitions) from the redshift of QSO down to $z=0$. In each case, if there are simultaneous detections of \lya, \lyb, and/or \lyg\ (and weaker transitions), we check that the equivalent width ratios are consistent. If there are any transitions left unidentified, we check whether it could be \ovi\ $\lambda\lambda$1031, 1037 as this doublet can be sometimes detected without any accompanying \hi\ \citep{tripp08}. Finally we check that the alignment in each absorber with multiple detected absorption lines is correct or whether it needs some additional adjustment.

In the region $R\la 1.1\rvir$ and for 84\% of the sample at any $R$, we believe the line identifications are reliable and accurate at the 98\% confidence level. In the Appendix, we provide some additional information regarding the line identification, in particular for the troublesome cases. We also make available in a machine-readable format the full line identification for all the targets listed in Table~\ref{t-sum} (see Appendix~\ref{a-lineid}).

\subsection{Determination of the Properties of the Absorption at $-700 \le v_{\rm LSR} \le -150$ \km }\label{s-prop}

Our systematic search window for absorption that may be associated with the CGM of M31 is \dvrange\ \citepalias{lehner15}.  The $-700$ \km\ cutoff corresponds to about $-100$ \km\ less than the most negative velocities from the rotation curve of M31 ($\sim -600$ \km, see \citealt{chemin09}). The $-150$ \km\ cutoff is set by the MW lines that dominate the absorption in the velocity range $-150 \la \vlsr\ \la +50$ \km. The $-100 \la \vlsr \la -50$ \km\ range is dominated by low and intermediate-velocity clouds that are observed in and near the Milky Way disk. Galactic high-velocity clouds (HVCs) down to velocities  $\vlsr \sim -150$ \km\ further above the MW disk have also been observed toward distant Galactic halo stars in the general direction of M31  \citep{lehner15,lehner12,lehner11a}. Since the M31 disk rotation velocities extend to about $-150$ \km\ in the northern tip of M31, there is a small window that is inaccessible for studying the CGM of M31 (see also \citealt{lehner15} and \S\ref{s-dwarfs-vel}).

To search for M31 CGM gas and determine its properties, we use  the following atomic and ionic transitions: \oi\ $\lambda$1302,  \cii\ $\lambda\lambda$1036, 1334, \civ\ $\lambda\lambda$1548, 1550, \siii\ $\lambda\lambda$1190, 1193, 1260, 1304, 1526 \siiii\ $\lambda$1206,  \siiv\ $\lambda\lambda$1393, 1402, \ovi\ $\lambda$1031, \feii\ $\lambda\lambda$1144, 1608, \alii\ $\lambda$1670. We also report results (mostly upper limits on column densities) for \nv\ $\lambda\lambda$1238, 1242, \nni\ $\lambda$1199 (\nni\ $\lambda\lambda$1200, 1201 being typically blended in the velocity range of interest $-700 \le \vlsr\ \le -150$ \km), \pii\ $\lambda$1301, \Siii\ $\lambda$1190, and \sii\ $\lambda\lambda$1250, 1253, 1259. 

To determine the column densities and velocities of the absorption, we use the apparent optical depth (AOD) method (see \S\ref{s-aod}), but in the Appendix~\ref{s-pf} we confront the AOD results with measurements from Voigt profile fitting (see also \S\ref{s-gen-comments}). As much as possible at COS resolution, we derive the properties of the absorption in individual components. Especially toward M31, this is important since along the same line of sight in the velocity window \dvrange, there can be multiple origins of the gas (including the CGM of M31 or MS, see Fig.~\ref{f-map} and \citetalias{lehner15}) as we detail in \S\ref{s-ms}. However, the first step to any analysis of the absorption imprinted on the QSO spectra is to model the QSO's continuum.

\subsubsection{Continuum Placement}\label{s-continuum}
To fit the continuum near the ions of interest, we generally use the automated continuum fitting method developed for the COS CGM Compendium (CCC, \citealt{lehner18}). Fig. 3 in \citet{lehner18} shows an example of an automatic continuum fit. In short, the continuum is fitted near the absorption features using Legendre polynomials. A velocity region of about $\pm$1000--2000 \km\ around the relevant absorption transition is initially considered for the continuum fit, but could be changed depending on the complexity of the continuum placement in this region. In all cases the interval for continuum fitting is never larger than $\pm$2000 \km\ or smaller than $\pm$250 \km. Within this pre-defined region, the spectrum is broken into smaller sub-sections and then rebinned. The continuum is fitted to all pixels that did not deviate by more than $2\sigma$ from the median flux, masking pixels from the fitting process that may be associated with small-scale absorption or emission lines. Legendre polynomials of orders between 1 and 5 are fitted to the unmasked pixels, with the goodness of the fit determining the adopted polynomial order. Typically the adopted polynomials are of orders between 1 and 3 owing to the relative simplicity of the QSO continua when examined over velocity regions of 500--4000 \km. The only systematic exception is \siiii\ where the polynomial order is always between 2--3 and 5 owing to this line being in the wing of the broad local \lya\ absorption profile.

This procedure is applied to our pre-defined set of transitions, with the continuum defined locally for each. Each continuum model is visually inspected for quality control. In a few cases, the automatic continuum fitting fails owing to a complex continuum (e.g., near the peak of an emission line or where many absorption lines were present within the pre-defined continuum window). In these cases, we first try to adjust the velocity interval of the spectrum to provide better-constrained fits; if that still fails, we manually select the continuum region to be fitted. 

\begin{figure*}[tbp]
\epsscale{1}
\plotone{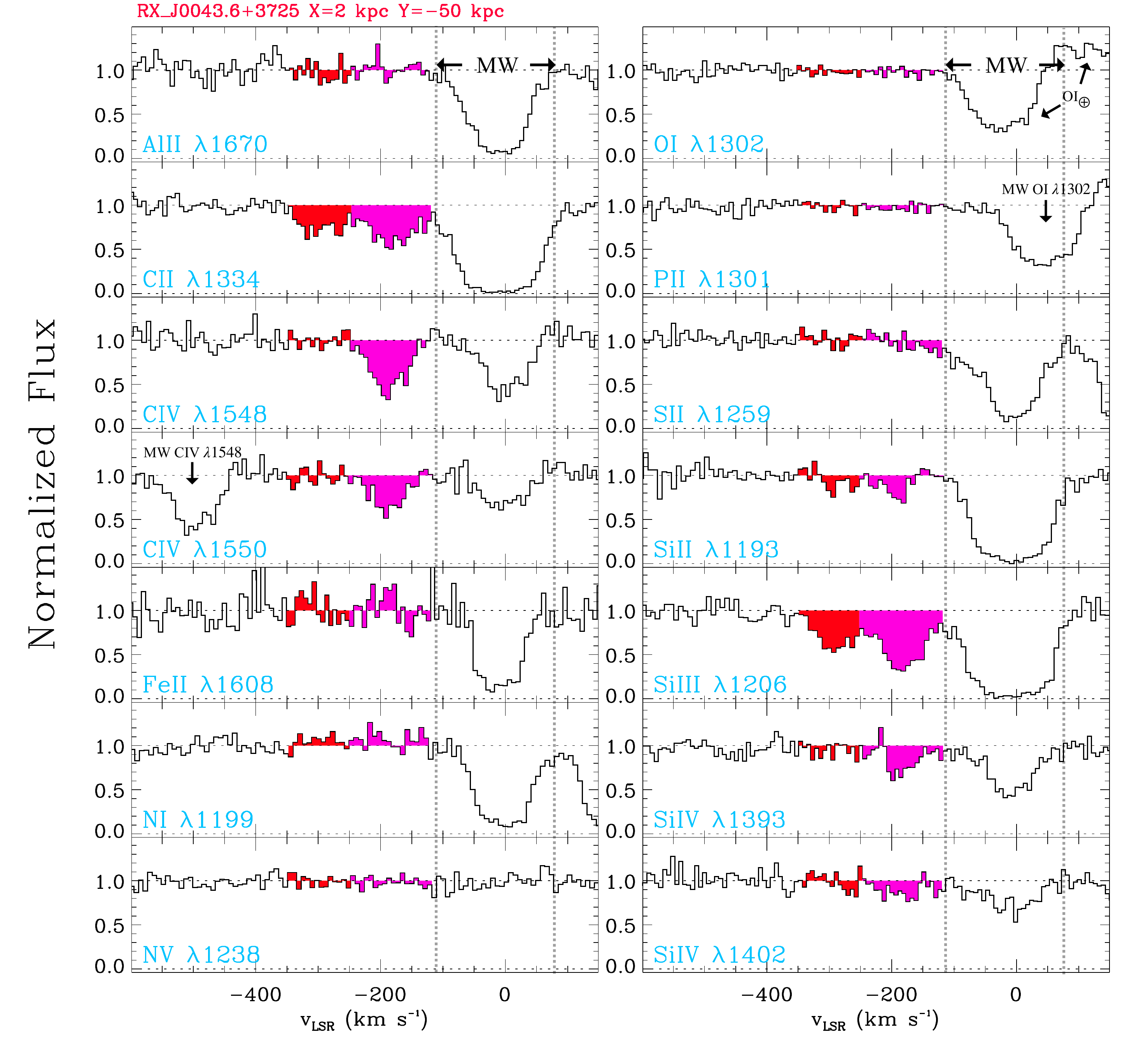}
\caption{Example of normalized absorption lines as a function of the LSR velocity toward RX\_J0043.6+3725 showing the typical atoms and ions probed in our survey. High negative velocity components likely associated with M31 are shown in colors, and each color represents a different component identified at the COS G130M-G160M resolution. In this case, significant absorption is observed in the two identified components  in \cii, \siii, and \siiii. Higher ions (\siiv, \civ) are observed in only one of the components, showing a change in the ionization properties with velocity. Some species are not detected, but their limits can still be useful in assessing the physical properties of the gas. The MW absorption is indicated between the two vertical dotted lines and is observed  in all the species but \nv. At $\vlsr \ga -100$ \km, airglow emission lines can contaminate \oi, and hence the MW absorption is contaminated, but typically that is not an issue for the surveyed velocity range \dvrange. 
\label{f-example-spectrum}}
\end{figure*}

\subsubsection{Velocity Components and AOD Analysis}\label{s-aod}

The next step of the analysis is to determine the velocity components and integrate them to determine the average central velocities and column densities for each absorption feature.  In Fig.~\ref{f-example-spectrum}, we show an example of the normalized velocity profiles. In the supplemental material, we provide a similar figure for each QSO in our sample. Although we systematically search for absorption in the full velocity range \dvrange, the most negative velocity of detected absorption in our sample is $\vlsr= -508$ \km; that is, we do not detect any M31 absorption in the range $-700 \la \vlsr \la -510 $ \km. In Fig.~\ref{f-example-spectrum}, MW absorption at $-100 \la \vlsr \la 100$ \km\ is clearly seen in all species but \nv. Absorption observed in the \dvobs\ that is not color-coded is produced by higher-redshift absorbers or other MW lines.

To estimate the column density in each observed component, we use the AOD method \citep{savage91}. In this method, the absorption profiles are converted into apparent optical depth per unit velocity, $\tau_a(v) = \ln[F_{\rm c}(v)/F_{\rm obs}(v)]$,  where $F_c(v)$ and $F_{\rm obs}(v)$ are the modeled continuum and observed fluxes as a function of velocity.  The AOD, $\tau_a(v)$, is related to the apparent column density per unit velocity, $N_a(v)$, through the relation $N_a(v) = 3.768 \times 10^{14} \tau_a(v)/(f \lambda(\mbox{\AA})$)  ${\rm cm}^{-2}\,({\rm km\,s^{-1}})^{-1}$, where $f$ is the oscillator strength of the transition and $\lambda$ is the wavelength in \AA. The total column density is obtained by integrating the profile over the pre-defined velocity interval, $N = \int_{v_1}^{v_2} N_a(v) dv $, where  $[v_1,v_2]$ are the boundaries of the absorption. We estimate the line centroids with the first moment of the AOD $v_a = \int v \tau_a(v) dv/\int \tau_a(v)dv$ \km.  As part of this process, we also estimate the equivalent widths, which we use mainly to determine if the absorption is detected at the $\ge 2\sigma$ level.  In cases where the line is not detected at $\ge 2\sigma$ significance, we quote a 2$\sigma$ upper limit on the column density, which is defined as twice the 1$\sigma$ error derived for the column density assuming the absorption line lies on the linear part of the curve of growth.

For features that are detected above the $2\sigma$ level, the estimated column densities are stored for further analysis. Since we have undertaken a full identification of the absorption features in each spectrum (see \S\ref{s-lineid}, Appendix \ref{a-lineid}), we can reliably assess if a given transition is contaminated using in particular the conflict plots described in the Appendix (see Appendix \ref{a-conflicplot}).  If there is evidence of some line contamination and several transitions are available for this ion (e.g., \siii, \siiv, \civ), we exclude it from our list. 

We find contamination affects the \siiii\ and \cii\ in the velocity range \dvrange\ in a few rare cases (6 components of \siiii\ and 3 components of \cii\ $\lambda$1334).\footnote{Toward RX\_J0048.3+3941, \cii\ $\lambda$1334 is contaminated in the third component, but \cii\ $\lambda$1036 is available to correct for it in this case.}. For all but one of these contaminated \siiii\ components, we can correct the contamination because the interfering line is a Lyman series line from a higher redshift and the other \hi\ transitions constrain the equivalent width of the contamination. The one case we cannot correct this way is the $-340$ \km\ component toward PHL1226 (see also Appendix~\ref{a-lineid}), which is associated with the MS.  In the footnote of Table~\ref{t-results}, we list the ions that are found to be contaminated at some level. For any column density that is corrected for contamination, the typical correction error is about 0.05--0.10 dex depending on the level of contamination as well as the SNRs of the spectrum in that region.

The last step is to check for any unresolved saturation. When the absorption is clearly saturated (i.e., the flux level reaches zero-flux in the core of the absorption), the line is automatically marked as saturated and a lower limit is assigned to the column density. In \S\ref{s-ms}, we will show how we separate the MS from the M31 CGM absorption, but we note that only the \siiii\ components associated with the MS and the MW have their absorption reaching zero-flux level, not the components associated with the CGM of M31.

When the flux does not reach a zero-flux level, the procedure for checking saturation depends on the number of transitions for a given ion or atom. We first consider ions with several transitions (\siii, \civ, \siiv, sometimes \cii) since they can provide information about the level of saturation for a given peak optical depth. For ions with several transitions, we compare the column densities with different $f\lambda$-values to determine whether there is a systematic decrease in the column density as $f\lambda$ increases. If there is not, we estimate the average column density using all the available measurements and propagate the errors using a weighted mean. For the \siii\ transitions, \siii\ $\lambda$1526 shows no evidence for saturation when detected based on the comparison with stronger transitions while \siii\ $\lambda$1260 or $\lambda$1193 can be saturated if the peak optical $\tau_a \ga 0.9$.  For doublets (e.g., \civ, \siiv), we systematically check if the column densities of each transition agree within $1\sigma$ error; if they do not and the weak transition gives a higher value (and there is no contamination in the weaker transition), we correct for saturation following the procedure discussed in \citet{lehner18} (and see also \citealt{savage91}). For \civ\ and \siiv, there is rarely any evidence for saturation (we only correct once for saturation of \civ\ in the third component observed in the MRK352 spectrum; in that component the peak optical $\tau_a \sim 0.9$). For single strong transitions (in particular \siiii\ and often \cii), if the peak optical depth is $\tau_a >0.9$, we conservatively flag the component as saturated and adopt a lower limit for that component. We adopt $\tau_a >0.9$ as the threshold for saturation based on other ions with multiple transitions (in particular \siii) where the absorption starts to show some saturation at this peak optical depth.

To estimate how the column density of silicon varies with $R$ (which has a direct consequence for the CGM mass estimates derived from silicon in \S\S\ref{s-nsi-vs-r} and \ref{s-mass}), it is useful to assess the level of saturation of \siiii, which is the only silicon ion that cannot be directly corrected for saturation\footnote{Some of the \siii\ transitions (especially, \siii\ $\lambda\lambda$1193, 1260) have evidence for saturation, but weaker transitions are always available (e.g., \siii\ $\lambda$1526), and therefore we can determine a robust value of the column density of \siii.}. The lower limits of the \siiii\ components associated with the CGM of M31 are mostly observed at $R\la 140$ kpc (only 2 are observed at $R>140$ kpc), but they do not reach zero-flux level; these components are conservatively marked as saturated because their peak apparent optical depth is $\tau_a > 0.9$ (not because $\tau_a \gg 2$) and because the comparison between the different \siii\ transitions show in some cases evidence for saturation (see above).  Hence the true values of the column densities of these saturated components is most likely higher than the adopted lower-limit values but are very unlikely to be overestimated by a factor $\gg 3$--4. We can estimate how large the saturation correction for \siiii\ might be using the strong \siii\ lines (e.g., \siii\ $\lambda$1193 or \siii\ $\lambda$1260) compared to the weaker ones (e.g., \siii\ $\lambda$1526). Going through the 8 sightlines showing some saturation in  the components of \siiii\ associated with the CGM of M31 (see Table~\ref{t-results}), for all the targets beyond 50 kpc, the saturation correction is likely to be small $<0.10$--0.15 dex based on the fact that many show no evidence of saturation in \siii\ $\lambda$1260 (when there is no contamination for this transition) or \siii\ $\lambda$1193. On the other hand, for the two most inner targets, the saturation correction is at least 0.3 dex and possibly as large as 0.6 dex based on the column density comparison between saturated \siii\ and weaker, unsaturated transitions. The latter would put $N_{\rm Si}\simeq 14.5$ close to the maximum values derived with photoionization modeling in the COS-Halos sample (see \S\ref{s-coshalos}). Therefore for the components associated with the CGM of M31 at $R>50$ kpc when we estimate the functional form of $N_{\rm Si}$ with $R$, we adopt an increase of 0.1 dex of the lower limits. For the two inner targets at $R<50$ kpc, we explore how an increase of 0.3 and 0.6 dex affects the estimation of $N_{\rm Si}(R)$.

\subsubsection{High Resolution Spectra and Profile Fitting Analysis}\label{s-gen-comments}

In the Appendices~\ref{s-comp-fit-aod} and \ref{s-pf} we explore the robustness of the AOD results by comparing high- and low-resolution spectra and by comparing to a Voigt profile fitting analysis. There is good overall agreement in the column densities derived from the STIS and COS data and our conservative choice of  $\tau_a \sim 0.9$ as the threshold for saturation in the COS data is adequate (see Appendix~\ref{s-comp-fit-aod}). For the profile fitting analysis, we consider the most complicated blending of components in our sample and demonstrate there are some small systematic differences between the AOD and PF derived column densities (see Appendix~\ref{s-pf}). However these difference are small and a  majority of our sample is not affected by heavy blending. Hence the AOD results are robust and are adopted for the remaining of the paper.

\subsection{Correcting for Magellanic Stream Contamination}\label{s-ms}
\begin{figure*}[tbp]
\epsscale{1.}
\plotone{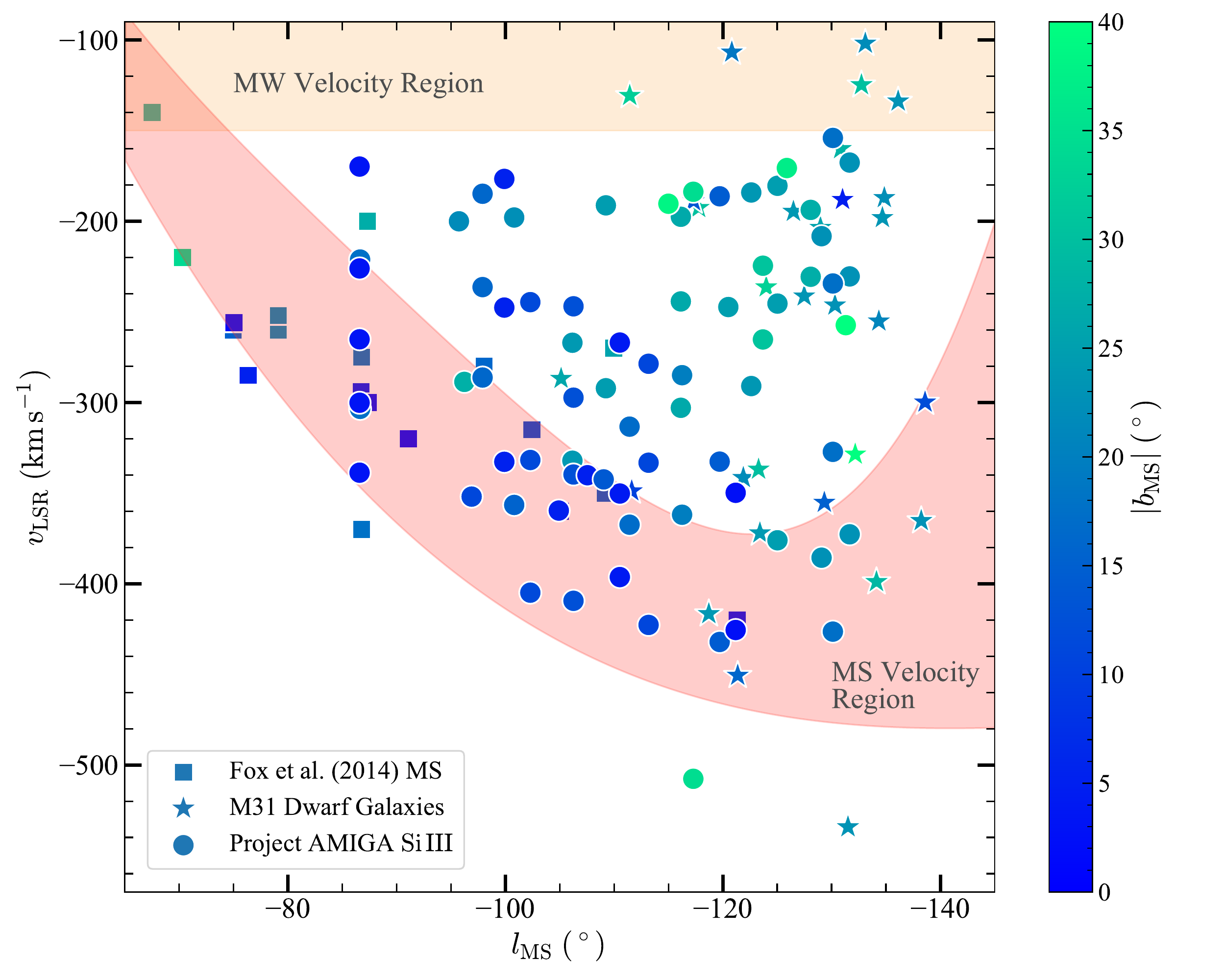}
\caption{The LSR velocity of the \siiii\ components (circles) observed in our sample as a function of the MS longitude \lms, color-coded according to the absolute MS latitude. Shaded regions show the velocities that can be contaminated by the MS and MW (by definition of our search velocity window, any  absorption at $\vlsr >-150$ \km\ was excluded from our sample). We also show the data (squares) from the MS survey from \citet{fox14} and the radial velocities of the M31 dwarf galaxies (stars). 
\label{f-nidever}}
\end{figure*}

Prior to determining the properties of the gas associated with the CGM of M31, we need to identify that gas and distinguish it from the MW and the MS. We have already removed from our analysis any contamination from higher redshift intervening absorbers and any contamination from the MW (defined as  $-150 \la \vlsr\ \la 100$ \km). However, as shown in Fig.~\ref{f-map} and discussed in \citetalias{lehner15}, the MS is another potentially large source of contamination: in the direction of M31, the velocities of the MS can overlap with those expected from the CGM of M31. The targets in our sample have MS longitudes and latitudes in the range $-132\degr \le \lms \le -86\degr$ and $-14\degr \le \bms \le +41\degr$. The \hi\ 21-cm emission GBT survey by \citet{nidever10} finds that the MS extends to about  $\lms \simeq -140\degr$. Based on this and previous \hi\ emission surveys, \citet{nidever08,nidever10} found a relation between the observed LSR velocities of the MS and \lms\ that can be used to assess contamination in our targeted sightlines based on their MS coordinates. Using Fig.~7 of \citet{nidever10}, we estimate the upper and lower boundaries of the \hi\ velocity range as a function of \lms, which we show in Fig.~\ref{f-nidever} by the curve colored area. The MS velocity decreases with decreasing \lms\ up to $\lms \simeq -120\degr$ where there is an inflection point where the MS LSR velocity increases. We note that the region beyond $\lms \la -135\degr$ is uncertain but cannot be larger than shown in Fig.~\ref{f-nidever} (see also \citealt{nidever10})---however, this does not affect our survey since all our data are at $\lms \ga -132\degr$.

We take a systematic approach to removing the MS contamination that does not reject entire sightlines based on their MS coordinates. Not all velocity components may be contaminated even on sightlines close to the MS. In Fig.~\ref{f-nidever}, we show LSR velocity of the \siiii\ components as a function of the MS longitude. We choose \siiii\ as this ion is the most sensitive to detect both weak and strong absorption and is readily observed the physical conditions of the MS and M31 CGM \citep{fox14,lehner15}. We consider the individual components as for a given sightline, several components can be observed falling in or outside the boundary region associated with the MS as illustrated in Fig.~\ref{f-nidever}. We find that $28/74\simeq 38\%$ of the detected \siiii\ components are within MS boundary region shown in Fig.~\ref{f-nidever}. We note that changing the upper boundary by $\pm 5$ \km\ would change this number by about $\pm 3\%$.

To our own sample, we also add data from two different surveys:  the \hst/COS MS survey by \citet{fox14} and the M31 dwarfs (\citealt{mcconnachie12} and see \S\ref{s-dwarfs}). For the MS survey, we  restrict the sample $-150\degr \le \lms \le -20\degr $, i.e., overlapping with our sample but also including higher \lms\ value while still avoiding the Magellanic Clouds region where conditions may be different. The origin of the sample for the M31 dwarf galaxies is fully discussed in \S\ref{s-dwarfs} . The larger galaxies M33, M32, and NGC\,205 are excluded here from that sample as their large masses are not characteristic. The LSR velocities of the M31 dwarfs as a function of \lms\ are plotted with a star symbol in Fig.~\ref{f-nidever}. For the MS survey, we select the LSR velocities of \siiii\ for the MS survey (note these are average velocities that can include multiple components), which are shown with squares in Fig.~\ref{f-nidever}. Most ($\sim90\%$) of the squares fall between the two curves in Fig.~\ref{f-nidever}, confirming the likelihood that these sightlines probe the MS (although we emphasize that this test was not initially used by \citealt{fox14} to determine the association with the MS).

The M31 dwarf galaxies are of course not affected by the MS, but can help us to determine how frequently they fall within the velocity range where MS contamination is likely. For $\lms \ga -132\degr$ (where all the QSOs are and to avoid the uncertain region), only 9\% (2/22) of the dwarfs are within the velocity region where MS contamination occurs. If the velocity distributions of the M31 dwarfs and M31 CGM gas are similar, this would strongly suggest that velocity components with the expected MS velocities are indeed more likely associated with the MS. We, however, note two additional dwarfs are close to the upper boundary, which would change the frequency of the dwarfs in the MS velocity-boundary region to 18\%.

\begin{figure}[tbp]
\epsscale{1.2}
\plotone{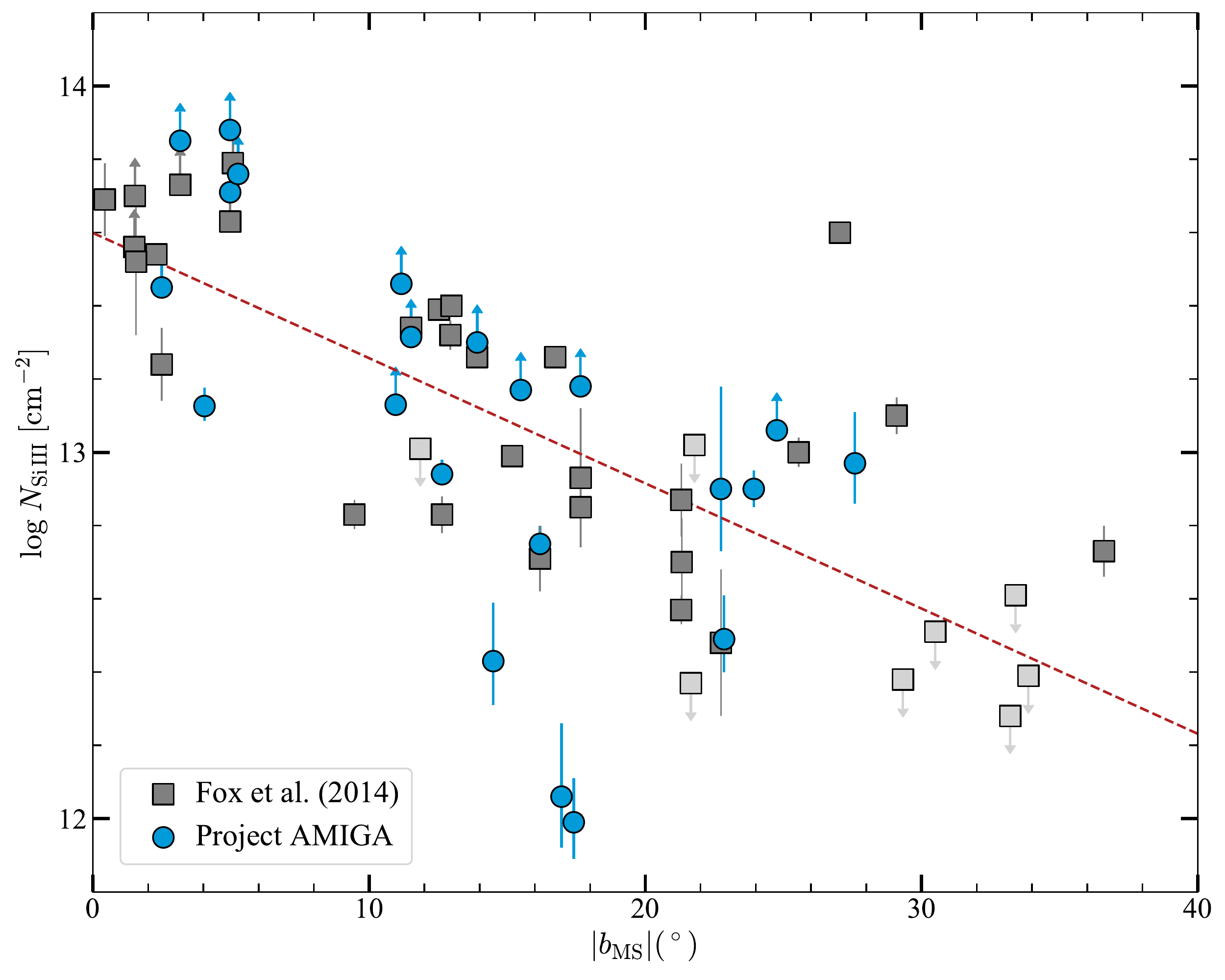}
\caption{The total column density of \siiii\ that are associated with the MS as a function of the absolute MS latitude. We also show the MS survey by \citet{fox14} restricted to data with $-150\degr \le \lms \le -20\degr $. The lighter gray squares with downward arrows are non-detections in the \citeauthor{fox14} sample. The dashed line is a linear fit to the data treating the limits as values. A Spearman ranking correlation test implies a strong anti-correlation with a correlation coefficient $r_{\rm S} = -0.72$ and $p\ll 0.1\%$.
\label{f-col-cont}}
\end{figure}

Observations of \hi\ 21-cm emission toward the QSOs observed with COS in MS survey \citep{fox14} and Project AMIGA \citep{howk17} show only \hi\ detections within $|\bms|\la 11\degr$. In the region defined by $-150\degr \le \lms \le -20\degr $, the bulk of the \hi\ 21-cm emission is observed within $|\bms|\la 5\degr$ \citep{nidever10}. We therefore expect the metal ionic column densities to have a strong absorption when  $|\bms|\la 10\degr$ and a weaker absorption as  $|\bms|$ increases. In Fig.~\ref{f-col-cont}, we show the total column densities of \siiii\ for the velocity components from the Project AMIGA sample found within the MS boundary region shown in Fig.~\ref{f-nidever}, i.e., we added the column densities of the components that are likely associated with the MS. We also show in the same figure the results from the \citet{fox14} survey. Both datasets show the same behavior of the total \siiii\  column densities with $|\bms|$, an overall decrease in $N_{\rm Si\,III}$ as $|\bms|$ increases. Treating the limits as values, combining the two samples,  and using the Spearman rank order, the test confirms the visual impression that there is a strong monotonic anti-correlation between $N_{\rm Si\,III}$ and $|\bms|$ with a correlation coefficient $r_{\rm S} = -0.72$ and a p-value $\ll 0.1\%$.\footnote{We note that if we increase the lower limits by 0.15 dex or more and similarly decrease the upper limits, the significance of the anti-correlation would be similar.} There is a large scatter (about $\pm 0.4$ dex around the dotted line) at any \bms, making it difficult to determine if any data points may not be associated with the MS (as, e.g., the three very low $N_{\rm Si\,III}$ at $12\degr<|\bms|<18\degr$  from our sample or the very high value at $|\bms|\sim 27\degr$  from the \citealt{fox14} sample).

\begin{figure}[tbp]
\epsscale{1.2}
\plotone{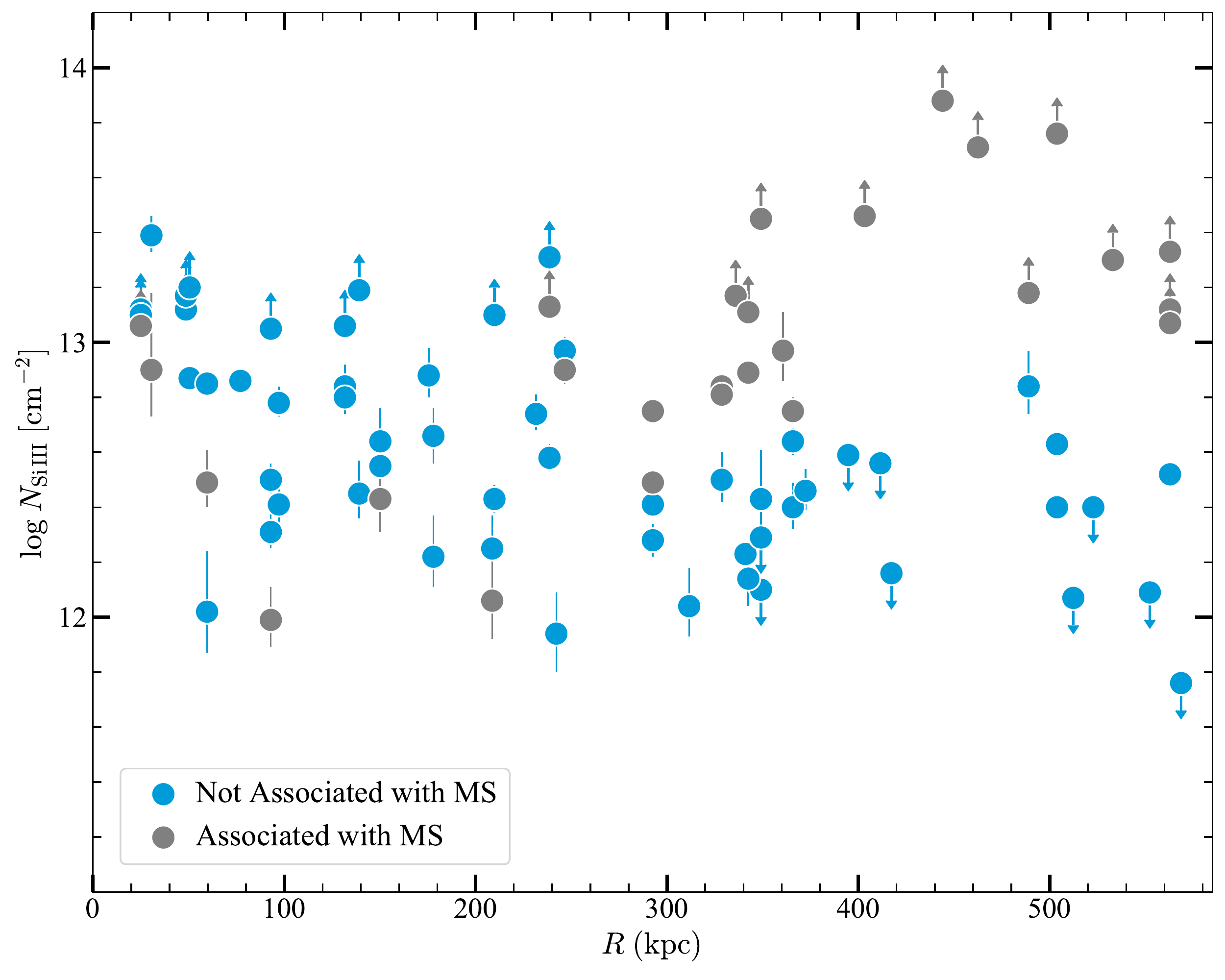}
\caption{Logarithm of the column densities of the individual components for \siiii\ as a function of the projected distances from M31 of the background QSOs where the separation is made for the components associated or not with the MS.
\label{f-col-vs-rho-ex}}
\end{figure}

In Fig.~\ref{f-col-vs-rho-ex}, we show the individual column densities of \siiii\ as a function of the impact parameter from M31 for the Project AMIGA sightlines where we separate components associated with the MS from those that are not. Looking at Figs.~\ref{f-map} and \ref{f-col-cont}, we expect the strongest column densities associated  with the MS to be at $|\bms|\la 10\degr$ and $R\ga 300$ kpc, which is where they are located on Fig.~\ref{f-col-vs-rho-ex}. We also expect a positive correlation between $N_{\rm Si\,III}$ and $R$ for the MS contaminated components while for uncontaminated components, we expect the opposite (see \citetalias{lehner15}). Treating again limits as values, the Spearman rank order test demonstrates a strong monotonic correlation between $N_{\rm Si\,III}$ and $R$ ($r_{\rm S} = 0.68$ with $p \ll 0.1\%$) while for uncontaminated components there is a strong monotonic anti-correlation ($r_{\rm S} = -0.57$ with $p \ll 0.1\%$), in agreement with the expectations. Based on these results, it is therefore reasonable to consider any absorption components observed in the COS spectra within the MS boundary region  defined in Fig.~\ref{f-nidever} as most likely associated with the MS. We therefore flag any of these components (28 out 74 components for \siiii) as contaminated by the MS and those are not included further in our sample.

Finally, we noted above that only a small fraction of the dwarfs are found in the MS contaminated region. While that fraction is small (9\%), this could still suggest that in the MS contaminated region, some of the absorption could be a blend between of both MS and M31 CGM components. However, considering the uncontaminated  velocities along  sightlines in (29 components) and outside (17 components) the contaminated regions, with  $p$-value of 0.74 the Kolmogorov-Smirnov (KS) comparison of the two samples cannot reject the  null-hypothesis that the distributions are the same. This strongly suggests that the correction from the MS contamination does not bias much the velocity distribution associated with the CGM of M31 (assuming that there is no strong change of the velocity with the azimuth $\Phi$; as we explore this further in \S\S\ref{s-dwarfs-vel} and \ref{s-map-vel}, there is, however, no strong evidence a velocity dependence with $\Phi$).

\section{M31 Dwarf Galaxy Satellites}\label{s-dwarfs}
While Project AMIGA is dedicated to understanding the CGM of M31, our survey also provides a unique probe of the dwarf galaxies found in the halo of M31. In particular, we have the opportunity to assess if the CGM of dwarf satellites plays an important role in the CGM of the host galaxy, as studied by cosmological and idealized simulations~\citep[e.g.][]{angles-alcazar17, hafen19a, hafen19b, bustard18}. When considering the dwarf galaxies in our analysis we have two main goals: 1) to determine if the velocity distribution of the dwarfs and the absorbers are similar, and 2) assess if some of absorption observed toward the QSOs could be associated directly with the dwarfs, either as gas that is gravitationally bound or recently stripped. 

The sample for the M31 dwarf galaxies is mostly drawn from the \citet{mcconnachie12} study of Local Group dwarfs, in which the properties of 29 M31 dwarf satellites were summarized. Four additional dwarfs (Cas\,II, Cas\,III, Lac\,I, Per\,I) are added from recent discoveries \citep{collins13,martin14,martin16,martin17}. M33 is excluded from that sample as its large mass is not characteristic of satellites.\footnote{In Appendix~\ref{a-m33}, we further discuss and present some evidence that the CGM of M33 is unlikely to contribute much to the observed absorption in our sample.} Table~\ref{t-dwarf} summarizes our adopted sample of M31 dwarf galaxies (sorted by increasing projected distance from M31), listing some of their key properties. As listed in this table, most of the M31 satellite galaxies are dwarf spheroidal (dSph) galaxies, which have been shown to have been stripped of most of their gas most likely via ram-pressure stripping \citep{grebel03}, a caveat that we keep in mind as we associate these galaxies with absorbers. 

\subsection{Velocity Transformation}\label{s-vel-trans}
So far we have used LSR velocity to characterize MW and MS contamination of gas in the M31 halo. However, as we now consider relative motions over $30\degr$ on the sky, we cannot simply subtract M31's systemic radius velocity to place these relative motions in the correct reference frame. Over such large sky areas, tangential motion must be accounted for because the ``systemic'' sightline velocity of the M31 system changes with sightline. To eliminate the effects of ``perspective motion", we follow \citet{gilbert18} (and see also \citealt{veljanoski14}) by first transforming the heliocentric velocity ($v_\sun$) into the Galactocentric frame, $v_{\rm Gal}$, which removes any effects the solar motion could have on the kinematic analysis. We converted our measured radial velocities from the heliocentric to the Galactocentric frame using the relation from \citet{courteau99} with updated solar motions from \citet{mcmillan11} and \citet{schonrich10}:
\begin{equation}\label{e-gal}
\begin{aligned}
v_{\rm Gal} = &  v_\sun + 251.24\, \sin(l)\cos(b) +\\
& 11.1\, \cos(l)\cos(b) + 7.25\, \sin(b)\,,
\end{aligned}
\end{equation}
where $(l,b)$ are the Galactic longitude and latitude of the object. To remove the bulk motion of M31 along the sightline to each object, we use the heliocentric systemic radial velocity for M31 of $-301$ \km\ \citep{vandermarel08,chemin09}, which is $v_{\rm M31,r}=-109$ \km\ in the Galactocentric velocity frame. The systemic transverse velocity of M31 is $v_{\rm M31,t}=-17$ \km\  in the direction on the sky given by the position angle $\theta_t = 287\degr$ \citep{vandermarel12}. The removal of M31's motion from the sightline velocities  resulting in peculiar line-of-sight velocities for each absorber or dwarf, $v_{\rm M31}$, is then given by \citep{vandermarel08}:
\begin{equation}\label{e-vm31}
\begin{aligned}
v_{\rm M31} = & v_{\rm Gal} - v_{\rm M31,r}\, \cos(\rho) + \\
            & v_{\rm M31,t}\, \sin(\rho)\cos(\phi -  \theta_t),
\end{aligned}
\end{equation}
where $\rho$ is the angular separation between the center of M31 to the QSO or dwarf position, $\phi$ the position angle of the QSO or dwarf with respect to M31's center. We note that the transverse term in Eqn.~\ref{e-vm31} is more uncertain \citep{vandermarel08,veljanoski14}, but its effect is also much smaller, and indeed including it or not would not quantitatively change the results; we opted to include that term in the velocity transformation. We apply these transformations to change the LSR velocities to heliocentric velocities to Galactocentric velocities to peculiar velocities for each component observed in absorption toward the QSOs and for each dwarf. With this transformation, an absorber or dwarf with no peculiar velocity relative to M31's bulk motion has $v_{\rm M31}=0$ \km, regardless of its position on the sky \citep{gilbert18}.

\subsection{Velocity Distribution}\label{s-dwarfs-vel}
\begin{figure}[tbp]
\epsscale{1.2}
\plotone{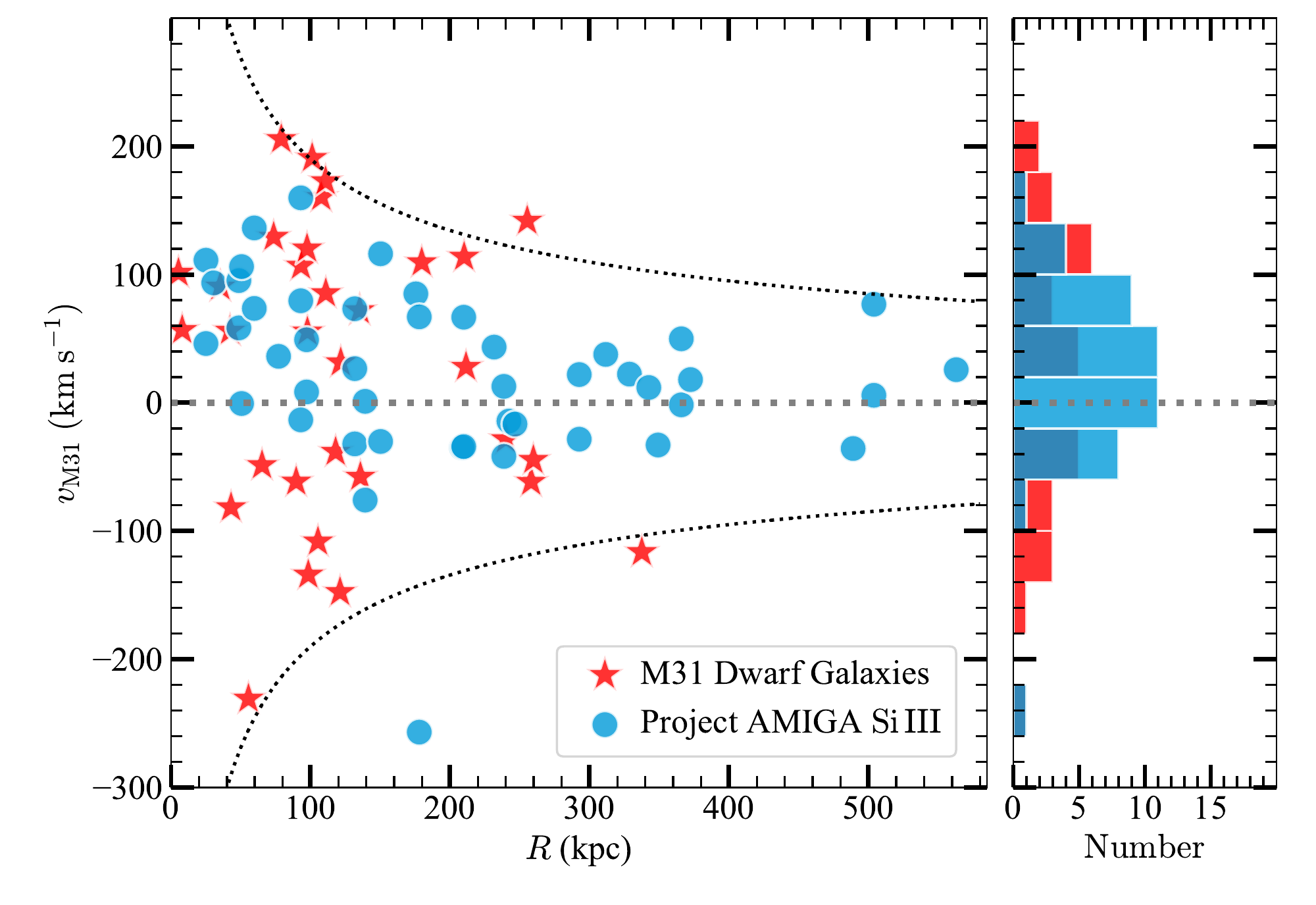}
\caption{{\it Left}:  The M31 peculiar velocity (as defined by Eqn.~\ref{e-vm31})  against the projected distances for the observed absorption components associated with M31 (using \siiii) and M31 dwarf galaxies. The dotted curves show the escape velocity divided by $\sqrt{3}$ to account for the unknown tangential motions of the absorbers and galaxies. {\it Right}: The M31 velocity distributions with the same color-coding definition. 
\label{f-v_vs_r_dwarfs}}
\end{figure}

In Fig.~\ref{f-v_vs_r_dwarfs}, we compare the M31 peculiar velocities of the absorbers using \siiii\ and dwarfs against the projected distance (see \S\ref{s-vel-trans}). In Fig.~\ref{f-v_vs_r_dwarfs}, we also show the expected escape velocity, $v_{\rm esc}$, as a function of $R$ for a $1.3\times 10^{12}$ M$_\sun$ point mass. We conservatively divide $v_{\rm esc}$ by $\sqrt{3}$ in that figure to account for remaining unconstrained projection effects. Nearly all the CGM gas traced by \siiii\ within $\rvir$ is found at velocities consistent with being gravitationally bound, and this is true even at larger $R$ for most of the absorbers. This finding also holds for most of the dwarf galaxies, and, as demonstrated by \citet{mcconnachie12}, it holds when the galaxies' 3D distances are used (i.e., using the actual distance of the dwarf galaxies, instead of the projected distances used in this work). Therefore, both the CGM gas and galaxies probed in our sample at both small and large $R$ are consistent with being gravitationally bound to M31.

Fig.~\ref{f-v_vs_r_dwarfs} also informs us that the dwarf satellite and CGM gas velocities overlap to a high degree but do not follow identical distributions. The mean and standard deviation of the M31 velocities for the dwarfs are $+34.2 \pm 110.0$ \km\ and $+36.6 \pm 68.0$ \km\ for the CGM gas. There is therefore a slight asymmetry favoring more positive peculiar motions. A simple two-sided KS test of the two samples rejects the null hypothesis that the distributions are the same at 95\% level confidence ($p=0.04$). And indeed while the two distributions overlap and the means are similar, the velocity dispersion of the dwarfs is larger than that of the QSO absorbers. For the QSO absorbers, all the components but one have their M31 velocities in the interval $-80 \le v_{\rm M31} \le +160$ \km, but 9/32 (28\%) of the dwarfs are outside that range. Four of the dwarfs are in the range $+160 < v_{\rm M31} \le +210$ \km, a velocity interval that cannot be probed in absorption owing to foreground MW contamination. The other five dwarfs have $v_{\rm M31}< -80$ \km, while only one out of 46 \siiii\ components (2\%) have $v_{\rm M31}< -80$ \km. Both the small fraction of dwarfs at $v_{\rm M31}> +160$ \km\ and $v_{\rm M31}< -80$ \km\ and the even smaller fraction of absorbers at $v_{\rm M31}< -80$ \km\ suggest that there is no important population of absorbers at the inaccessible velocities  $v_{\rm M31}> +160$ \km\ (see also \S\ref{s-ms}).

\begin{figure*}[tbp]
\epsscale{1}
\plotone{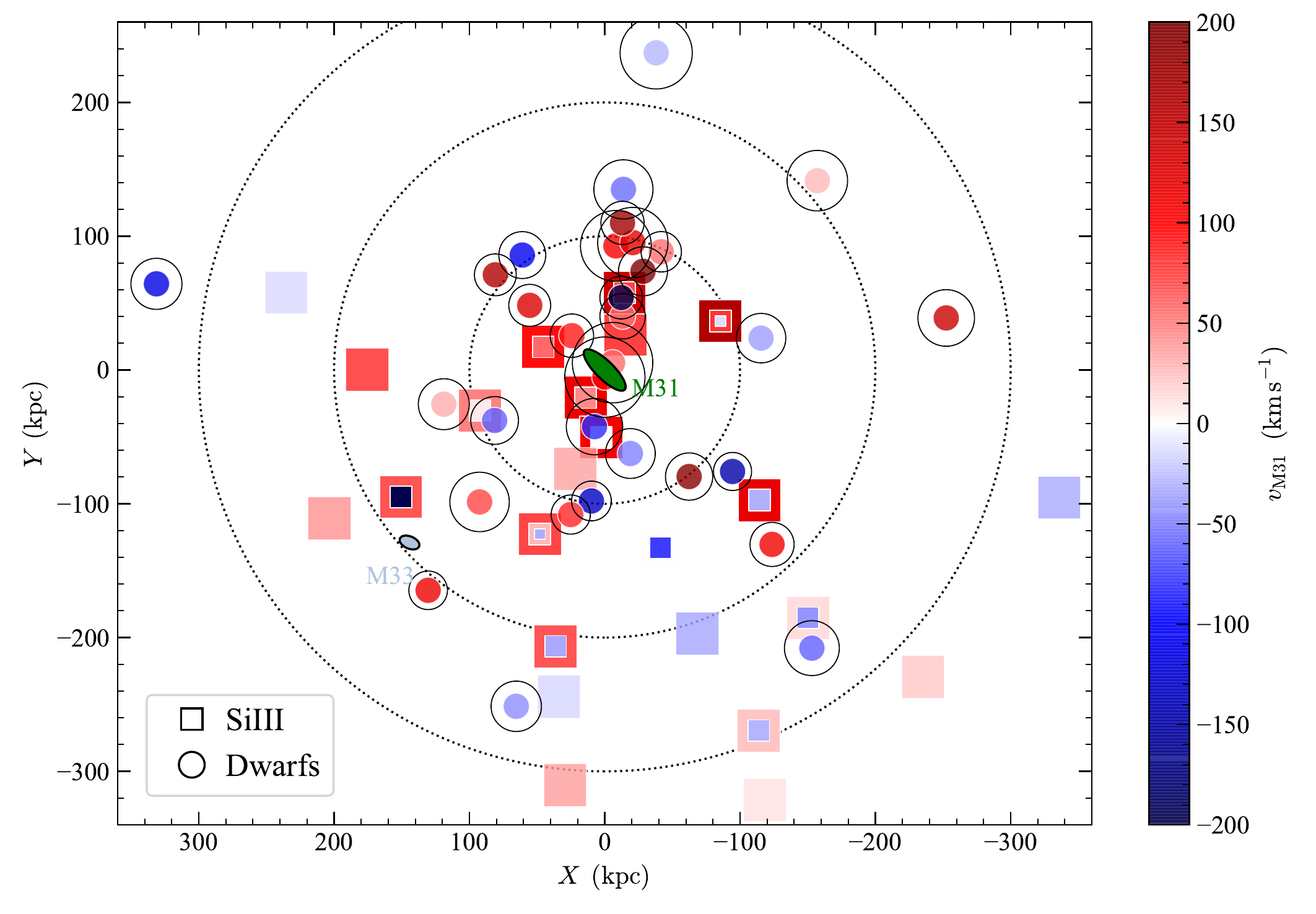}
\caption{Locations of the QSOs ({\it squares}) and dwarfs ({\it circles}) relative to M31 (see Fig.~\ref{f-map}). The data are color-coded according to the relative velocities of the detected \siiii\ (multiple colors in a symbol indicate multiple detected components) or the dwarfs. The black circles centered on the dwarfs indicate their individual $R_{200}$. 
\label{f-velmap-dwarfs}}
\end{figure*}

\subsection{The Associations of Absorbers with Dwarf Satellites}\label{s-dwarfs-cgm}

Using the information from Table~\ref{t-dwarf}, we cross-match the sample of dwarf galaxies and QSOs to determine the QSO sightlines that are passing within a dwarf's $R_{200}$ radius. There are 11 QSOs (with 58 \siiii\ components) within $R_{200}$ of 16 dwarfs. In Table~\ref{t-xmatch-dwarf}, we summarize the results of this cross-match. Fig.~\ref{f-velmap-dwarfs}, we show the map of the QSOs and dwarf locations in our survey where the M31 velocities of the \siiii\ components and dwarfs are color coded on the same scale and the circles around each dwarf represent their $R_{200}$ radius. 

Table~\ref{t-xmatch-dwarf} and Fig.~\ref{f-velmap-dwarfs} show that several absorbers can be found within $R_{200}$ of several dwarfs when \siiii\ is used as the gas tracer. For example, the two components observed in \siiii\ toward Zw535.012 are found within $R_{200}$ of 6 dwarf galaxies. In Table~\ref{t-xmatch-dwarf}, we also list the escape velocity ($v_{\rm esc}$) at the observed projected distance of the QSO relative to the dwarf as well as the velocity separation between the QSO absorber and the dwarf ($\delta v \equiv |v_{\rm M31, Si\,III} - v_{\rm M31,dwarf}|$). So far we have not considered the velocity separation $\delta v$ between the dwarf and the absorber, but it is likely that if $\delta v \gg v_{\rm esc}$ then the observed gas traced by the absorber is unlikely to be bound to the dwarf galaxy even if $\Delta_{\rm sep} = R/R_{200}<1$. 

If we set $\delta v <v_{\rm esc}/\sqrt{3}$, then the sample of components would be reduced to 31 instead of 58. The sample is reduced still further down to 12 if the two most massive dwarfs (M32 and NGC\,205) are removed from the sample and down to 6 if the most massive dwarfs with $M_{\rm h}>3.9\times 10^{10}$ M$_\sun$ are removed from the sample. Applying a cross-match where $\delta v < v_{\rm esc}$ and $\Delta_{\rm sep}<1$  can reduce the degeneracy between different galaxies, especially if one excludes the four most massive galaxies. For example, RXS\_J0118.8+3836 is located at $0.40 R_{200}$ and $0.72 R_{200}$ from Andromeda\,XV and Andromeda\,XXIII, but only in the latter case $\delta v \ll v_{\rm esc}$ (and in the former case $\delta v > v_{\rm esc}$), making the two components observed toward RXS\_J0118.8+3836 more likely associated with Andromeda\,XXIII.

Several sightlines therefore pass within $\Delta_{\rm sep}<1$ of a dwarf galaxy and  show a velocity absorption within the escape velocity.  This gas could be gravitationally bound to the dwarf. However, there are also  5 absorbers where $\delta v <v_{\rm esc}/\sqrt{3}$, but the QSO is at  $1 < \Delta_{\rm sep} \le 2$ from the dwarf, i.e., the velocity separation is small, but the spatial projected separation makes unlikely to be bound to the dwarf. Here the velocity match may be a coincidence or a result of the relative proximity of the dwarfs and QSOs in the CGM of M31 assuming that the gas and dwarfs both follow the same global velocity motion of the M31 CGM. As illustrated in Fig.~\ref{f-velmap-dwarfs}, there are, however, some dwarfs with $\Delta_{\rm sep}\la 2$ with a radial velocity very different from that observed in absorption toward the QSO or vice-versa, implying not all the dwarfs and gas velocities are tightly connected.

In summary, it is plausible that absorbers with $\Delta_{\rm sep}<1$ and  $\delta v\ll v_{\rm esc}$ trace gas associated with the CGM of a dwarf, but we cannot confirm unambiguously this association.  We inspected a number of gas properties (e.g., column densities, ionization levels, kinematics), but did not find any that can differentiate clearly between a dwarf CGM origin from a M31 CGM origin. Nothing in the properties of the components found within $\Delta_{\rm sep}<1$ of a dwarf and having $\delta v\ll v_{\rm esc}$ make them outliers. This is certainly not surprising since  any association assumes that the dwarf galaxies have a rich gas CGM. Yet all the satellites listed in the cross-matched Table~\ref{t-xmatch-dwarf} are  dSph galaxies, which are known to be neutral gas poor  \citep{grebel03}. The dSph galaxies are also likely ionized gas deficient since the favor mechanism to strip their gas is ram-pressure, a stripping mechanism efficient on both the neutral and ionized gas  \citep{grebel03,mayer06}. Therefore these galaxies are unlikely to have gas rich CGM and 
based  our observations we do not find any persuasive evidence that gas associated with M31 satellites causes the absorption we see in the M31 CGM. 

\section{Properties of the M31 CGM}\label{s-properties}
We now focus on determining the properties of the CGM of M31 using only the velocity components that are not contaminated by the MS (see \ref{s-ms}). We use the following atoms and ions to characterize the M31 CGM: \oi, \siii, \siiii, \siiv, \cii, \civ, \ovi, \feii. \oi\ and \feii\ are not commonly detected, but even so they are useful in assessing the ionization and depletion levels of the CGM gas. We use the terminology ``low ions'' for singly ionized species, ``intermediate ions" for \siiii\ and \siiv, and ``high ions" for \civ\ and \ovi.  Also note that we adopt here the solar relative abundances from \citet{asplund09}.

\subsection{Metallicity of the CGM}\label{s-metallicity}

Radio observations have not detected any \hi\ 21-cm emission toward any of the QSO targets in Project AMIGA down to a $5\sigma$ level of $\mlnhi \ga 17.6$ \citepalias{howk17}; many sightlines could therefore have  $\mlnhi \ll 17.6$. As a consequence of this, we cannot directly estimate the metallicities of the CGM in our sample. However, we have some weak detections of \oi\ in 4 components at better than the $3\sigma$ level. Since \oi\ and \hi\ have nearly identical ionization potentials and are strongly coupled through charge exchange reactions \citep{field71}, \oi\ is an excellent proxy for \hi, requiring no or very small ionization correction as long as the photoionization spectrum is not too hard \citep[e.g.,][]{lehner03}. Therefore \oi\ can be compared to the limit of \hi\ to put a lower limit on the metallicity. The \oi\ logarithmic column densities are in the range 13.3 to 13.7 dex (see Table~\ref{t-results}), with a mean of 13.5 dex. This implies $[$\oi/\hi $] = \log (N_{\rm O\,I}/N_{\rm H\,I}) - \log ({\rm O/H})_\sun \ga -0.7$ or a metallicity $Z\ga 0.2 Z_\sun$. This lower limit, however, assumes that there is no beam dilution effect, i.e., we assume the limit on the \hi\ column density in the 2 kpc beam (at the distance of M31) would be the same than in a pencil-beam observed in absorption. Any beam dilution would increase the limit on \hi, and therefore the metallicity limit could be less stringent. We therefore caution the reader not to take this limit as a hard lower limit. 

\subsection{Relative Abundances}\label{s-abund}
While the metallicity remains quite uncertain, from the relative abundances of detected ions, we can assess the level of ionization, dust depletion, nucleosynthetic history. For assessing depletions, we can compare refractory elements like Fe to less refractory elements like Si \citep[e.g.,][]{savage96,jenkins09}. \feii\ and \siii\ have similar ionization energies (8--16 eV) and their observed ratio should be minimally affected by differential ionization. Hence the ratio $[$\feii/\siii$] = \log (N_{\rm Fe\,II}/N_{\rm Si\,II}) - \log ({\rm Fe/Si})_\sun $ traces dust depletion levels. Unfortunately (but perhaps not surprisingly), \feii\ is only detected in the sightline closest to M31, and in that sightline, we derive $[$\feii/\siii$] = -0.13 \pm 0.16$. In ten other sightlines, we place upper limits on that ratio where the two smallest upper limits imply  $[$\feii/\siii$] \la 0$, while all the others are above 0 dex. While the information is minimal, this still demonstrates that there is no evidence for significant dust depletion in the CGM of M31. As depletions get stronger in denser gas, it is perhaps not surprising that we find little evidence for it in a sample where the sightlines all have $\mlnhi \la 17.6$ and low ions are not commonly detected. While we assume that dust would be the major factor to deplete Fe relative to Si, the lack of evidence for depletion of Fe also points to a negligible nucleosynthesis effect on that ratio that would produce a non-solar $\alpha$-particle (e.g. Si) enhancement relative to Fe \citep[e.g.,][]{welty97}.

Using ratios of elements with different nucleosynthetic origins, we can assess the chemical enrichment history of the M31 halo gas by measuring departures from a solar relative abundance ratio in elements of different nucleosynthetic origin. For instance, the \ca\ ratio should be sensitive to nucleosynthesis effects since there is a time-lag between the production of $\alpha$-elements and carbon (see, e.g., \citealt{cescutti09,mattsson10}). This analysis would be complicated by large depletions, but as we have shown above the Fe/Si ratios show little if any evidence of large depletions. As Fe is typically the most depleted element in these conditions \citep{savage96,welty99b,jenkins09}, we can reliably assume that \ca\ does not suffer large depletions and can therefore be used as a nucleosynthetic indicator.  

However, we must consider ionization effects in addition to depletions. Differential ionization can affect the \cii/\siii\  (i.e., \ca) ratio because \cii\ has a higher ionization energy range (12--25 eV) than \siii\ (8--16 eV). To assess this, we use the nine absorbers with detections of both \siii\ and \cii\ to estimate $[$\cii/\siii$] = \log (N_{\rm C\,II}/N_{\rm Si\,II}) - \log ({\rm C/Si})_\sun $. Since this subsample includes both detections and lower limits owing to saturation of \cii, we use a survival analysis where the four censored lower limits are included \citep{feigelson85,isobe86}. We find that the mean $[$\cii/\siii$] = 0.07 \pm  0.09 $ (where the error is the error on the mean from the Kaplan-Meier estimator) and the $1\sigma$ dispersion is $0.19$ dex. This ratio is consistent with a solar value. If non-detections of \siii\ are included the mean rises to $[$\cii/\siii$] = 0.52 \pm 0.11 $, strongly indicating ionization affects this ratio owing to photons ionizing \siii\ into \siiii.

Therefore based on the relative abundances of Fe and C to Si, there is no evidence for strong dust depletion or non-solar nucleosynthesis effects in the CGM of M31. We emphasize that this does not mean there is no dust in the CGM of M31, and indeed several studies have shown that the CGM of galaxies can have a substantial mass of dust \citep[e.g.,][]{menard10,peek15}. However, its effect on elemental abundances must be smaller than in the dense regions of galaxies. The lack of nucleosynthesis effects on the  abundance of Fe or C relative to Si strongly suggests that the overall metallicity of the gas is not extremely low, as enhancements of $\alpha$ elements are seen in low-metallicity MW halo stars and in low-metallicity gas in CGM absorbers over a range of redshift. For a sample of \hi-selected absorbers with $15\la \mlnhi \la 18$ at $0.2\la z\la 1$, \citet{lehner19} found little correlation between the \ca\ and the metallicity. However in stars and \hii\ regions in the local universe, there is evidence of a trend between \ca\ and the metallicity where  $\ca \simeq -0.6$ at $-2\la \log Z/Z_\sun \la -0.5$ and $\ca \simeq 0$ near solar metallicities \citep[e.g.,][]{akerman04,fabbian09}.  Therefore the metallicity of the M31 CGM could still be sub-solar, but is unlikely to be much below $1/3\;Z_\sun$. This is consistent with the rough metallicity estimate set in \S\ref{s-metallicity}. 

\subsection{Ionization Fractions}\label{s-ionization}

The ionization fraction of the CGM gas can be estimated directly by comparing the column densities of \oi\ to those of \siii, \siiii, and \siiv\ \citep[e.g.,][]{lehner01b,zech08}. \oi\ is an excellent proxy for neutral gas (see \S\ref{s-metallicity}). \siii\ is found in both neutral and ionized gas, and \siiii\ and \siiv\ arise only in ionized gas. O and Si are both $\alpha$-elements with similar nucleosynthetic origins and have similar levels of dust depletion in the diffuse gas \citep{savage96,jenkins09}. Therefore if the ratio $[$\oi/Si$] = \log (N_{\rm O\,I}/N_{\rm Si}) - \log ({\rm O/Si})_\sun $ is sub-solar, ionization is important in the M31 CGM. 

To obtain the total Si column density we use the individual ion columns listed in Table~\ref{t-results}. In the case of non-detections of the Si ions or \oi, we conservatively add the upper limits to the column densities. When there are lower limits present, we add the column densities using the lower limit values. When both detections and non-detections are present, we consider the two extreme possibilities where we set the column density of the non-detection either to the upper limit value (i.e., the absorption is nearly detected--case 1) or we neglect the upper limit (i.e., it is a true non-detection--case 2). For 28 targets, we can estimate the  $[$\oi/Si$]$ ratio. Considering case 1, we find that the mean and dispersion is $[$\oi/Si$]<-0.95 \pm 0.38$ and the full range is $[<-1.78, <-0.34]$, i.e., on average the gas is ionized at the $>89\%$ level. In case 2, the mean and dispersion is $[$\oi/Si$]<-0.74 \pm 0.51$, so that the ionization fraction is still $>81\%$ on average. These are upper limits because typically \oi\ is not detected. However, even in the 5 cases where \oi\ is detected, 4/5 are upper limits too because \siiii\ is saturated and hence only a lower limit on the column density of Si can be derived. In that case,  $[$\oi/Si$]$ ranges from $<-1.78$ to $-0.43$ (or to $<-0.85$ if we remove the absorber where the \oi\ absorption is just detected at the $2\sigma$ level), i.e., even when \oi\ is detected to more than $3\sigma$, the gas is still ionized at levels $>86\%$--$98\%$.

The combination of \siii, \siiii, and \siiv\ allows us to probe gas within the ionization energies 8--45 eV, i.e., the bulk of the photoionized CGM of M31. The high ions, \civ\ and \ovi, have ionization energies 48--85 eV and 114--138 eV, respectively, and are not included in the above calculation. The column density of H can be directly estimated in the ionization energy 8--45 eV range from the observations via $\mlnh  = \log N_{\rm Si} - \log Z/Z_\sun$. As we show below, Si varies strongly with $R$ with values $\log N_{\rm Si}\ga 13.7$ at $R\la 100$ kpc and $\log N_{\rm Si} \la 13.3$ at $R\ga 100$ kpc, which implies $\mnh \ga 1.5\times 10^{18} (Z/Z_\sun)^{-1}$ \cmm\ and $\la 0.6\times 10^{18} (Z/Z_\sun)^{-1}$ \cmm, respectively. For the high ions, a ionization correction needs to be added, and, e.g., for \ovi, $\mlnh  = \log N_{\rm O\,VI} - \log Z/Z_\sun - \log f^i_{\rm O\,VI }$ where $f^i_{\rm O\,VI }\la 0.2$ is the ionization fraction of \ovi\ that peaks around 20\% for any ionizing models \citep[e.g.][]{oppenheimer13,gnat07,lehner14}. As discussed below, there is little variation of \novi\ with $R$ and is always such that $\mlnovi \ga 14.4$--$14.9$ within 300 kpc from M31, which implies $\mnh \ga (2.5$--$8.1)\times 10^{18} (Z/Z_\sun)^{-1}$ \cmm. Therefore the CGM of M31 is not only mostly ionized (often at levels close to 100\%), but it also contains a substantial fraction of highly ionized gas with higher column densities than the weakly photoionized gas.

\subsection{Ion Column Densities versus $R$}\label{s-n-vs-r}

\begin{figure*}[tbp]
\epsscale{1}
\plotone{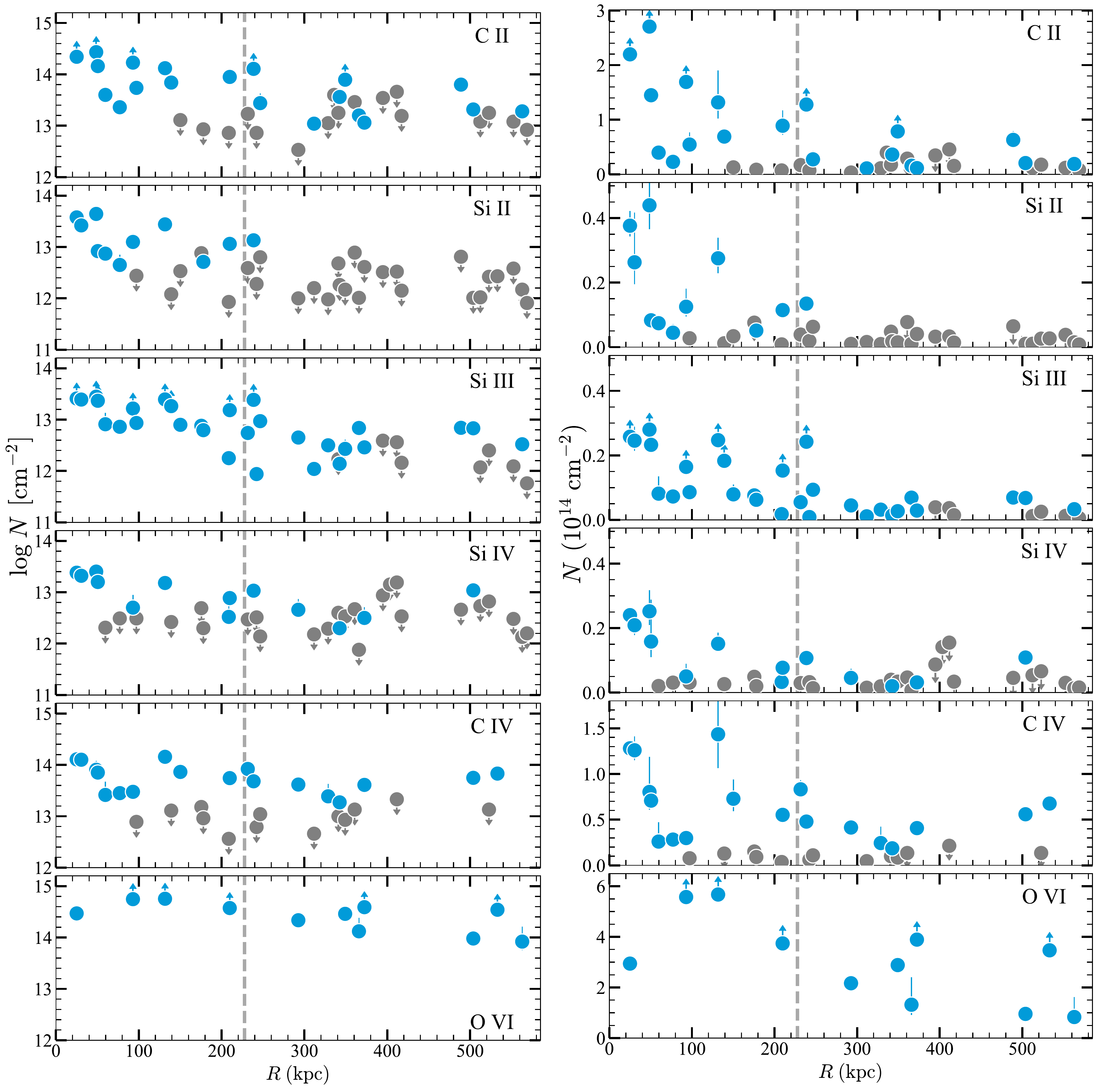}
\caption{Total column densities of the ions as a function of $R$ with ionization potential increasing from top to bottom panels. The column densities are shown in logarithm values (with the same relative vertical scale of about 3 dex in each panel) on the left and in linear units on the right. Blue circles are detections, while gray circles with downward arrows are non-detections. A blue circle with an upward arrow denotes that the absorption is saturated, resulting into a lower limit. The components associated with the MS have been removed. The dashed vertical line marks $R_{200}$. Note how \siiii\ and \ovi\ are detected at high frequency well beyond $R_{200}$.
\label{f-coltot-vs-rho}}
\end{figure*}

In Fig.~\ref{f-coltot-vs-rho}, we show the logarithmic (left) and linear (right) values of the total column densities of the components associated with M31 for \cii, \siii, \siiii, \siiv, \civ, and \ovi\ as a function of the projected distances from M31. Gray data points are upper limits, while blue data with upward arrows are lower limits owing to saturated absorption. Overall, the column densities decrease at higher impact parameter. As the ionization potentials of the ions increase, the decrease in the column densities becomes shallower; \ovi\ is almost flat. These conclusions were already noted in \citetalias{lehner15}, but now that the region from 50 to 350 kpc is filled with data, these trends are even more striking. However, our new sample shows also an additional feature with a remarkable change around $R_{200} \simeq 230$ kpc of M31 notable especially for the low and intermediate ions whereby high \cii, \siii, \siiii, and \siiv\ column densities are observed solely at $R\la R_{200}$. Low column densities \cii, \siii, \siiii, and \siiv\ are observed at all $R$, but strong absorption is observed only at $R\la R_{200}$. The frequency of strong absorption is also larger at $R\la 0.6 R_{200}$ than at larger $R$ for all ions. A similar pattern is observed for \civ\ and \ovi, but the difference between the low and high column densities is smaller: for \cii, \siii, \siiii, and \siiv, the difference between low and high column densities is a factor $\ga 5$--10, while it drops to a factor 2--4 for \civ, and  possibly even less for \ovi.

\begin{figure}[tbp]
\epsscale{1.2}
\plotone{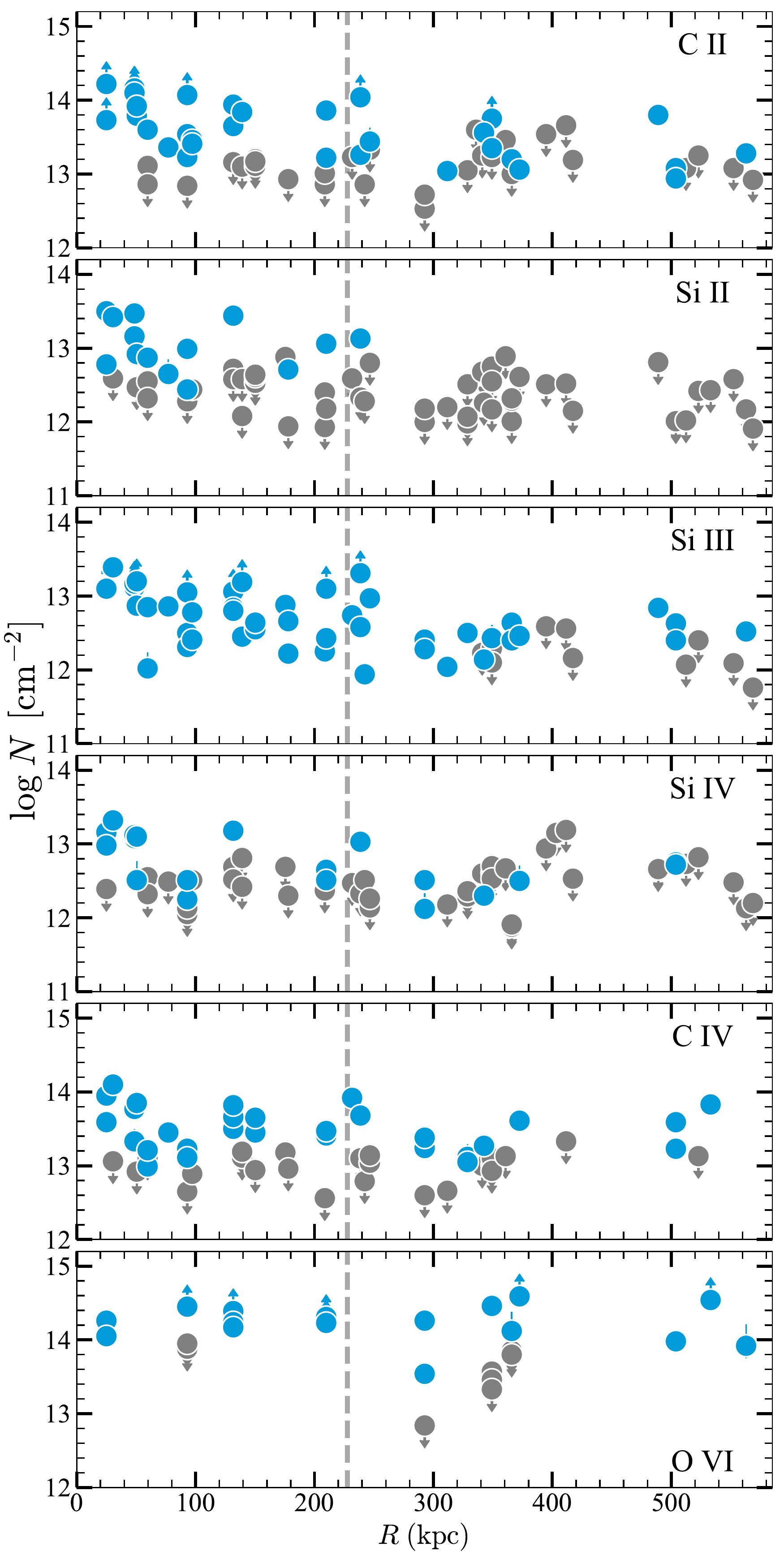}
\caption{Logarithm  of the  column densities for the individual components of various ions (low to high ions from top to bottom) as a function of the projected distances from M31 of the background QSOs. Blue circles are detections, while gray circles with downward arrows are non-detections. A blue circle with an upward arrow denotes that the absorption is saturated, resulting into a lower limit.  The components associated with the MS have been removed.  The dashed vertical lines shows the $R_{200}$ location. The same relative vertical scale of about 3 dex is used in each panel for comparison between the different ions. 
\label{f-col-vs-rho}}
\end{figure}

In Fig.~\ref{f-col-vs-rho}, we show the logarithmic values of the column densities  derived from the individual components for \cii, \siii, \siiii, \siiv, \civ, and \ovi\ as a function of the projected distances from M31. Similar trends are observed in Fig.~\ref{f-coltot-vs-rho}, but Fig.~\ref{f-col-vs-rho} additionally shows that 1) more complex velocity structures (i.e., multiple velocity components) are predominantly observed at  $R\la R_{200}$, and 2) factor $\ga 2$--10 changes in the column densities are observed across multiple velocity components along a given sightline.

\subsection{Silicon Column Densities versus $R$}\label{s-nsi-vs-r}

\begin{figure}[tbp]
\epsscale{1.2}
\plotone{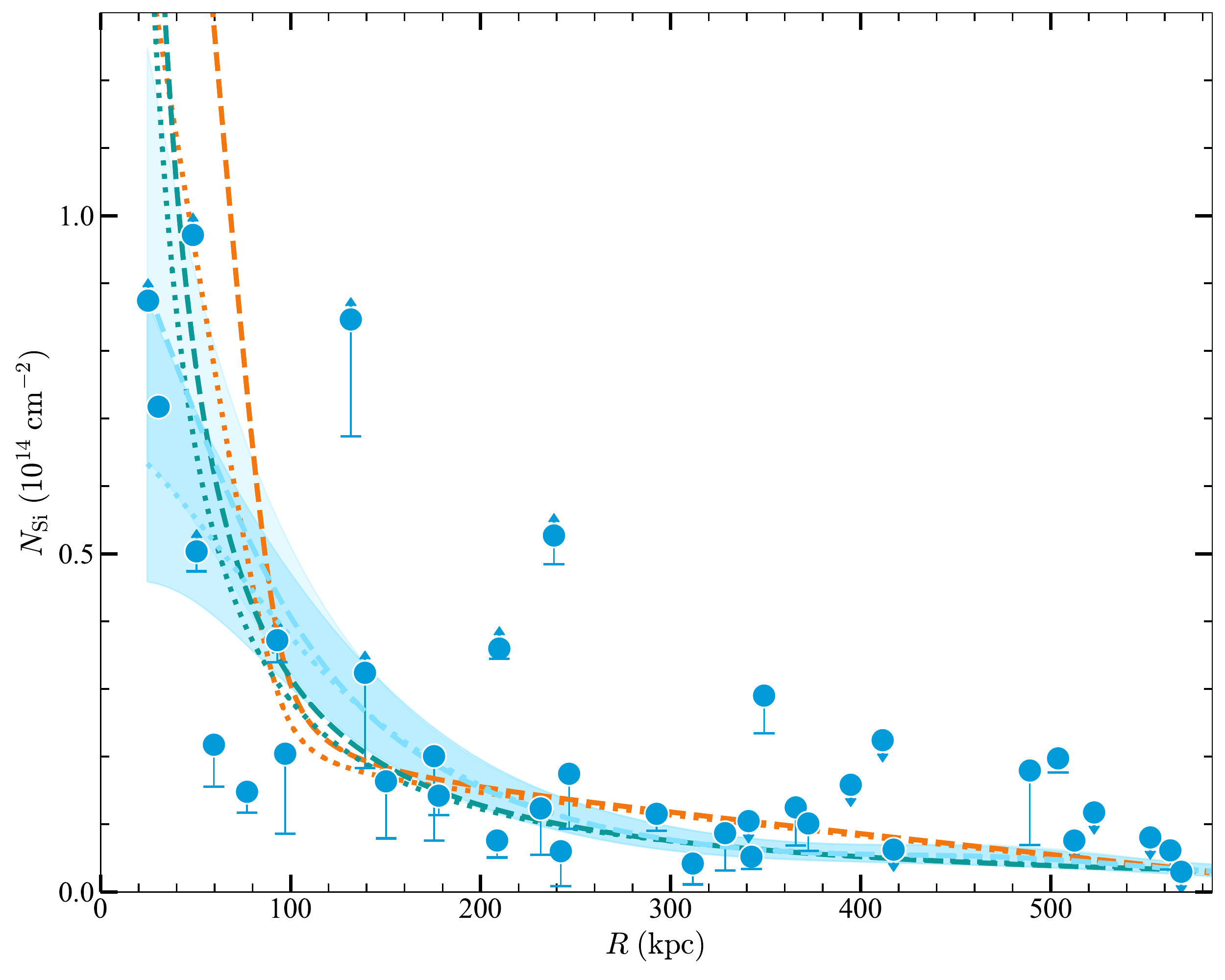}
\caption{Total column densities of Si (i.e., $N_{\rm Si} =  N_{\rm Si\,II} +  N_{\rm Si\,III} + N_{\rm Si\,IV}$) as a function of the projected distances from M31 of the background QSOs. The vertical-ticked bars show the range of values allowed if the upper limit of a given Si ion is negligible or not. The lower limits have upward arrows, and the upper limits are flagged using downward arrows. The orange, green, and blue curves are the H, SPL, GP-derived models to the data, respectively (see text for details regarding how censoring is treated in each model). The dotted and dashed curves correspond to model where the lower limits at $R<50$ kpc are increased by 0.3 or 0.6 dex. The blue areas correspond to the dispersion derived from the GP models (see Appendix~\ref{a-model} for more detail).
\label{f-coltotsi-vs-rho}}
\end{figure}

With \siii, \siiii, and \siiv, we can estimate the total column density of Si within the ionization energy range 8--45 eV without any ionization modeling. Gas in this range should constitute the bulk of the cool photoionized CGM of M31 (see \S\ref{s-ionization}). In Fig.~\ref{f-coltotsi-vs-rho}, we show the total column density of Si (estimated following \S\ref{s-ionization}) against the projected distance $R$ from M31. The vertical-ticked bar in Fig.~\ref{f-coltotsi-vs-rho} indicate data with some upper limits, and the length of the vertical bar represents the range of $N_{\rm Si}$ values allowed between cases 1 and 2 (see \S\ref{s-ionization}).

Fig.~\ref{f-coltotsi-vs-rho} reinforces the conclusions observed from the individual low ions in Figs.~\ref{f-coltot-vs-rho} and \ref{f-col-vs-rho}. Overall there is a decrease of the column density of Si at larger $R$. This decrease has a much stronger gradient in the inner region of the M31 CGM between ($R\la 25$ kpc) and about $R\sim 100$--$150$ kpc than at $R\ga 150$ kpc. $N_{\rm Si}$ changes by a factor $>5$--$10$ between about 25 kpc and 150 kpc while it changes by a factor $\la 2$ between 150 kpc and 300 kpc. The scatter in $N_{\rm Si}$ is also larger in the inner regions of the CGM than beyond $\ga 120$--150 kpc.

To model this overall trend (which is also useful to determine the baryon and metal content of the CGM, see \S\ref{s-mass}), we consider three models, a hyperbolic (H) model, single-power law (SPL) model, and a Gaussian Process (GP) model. We refer the refer to Appendix~\ref{a-model} where we fully explain the modeling process and how lower and upper limits are accounted for in the modeling. Fig.~\ref{f-coltotsi-vs-rho} shows these 3 models greatly overlap.  The non-parametric GP model overlaps more with the SPL model than with the H fit in the range $250\la R\la 400$ kpc and at $R<90$ kpc (especially for the high H fit, see Fig.~\ref{f-coltotsi-vs-rho}). While there are some differences between these models (and we will explore in \S\ref{s-mass} how these affect the mass estimates of the CGM), they all further confirm the strong evolution of the column density of Si with $R$ between $\la 25$ and $90$--150 kpc and a much shallower evolution with $R$ beyond 200 kpc. In \S\ref{s-mass}, we  use these models to constrain the metal and baryon masses of the cool CGM gas probed by \siii, \siiii, \siiv.

\subsection{Covering Factors}\label{s-fc}
As noted in \S\ref{s-n-vs-r}, the diagnostic ions behave differently with $R$ in a way that probably reflects the underlying physical conditions. For example, \siii\ has a high detection rate within  $R<100$ kpc, a sharp drop beyond $R>100$ kpc, and a total absence at $R\ga 240$ kpc (see Figs.~\ref{f-coltot-vs-rho} and \ref{f-col-vs-rho}). On the other hand, \siiii\ and \ovi\ are mostly detected at all $R$, but the column densities of \siiii\ fall significantly with $R$ while \ovi\ remains relatively flat. In this section, we quantity further the detection rates, or the covering factors, for each ion.

\begin{figure*}[tbp]
\epsscale{1}
\plotone{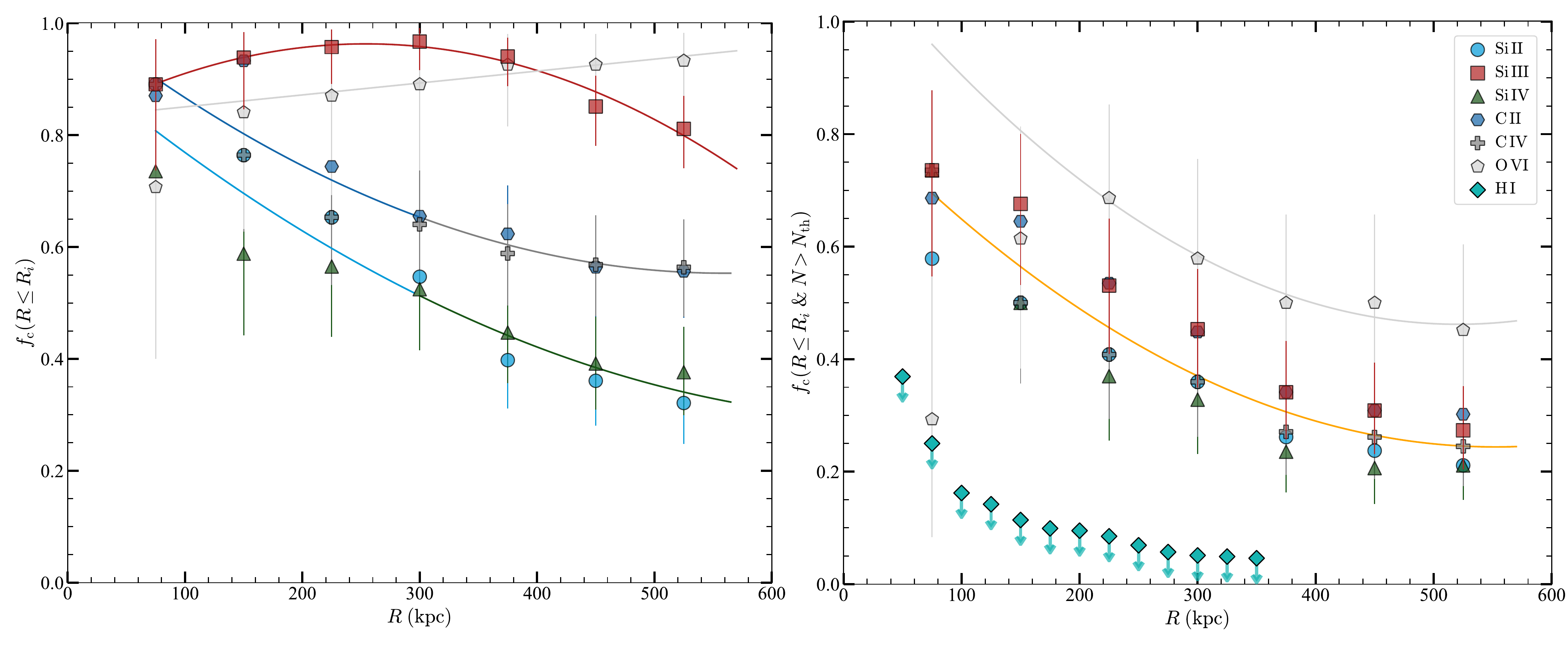}
\caption{Cumulative covering factors for impact parameters less than $R$ without ({\it left}) and with ({\it right}) some threshold cut on the column densities (for Si ions, $\log N_{\rm th} = 13$; for C ions, $\log N_{\rm th} = 13.8$; for \ovi, $\log N_{\rm th} = 14.5$, and for \hi, $\log N_{\rm th} = 17.6$, see text for more detail and \citetalias{howk17}). Confidence intervals (vertical bars) are at the 68\% level and data points are the median values. On the left panel, the solid lines are polynomial fits to median values of \fc\ for \siiii, \ovi, \cii--\civ, \siii--\siiv\ (i.e., taking the mean value of \fc\ between these two ions  at a given $R$). On the right panel, the orange line is  a polynomial fit to the mean values of \fc\ for \cii, \civ, \siii, \siiii, and \siiv\ while the gray line is polynomial fit to the median values of \fc\ for \ovi.
\label{f-fccum-vs-rho}}
\end{figure*}

To calculate the covering factors of the low and high ions, we follow the methodology described in \citetalias{howk17} for \hi\ by assuming a binomial distribution. We assess the likelihood function for values of the covering factor given the number of detections against the total sample, i.e., the number of targets within a given impact parameter range (see \citealt{cameron11}). As demonstrated by \citet{cameron11}, the normalized likelihood function for calculating the Bayesian confidence intervals on a binomial distribution with a non-informative (uniform) prior follows a $\beta$-distribution.

In Fig.~\ref{f-fccum-vs-rho}, we show the {\em cumulative} covering factors (\fc) for the various ions, where each point represents the covering factor for all impact parameters less than the given value of $R$. The vertical bars are 68\% confidence intervals. As discussed in \S\S\ref{s-n-vs-r}, \ref{s-nsi-vs-r}, for all the ions but \ovi, the highest column densities are only observed at $R\la 100$--150 kpc, with a sharp decrease beyond that. For the covering factors, we therefore consider 1) the entire sample (most of the upper limits---non-detections---are at the level of lowest column densities of a detected absorption, so it is adequate to do that), and 2) the sample where we set a  threshold column density ($N_{\rm th}$) to be included in the sample. In the left panel of Fig.~\ref{f-fccum-vs-rho}, we show the first case, while in the right panel, we focus on the strong absorbers only. For the Si ions, we use $\log N_{\rm th} = 13$; for the C ions, $\log N_{\rm th} = 13.8$; for \ovi, $\log N_{\rm th} = 14.5$. These threshold column densities are chosen based on Fig.~\ref{f-coltot-vs-rho}. We also show in the right panel of Fig.~\ref{f-fccum-vs-rho}, the  results for the \hi\ emission from \citetalias{howk17}.

These results must be interpreted in light of the fact that the intrinsic strength of the diagnostic lines varies by ion. The oscillator strength, $f \lambda$, of these different ions are listed in Table~\ref{t-strength} along with the solar abundances of these elements. The optical depth is such that $\tau \propto f\lambda N $ (see \S\ref{s-aod}) and $f\lambda $ is a good representation of the strength of a given transition. For the Si ions, \siiii\ has the strongest transition, a factor 2.7--5.5 stronger than \siiv\ and a factor 1.3--5.7 stronger than \siii\ (the weaker \siii\ $\lambda$1526 is sometimes used but mostly to better constrain the column density of \siii\ if the absorption is strong). \siii\ and \siiv\ have more comparable strength, which is also the case between \cii\ and \civ.  Comparing between different species, while $(f\lambda)_{\rm Si\,III} \simeq 14.4 (f\lambda)_{\rm O\,VI}$, but this is counter-balanced by oxygen being 15 times more abundant than silicon (and a similar conclusion applies comparing \siiii\ with \cii\ or \civ).

With that in mind, we first consider the left panel of Fig.~\ref{f-fccum-vs-rho}, i.e., where all the absorbers irrespective of their absorption strengths are taken into account to estimate the cumulative covering factors. We fitted 4 low-degree polynomials to the data: \siiii, \ovi, and treating in pair \cii-\civ\ and \siii-\siiv\ as they appear to follow each other reasonably well, respectively. For \cii-\civ\ and \siii-\siiv, we fit the mean covering factors of each ionic pair. For \ovi, we only fitted data beyond 200 kpc owing to the smaller size sample (there are only 3 data points within 200 kpc and 11 in total, see Fig.~\ref{f-coltot-vs-rho}). It is striking how the cumulative covering factors of \siiii\ and \ovi\ vary with $R$ quite differently from each other and from the other ions. The cumulative covering factor of \siiii\ increases with $R$, reaches a maximum somewhere between 250 and 300 kpc, and then decreases, but still remains much higher than \fc\ of  \cii-\civ\ or \siii-\siiv. The cumulative covering factor of \ovi\ monotonically increases with $R$ up to $R\sim 569$ kpc.  In contrast, while the cumulative covering factors of \cii-\civ\ or \siii-\siiv\ are offset from each other, they both monotonically decrease with $R$. There seems to be a plateau in \cii-\civ\ covering factor beyond 400 kpc, which is not observed for \siii\ or \siiv.

Turning to the right panel of Fig.~\ref{f-fccum-vs-rho} where we show \fc\ for a given column density threshold that changes with species (see above), the relation between \fc\ and $R$ is quite different. For all the ions, the cumulative covering factors monotonically decrease with increasing $R$. For  \cii, \civ, \siii, \siiii, and \siiv, the covering factors are essentially the same within $1 \sigma$, and the orange line in Fig.~\ref{f-fccum-vs-rho} shows a second-degree polynomial fit to the mean values of \fc\ between these different ions. Ignoring data at $R<200$ kpc owing to the small sample size, \ovi\ has a similar evolution of \fc\ with $R$, but overall \fc\ is tentatively a factor $\sim$1.5 times larger than for the other ions at any $R$.

The contrast between the two panels of Fig.~\ref{f-fccum-vs-rho}  strongly suggests that the CGM of M31 has three main populations of absorbers: 1) the strong absorbers that are found mostly at $R \la 100$--$150$ kpc ($0.3$--$0.5 \rvir$) probing the denser regions and multiple gas-phase (singly to highly ionized gas) of the CGM, 2) weak absorbers probing the diffuse CGM traced principally by \siiii\ (but also observed in higher ions and more rarely in \cii) that are found at any surveyed $R$ but more frequent at $R\la \rvir$, and 3) hotter, more diffuse CGM probed by \ovi, \ovi\ having the unique property compared to the ions that its column density remains largely invariant with the radius of the M31 CGM.

\subsection{Ion ratios and their Relation with $R$}\label{s-ratio-vs-r}

\begin{figure}[tbp]
\epsscale{1.2}
\plotone{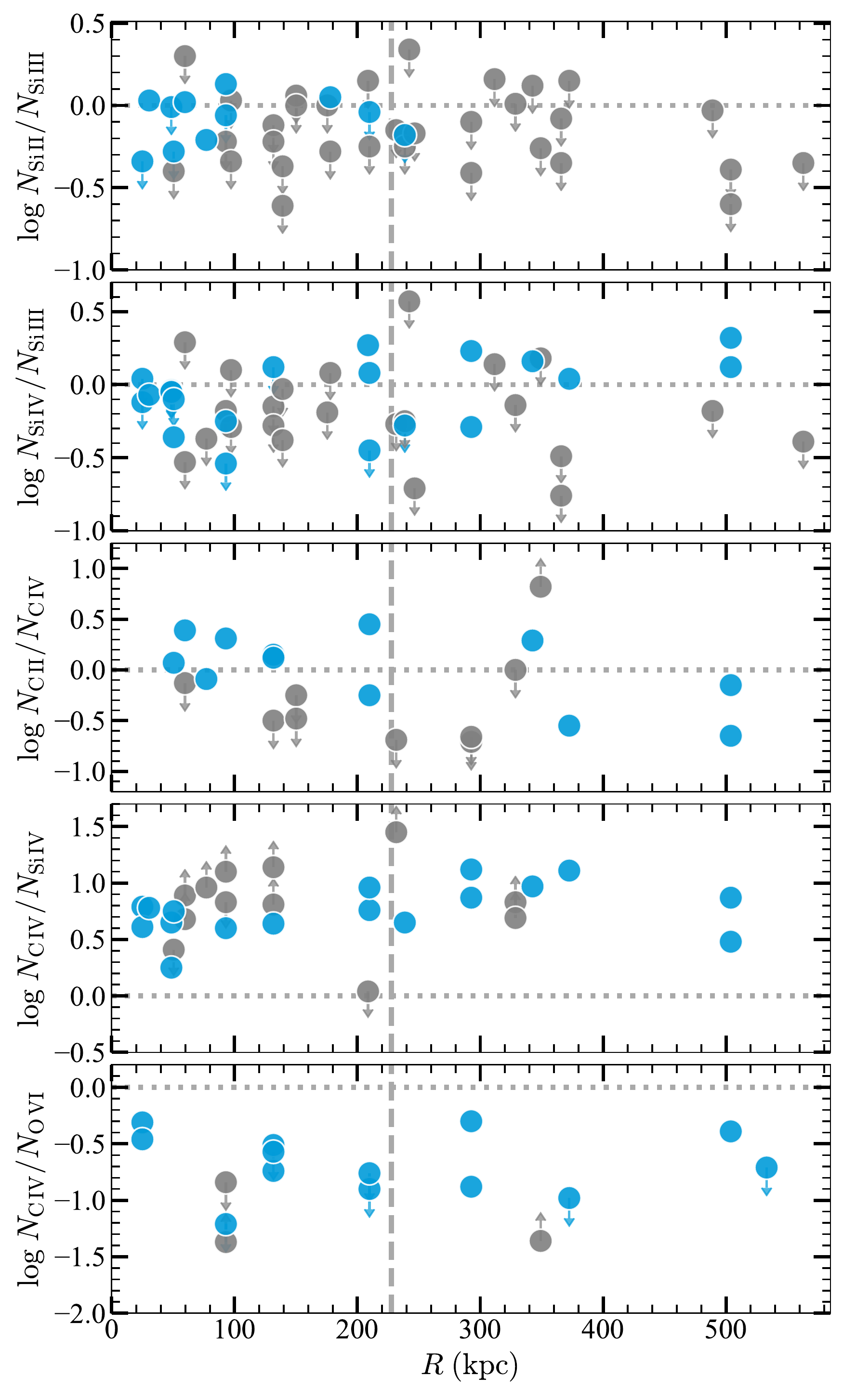}
\caption{Logarithmic column density ratios of different ions as a function of the projected distances from M31 of the background QSOs. The column densities in individual components are compared to estimate the ionic ratios. Blue symbols indicate that both ions in the ratio are detected. Blue down or up arrows indicate that the absorption is saturated for the ion in the denominator or numerator of the ratio. Gray symbols indicate that one of the ions in the ratio is not detected at $>2\sigma$. The components associated with the MS have again been removed. The dashed vertical lines mark $R_{200}$.
\label{f-ratio-vs-rho}}
\end{figure}

In \S\ref{s-ionization}, we show that the ratio of \oi\ to Si ions provides a direct estimate of the ionization fraction of the CGM gas of M31. Using ratios of the main ions studied here (\cii, \civ, \siii, \siiii, \siiv, \ovi), we can further constrain the ionization and physical conditions in the CGM of M31 and how they may change with $R$. To estimate the ionic ratios, we consider the component analysis of the absorption profiles, i.e., we compare the column densities estimated over the same velocity range. However, coincident velocities do not necessarily mean that they probe the same gas, especially if their ioniziation potentials are quite different (such as for \cii\ and \civ).  In Fig.~\ref{f-ratio-vs-rho}, we show the results for several ion ratios as a function of $R$.

\subsubsection{The \siii/\siiii\ and \siiv/\siiii\ ratios}\label{s-si2-si3-si4}
The \siii/\siiii\ and \siiv/\siiii\ ratios are particularly useful because they trace different ionization levels independently of relative elemental abundances. The ionization potentials for these ions are 8.1--16.3 eV for \siii, 16.3--33.5 eV for \siiii, and 33.5--45.1 eV for \siiv. The top two panels of Fig.~\ref{f-ratio-vs-rho} show the ratios \siii/\siiii\ and \siiv/\siiii\ as a function of $R$. In both panels, there are many upper limits and there is no evidence of any correlation with $R$, except perhaps for the for the \siii/\siiii\ ratio where the only detections of \siii\ are at $R\la R_{200}$ (see also Fig.~\ref{f-col-vs-rho}).

With so many upper limits, we use the Kaplan-Meier estimator (see \S\ref{s-abund}) to estimate the mean of these ratios: $\langle \log N_{\rm Si\,II}/N_{\rm Si\,III}\rangle = (-0.50 \pm 0.04) \pm 0.23$ (mean, error on the mean from the Kaplan-Meier estimator, and standard deviation for 44 data points with 38 upper limits) and $\langle \log N_{\rm Si\,IV}/N_{\rm Si\,III}\rangle = (-0.49 \pm 0.07) \pm 0.20$ (43 data points with 32 upper limits). There are only 4/44 components where $ \log N_{\rm Si\,II}/N_{\rm Si\,III} \simeq 0$ and 8/43  where $ \log N_{\rm Si\,IV}/N_{\rm Si\,III} \ga 0$. In the latter cases, \siiv\ could be produced by another mechanism such as collisional ionization.  Among the three Si ions in our survey, \siiii\ is the dominant ion at any $R$ from M31 in the ionizing energy range 8.1--45.1 eV. Ions (of any element) with ionizing energies in the range 16.3--33.5 eV are therefore expected to be dominant ions at least for processes that are dominated by photoionization. 

The \siii/\siiii\ ratio has previously been used to constrain the properties of the photoionized gas. According to photoionization modeling produced by \citet{oppenheimer18a}, an ionic ratio of $\langle \log N_{\rm Si\,II}/N_{\rm Si\,III}\rangle = (-0.50 \pm 0.04) \pm 0.23$ would imply gas density in the range $-3 \la \mlnnh \la -2.5 $ and a temperature of the gas around $10^4$ K (see Fig.~16 in \citealt{oppenheimer18a}). 

\subsubsection{The \cii/\civ\ ratio}\label{s-c2-c4}
For the \cii/\civ\ ratio the ionizing energy ranges are well separated with 11.3--24.3 eV for \cii\ and 47.9--64.4 eV for \civ. In fact with an ionization potential above the \heii\ ionization edge at 54.4 eV, \civ\ can be also produced not just photoionization but also by collisional ionization. Therefore \cii\ and \civ\ are unlikely to probe the same ionization mechanisms or be in a gas with the same density. We note that \cii\ has ionization energies that overlap with \siiii\ and larger than those of \siii, which certainly explain the presence of \cii\ beyond $R_{200}$ where \siii\ is systematically not detected.

The third panel of Fig.~\ref{f-ratio-vs-rho} shows the \cii/\civ\ ratios. There is again no strong relationship between \cii/\civ\ and $R$, but $ \log N_{\rm C\,II}/N_{\rm C\,IV} \ga 0$ is more frequently observed at $R<R_{200}$ (6/12) than at $R>R_{200}$ (2/9), consistent with the observation made in \S\ref{s-n-vs-r} that the gas becomes more highly ionized as $R$ increases. With the survival analysis (considering the only lower limit as a detection), we find $\langle \log N_{\rm C\,II}/N_{\rm C\,IV}\rangle = (-0.21 \pm 0.11) \pm 0.40$ (21 data points with 8 upper limits). Considering data at $R<R_{200}$, we have $\langle \log N_{\rm C\,II}/N_{\rm C\,IV}\rangle = (-0.07 \pm 0.10) \pm 0.28$ (12 data points with 4 upper limits), while at $R\ge R_{200}$, we find  $\langle \log N_{\rm C\,II}/N_{\rm C\,IV}\rangle = (-0.33 \pm 0.18) \pm 0.22$ (9 data points with 4 upper limits), confirming again that the gas is more ionized and also more highly ionized at  $R>R_{200}$.

\subsubsection{The \civ/\siiv\ ratio}\label{s-c4-si4}
For the \civ/\siiv\ ratio, different species are compared, but as we discuss in \S\ref{s-abund}, the relative abundances of C and Si are consistent with the solar ratio owing to little evidence of any strong dust depletion or nucleosynthesis effects, i.e., these effects should not affect the observed ratio of \civ/\siiv. \siiv\ and \civ\ have near adjacent ionization energies 33.5--45.1 eV to 47.9--64.4 eV, respectively. Both photoionization and collisional ionization processes can be important at these ionizing energies, but if $\log N_{\rm C\,IV}/N_{\rm Si\,IV}>0$, then the ionization from hot stars is unimportant (see Fig.~13 in \citealt{lehner11b}), which is nearly always the case, as illustrated in Fig.~\ref{f-ratio-vs-rho}. A harder photoionizing spectrum or collisional ionization must be at play to explain the origin of these ions.

Fig.~\ref{f-ratio-vs-rho} suggest a moderate correlation between $\log N_{\rm C\,IV}/N_{\rm Si\,IV}$ and $R$. If the two data points beyond 400 kpc are removed (and treating the limits as actual values), a  Spearman rank order implies a monotonic correlation between $\log N_{\rm C\,IV}/N_{\rm Si\,IV}$ and $R$ with a correlation coefficient $r_{\rm S} = +0.45$ and  $p=0.019$ for the gas at $R<1.2\rvir$. Considering the entire sample, the Spearman rank test yield $r_{\rm S} = 0.34$ and  $p=0.07$. This is again consistent with our earlier conclusion that the gas becomes more highly ionized as $R$ increases.  With the survival analysis (considering the 3 upper limits as detections),\footnote{If these 3 upper limits are included or excluded from the sample, the means are essentially the same. } we find $\langle \log N_{\rm C\,IV}/N_{\rm Si\,IV}\rangle = (+0.87 \pm 0.07) \pm 0.24$ (29 data points with 9 lower limits). This is about a factor 1.9 larger than the mean derived for the broad \civ\ and \siiv\ components in the Milky Way disk and low-halo \citep{lehner11b}, which is about $1\sigma$ larger.

\subsubsection{The \civ/\ovi\ ratio}\label{s-c4-o6}
Finally, in the last panel of Fig.~\ref{f-ratio-vs-rho}, we show the \civ/\ovi\ ratio as a function of $R$. As for the \civ/\siiv\ ratio, different species are compared, and for the same reasons, the relative dust depletion or nucleosynthesis effects should be negligible. With 113.9--138.1 eV ionizing energies needed to produce \ovi, this is the highest high ion in the sample and as we demonstrated in the previous section the \ovi\ properties (covering factor and column density as a function of $R$) are quite unique.  Not surprisingly Fig.~\ref{f-ratio-vs-rho} does not reveal any relation between $\log N_{\rm C\,IV}/N_{\rm O\,VI}$ and $R$.

If we treat the two lower limits as detections, then the survival analysis yields $\langle \log N_{\rm C\,IV}/N_{\rm O\,VI}\rangle = (-0.93 \pm 0.11) \pm 0.32$ (16 data points with 6 upper limits). The mean and range of $\log N_{\rm C\,IV}/N_{\rm O\,VI}$ are smaller than observed in the Milky Way disk and low halo where the full range varies from $-1$ to $+1$ dex (see, e.g., Fig.~14 of \citealt{lehner11b}). This demonstrates that the highly ionized gas in the  113.9--138.1 eV range is much more important than in the 47.9--64.4 eV range at any $R$ of the M31 CGM.

\subsection{Metal and Baryon Mass of the M31 CGM}\label{s-mass}

With a better understanding of the column density variation with $R$, we can estimate with more confidence the metal and baryon mass of the M31 CGM than in our original survey where we had very little information between 50 and 300 kpc \citepalias{lehner15}. The metal mass can be directly estimated from the column densities of the metal ions. With the silicon ions, we have information on its three dominant ionization stages in the $T<7\times 10^4$ K ionized gas (ionizing energies in the range 8--45 eV, see \S\ref{s-nsi-vs-r}), so we can obtain a direct measured metal mass  without any major ionization corrections.  Following \citetalias{lehner15} (and see also \citealt{peeples14}),  the metal mass of the cool photoionized CGM is
$$
M^{\rm cool}_{\rm Z} =  2\pi\, \mu_{\rm Si}^{-1}\, m_{\rm Si}\, \int R \,N_{\rm Si}(R) \,dR\,,
$$
where  $\mu_{\rm Si}=0.064$ is the solar mass fraction of metals in silicon (i.e., $12+\log ({\rm Si/H})_\sun = 7.51$ and $Z_\sun = 0.0142$ from \citealt{asplund09}), $m_{\rm Si} = 28 m_{\rm p}$, and for $N_{\rm Si}(R)$ we use the hyperbola (``H model'', Eqn.~\ref{e-colsi-vs-r}), single power-law (``SPL model", Eqn.~\ref{e-colsi-vs-r1}), and GP models that we determine in \S\ref{s-nsi-vs-r} and Appendix~\ref{a-model} (see  Fig.~\ref{f-coltotsi-vs-rho}).

A direct method to estimate the total mass is to convert the total observed column density of Si  to total hydrogen column density via $N_{\rm H}  = N_{\rm H\,I} + N_{\rm H\,II} =   N_{\rm Si}\, ({\rm Si}/{\rm H})_\sun^{-1}\, ({\rm Z/Z}_\sun)^{-1}$. The baryonic mass of the CGM of M31 is then:
$$
M^{\rm cool}_{\rm g} =  2\pi\, m_{\rm H}\, \mu\, \fc \,  \Big(\frac{\rm Si}{\rm H}\Big)_\sun^{-1}  \Big(\frac{Z}{Z_\sun}\Big)^{-1}\, \int R\, N_{\rm Si}(R)\, dR\,,
$$
where  $\mu \simeq 1.4$ (to correct for the presence of He), $m_{\rm H} = 1.67\times 10^{-24}$ g is the hydrogen mass, \fc\  is covering fraction (that is 1 over the considered radii), and $\log ({\rm Si/H})_\sun = -4.49$ is the solar abundance of Si. Inserting the values for each parameter, $M^{\rm cool}_{\rm g}$ can be simply written in terms of $M^{\rm cool}_{\rm Z}$: $M^{\rm cool}_{\rm g} \simeq 10^2 (Z/Z_\sun)^{-1} M^{\rm cool}_{\rm Z}$.

In Table~\ref{t-mass}, we summarize the estimated metal mass over different regions of the CGM for the three models of $N_{\rm Si}(R)$, within $R_{\rm 200}$ (first entry), within \rvir\ (second entry), within $1/2 \rvir$ (third entry), between $1/2 \rvir$ and \rvir\ (fourth entry), and within 360 kpc (fifth entry), which corresponds to the radius where at least one of the Si ions is always detected (beyond that, the number of detections drastically plummets). A key difference between the H/SPL models and the GP model is that the range of values for the H/SPL models is derived using the low (dotted) and high (dashed) curves in Fig.~\ref{f-coltotsi-vs-rho} while for the GP models we actually use the standard deviations from the low and high models (i.e., the top and bottom of the shaded blue curve in Fig.~\ref{f-coltotsi-vs-rho}). Hence it is not surprising that the mass ranges for the GP model are larger. Nevertheless there is a large overlap between the three models. As the GP results overlap with the other models and provide empirical confidence intervals, we adopt them for the remaining of the paper. At \rvir, the metal and cool gas masses are therefore $(2.0 \pm 0.5)  \times 10^7$ and $2\times 10^9\, (Z/Z_\sun)^{-1}$ M$_\sun$, respectively. Owing to the new functional form of $N_{\rm Si}(R)$ and how the lower limits are treated, this explains the factor 1.4 times increase in the metal mass compared to that derived in \citetalias{lehner15}.

These masses do not include the more highly ionized gas traced by \ovi\ or \civ. Even though the sample with \ovi\ is smaller than \civ, we use \ovi\ to probe the higher ionization gas phase because as we show above the properties of \ovi\ (column density and covering fraction as a function of $R$) are quite different from all the other ions, including \civ, which behaves more like the other, lower ions. Furthermore, \citet{lehner11b} using 1.5--3 \km\ resolution UV spectra show that \civ\ can probe cool and hotter gas while the profiles of \nv\ and \ovi\ are typically broad and more consistent with hotter gas. Since \ovi\ is always detected and there is little evidence for variation with $R$ (see Fig.~\ref{f-coltot-vs-rho}), we can simply use the mean column density $\mlnovi = 14.46 \pm 0.10$ (error on the mean using the survival method for censoring) to estimate the baryon mass assuming a spherical distribution:
$$
M^{\rm warm}_{\rm g} = \pi r^2 \, m_{\rm H}\, \mu\, \fc \, \frac{N_{\rm O\,VI}}{f^i_{\rm O\,VI}}\, \Big(\frac{\rm O}{\rm H}\Big)_\sun^{-1}\,  \Big(\frac{Z}{Z_\sun}\Big)^{-1},
$$
where the \ovi\ ionization fraction is $f^i_{\rm O\,VI} \la 0.2$ (see \S\ref{s-ionization}), $\fc =1$ for \ovi\ at any $R$ (see Fig.~\ref{f-coltot-vs-rho}). At \rvir, we find $M^{\rm warm}_{\rm g}\ga  9.3\times 10^9\,(Z/Z_\sun)^{-1}$ M$_\sun$ or  $M^{\rm warm}_{\rm g} \ga 4.4 \,M^{\rm cool}_{\rm g}$ (assuming the metallicity is about similar in the cooler and hotter gas-phases).  At $R_{200}$, we find $M^{\rm warm}_{\rm g}\ga  5.5\times 10^9\,(Z/Z_\sun)^{-1}$ M$_\sun$ (assuming the metallicity is similar in the cooler and hotter gas-phases). These are lower limits because the fraction of \ovi\ could be much smaller than 20\% and the metallicity of the cool or warm ionized gas is also likely to less than solar (see below). In terms of metal mass in the highly ionized gas-phase, we have $M^{\rm warm}_{\rm g} \simeq 10^2 (Z/Z_\sun)^{-1} M^{\rm warm}_{\rm Z}$ and hence also $ M^{\rm warm}_{\rm Z} \ga 4.4 \, M^{\rm cool}_{\rm Z}$. Since \ovi\ is detected out to the maximum surveyed radius of 569 kpc, and at that radius (i.e., $1.9 \rvir$), $M^{\rm warm}_{\rm g}\ga  34 \times 10^9\,(Z/Z_\sun)^{-1}$ M$_\sun$.


By combining both the cool and hot gas-phase masses, we can find the baryon mass for gas in the temperature range $\sim 10^3$--$10^{5.5}$ K at \rvir:
$$
\begin{aligned}
M_{\rm g} & = M^{\rm cool}_{\rm g} + M^{\rm warm}_{\rm g}\\
& \ga 1.1\times 10^{10}\,   \Big(\frac{Z}{Z_\sun}\Big)^{-1} \; {\rm M_\sun}\,.
\end{aligned}
$$
Within $R_{200}$, the total mass $M_{\rm g} \ga 7.2\times 10^9$ M$_\sun$. As the stellar mass of M31 is about $10^{11}$ M$_{\sun}$ \citep[e.g.,][]{geehan06,tamm12}, the mass of the diffuse weakly and highly ionized CGM of M31 within $1\rvir$ is therefore at least 10\% of the stellar mass of M31 and could be significantly larger than 10\%.

This estimate does not take into account the hot ($T\ga 10^6$ K) coronal gas. The diffuse X-ray emission is observed to extend to about 30--70 kpc around a handful of massive, non-starbursting galaxies \citep{anderson11,bregman18} or in stacked images of galaxies \citep{anderson13,bregman18}, but beyond $50$ kpc, the CGM is too diffuse to be traced with X-ray imaging, even though a large mass could be present. Using the results summarized recently by \citet{bregman18}, the hot gas mass of spiral galaxy halos is in the range $M^{\rm hot}_{\rm g}\simeq 1$--$10 \times 10^9$ M$_\sun$ within 50 kpc. For M31, $M_{\rm g} = M^{\rm cool}_{\rm g} + M^{\rm warm}_{\rm g} \ga 0.4\times 10^9$ M$_\sun$ within 50 kpc. Extrapolating the X-ray results to \rvir, \citet{bregman18} find masses of the hot X-ray gas similar to the stellar masses of these galaxies in the range $M^{\rm hot}_{\rm g}\simeq1$--$10\times 10^{11}$ M$_\sun$. For the MW hot halo within $1\rvir$, \citet{gupta17} (but see also \citealt{gupta12,gupta14,wang12,henley14}) derive $3$--$10\times 10^{10}$   M$_\sun$, i.e., on the low side of the mass range listed in \citet{bregman18}. The hot gas could therefore dominate the mass of the CGM of M31. There are, however, two caveats to that latter conclusion. First, if $f_{\rm O\,VI}\ll 0.2$, then $M^{\rm warm}_{\rm g}$ could be become much larger. Second, the metallicity of the hot X-ray gas ranges from 0.1 to $0.5 Z_\sun$ with a mean metallicity of $0.3 Z_\sun$ \citep{bregman18,gupta17}, while for the cooler gas we have conservatively adopted a solar abundance. If instead we adopt a  $0.3 Z_\sun$  metallicity (consistent with the rough limits set in \S\S\ref{s-metallicity}, \ref{s-abund}), then $M_{\rm g} \simeq 3.7\times 10^{10}$ M$_\sun$ at \rvir, which is now comparable to the hot halo mass of the MW. If we adopt the average metallicity derived for the X-ray gas, then $M^{\rm cool}_{\rm g} + M^{\rm warm}_{\rm g}$ would be comparable to the hot gas mass if $M^{\rm hot}_{\rm g} \sim 5\times 10^{10}$ M$_\sun$ at \rvir\ for M31. Depending on the true metallicities and the actual state of ionization, the cool and warm gas in the M31 halo could therefore contribute to a substantial enhancement of the total baryonic mass compared to our conservative assumptions.

\subsection{Mapping the Metal Surface Densities in the CGM of M31}\label{s-map-metal}

\begin{figure*}[tbp]
\epsscale{1.}
\plotone{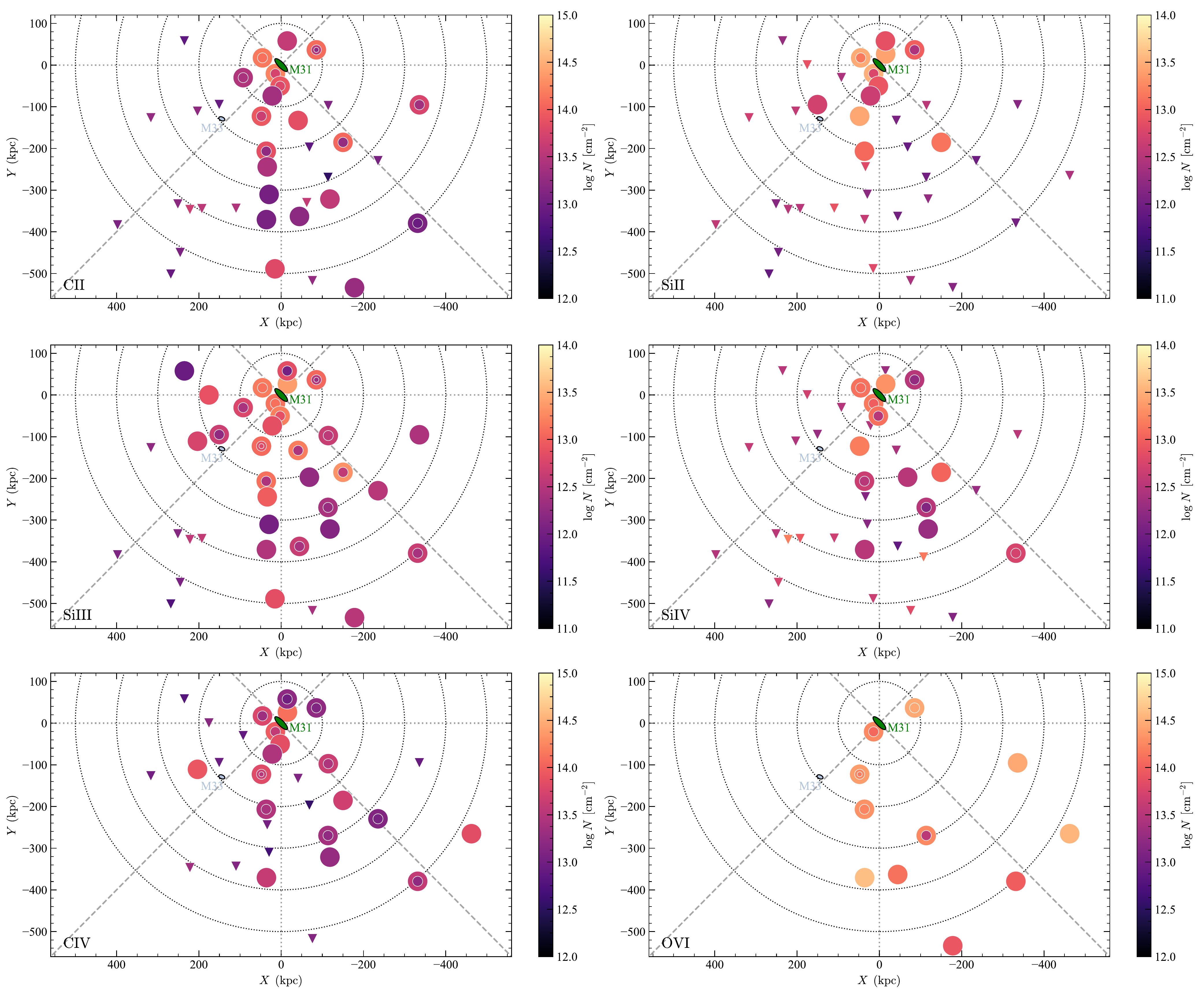}
\caption{Positions of the Project AMIGA targets relative to the M31, where the axes show the physical impact parameter from the center of M31 (north is up, east to the left). Dotted circles are centered on M31 to mark 100 kpc intervals. The dashes lines represent the projected minor and major axes of M31 and the thin dotted lines are $\pm 45\degr$ from the major/minor axes (which by definition of the coordinate systems also correspond to the vertical and horizontal zero-axis). Each panel corresponds to a different ion. In each panel, the column densities of each velocity component are shown and color coded according the vertical color bar.  Circles represent detections while triangles are non-detections. Circles with several color indicate the observed absorption along the sightlines have more than one component. 
\label{f-colmap}}
\end{figure*}

Thus far, we have ignored the distribution of the targets in azimuthal angle ($\Phi$) relative to the projected minor and major axes of M31, where different physical processes may occur. In Fig.~\ref{f-colmap}, we show the distribution of the column densities of each ion in the X--Y plane near M31 where the circles represent detections and downward triangles are non-detections. Multiple colors in a given circle indicate several components along that sightline for that ion. In that figure, we also show the projected minor and major axes of M31 (dashed lines). The overall trends that are readily apparent from Fig.~\ref{f-colmap} are the ones already described in the previous sections: 1) overall the column density decreases with increasing $R$, 2)  the decrease in $N$ is much stronger for low ions than high ions, 3) \siiii\ and \ovi\ are observed at any $R$ while singly ionized species tend to be more frequently observed at small impact parameters. This figure (and Fig.~\ref{f-col-vs-rho}) also reveals that absorption with two or more components is observed more frequently at $R<200$ kpc: using \siiii, 64\%--86\% of the sightlines have at least 2 velocity components at $R<200$ kpc, while this drops to 14\%--31\% at $R>200$ kpc (68\% confidence intervals using the Wilson score interval); similar results are found using the other ions. However, the complexity of the velocity profiles does not change with $\Phi$.

Considering various radius ranges (e.g., 25--50 kpc, 50--100 kpc, etc.) up to $1\rvir$, there is no indication that the column densities strongly depend on $\Phi$. Considering \siiii\ first, it is equally detected along the projected major and minor axes and in-between (wherever there is a sigthline) and overall the strength of the absorption mostly depends on $R$, not $\Phi$. Considering the other ions, they all show a mixture of detections and non-detections, and the non-detections (that are mostly beyond 50 kpc) are not preferentially observed along a certain axis or one of the regions shown in Fig.~\ref{f-colmap}. We therefore find no strong evidence of an azimuthal dependence in the column densities. 

Beyond $\ga 1.1 \rvir$, the situation is different with all but one detection (in \civ\ and \ovi\ only) being near the southern projected major axis and about $52\degr$ east off near the $X=0$ kpc axis. There is detection in this region of \siiii, \civ, \siiv, \ovi, and also \cii. That is the main region where \cii\ is detected beyond 200 kpc. In contrast, between the $X=0$ kpc axis and southern projected minor axis, the only region where there are several QSOs beyond \rvir, there is no detection in any of the ions (excluding \ovi\ because there is no \fuse\ observations in these directions). Although that direction is suspiciously in the direction of the MS, it is very unlikely to be additional contamination from the MS because 1) the velocities would be off from those expected of the MS in these directions (see Fig.~\ref{f-nidever} and also \S\ref{s-map-vel}), and 2) there is no overall decrease of the column densities as $|b_{\rm MS}|$ increases, a trend observed for the components identified as the MS components (see Fig.~\ref{f-col-cont}). In fact, regarding the second point, the opposite trend is observed with the highest column densities being more frequently at $|b_{\rm MS}|\ga 15 \degr$ than near the MS main axis ($b_{\rm MS}\sim 0\degr$). Therefore while at $R<\rvir$ there is no apparent trend between $N$ and $\Phi$ for any ions (although we keep in mind that the azimuthal information for \ovi\ is minimal), most of the detections at $R>\rvir$ are near the southern projected major axis and $52\degr$ east off of that axis. 

The fact that the gas is observed mainly in a specific region of the CGM beyond \rvir\ suggests an IGM filament feeding the CGM of M31, as is observed in some cosmological simulations. In particular, \citet{nuza14} study the gas distribution in simulated recreations of MW and M31 using a constrained cosmological simulation of the Local Group from the Constrained Local UniversE Simulations (CLUES) project. In their Figures 3 and 6, they show different velocity and density projection maps where the central galaxy (M31 or MW) is edge-on. They find that some of the gas in the CGM can flow in a filament-like structure, coming from outside the virial radius all the way down to the galactic disk. 

\begin{figure*}[tbp]
\epsscale{1.}
\plotone{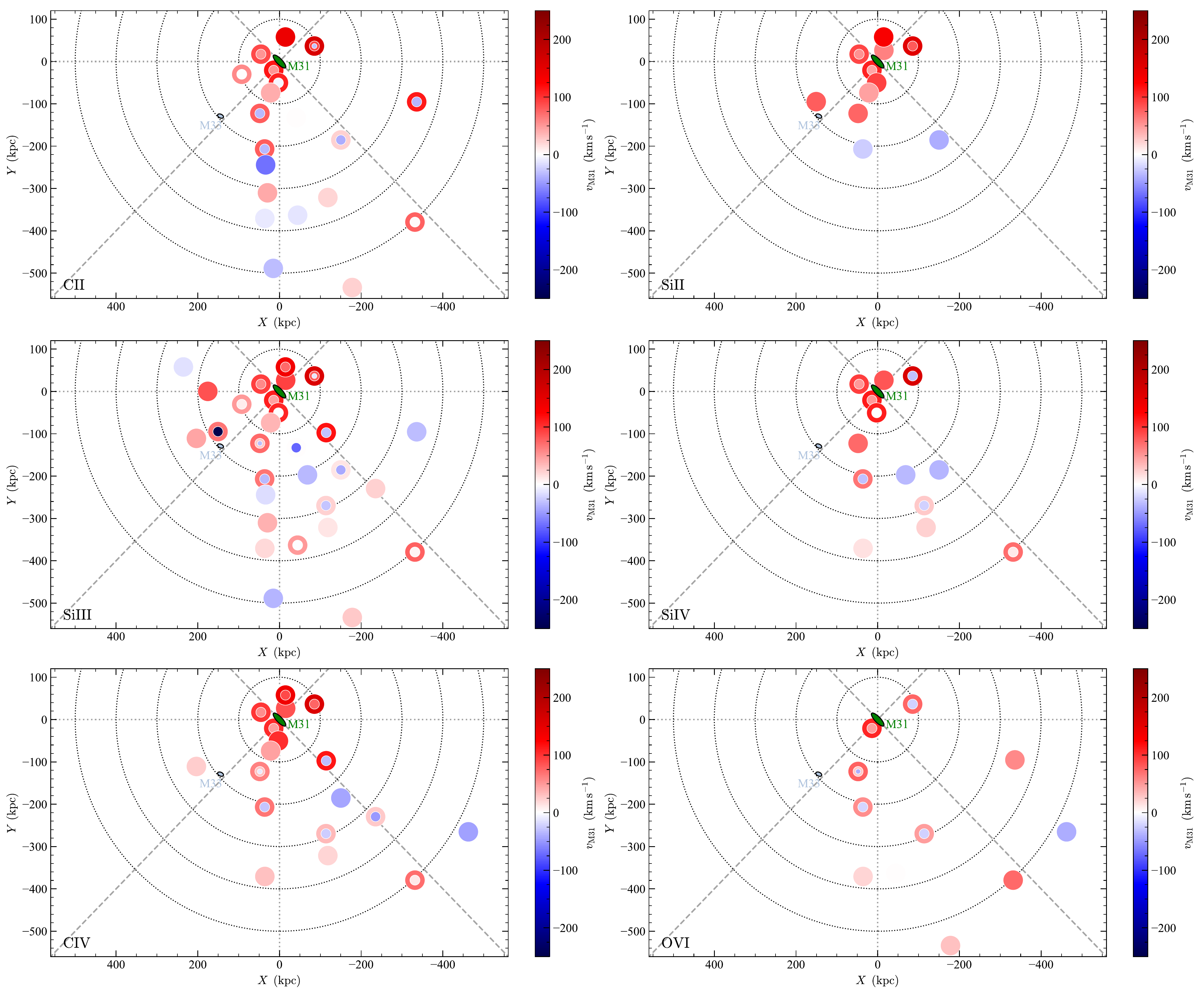}
\caption{Similar to Fig.~\ref{f-colmap}, but we now show the distribution of the M31 velocities for each component observed for each ion. Circles with several color indicate the observed absorption along the sightlines have more than one components. By definition, in the M31 velocity frame, an absorber with no peculiar velocity relative to M31's bulk motion has $v_{\rm M31}=0$ \km.
\label{f-velmap}}
\end{figure*}

\subsection{Mapping the Velocities in the CGM of M31}\label{s-map-vel}
\begin{figure*}[tbp]
\epsscale{1.}
\plotone{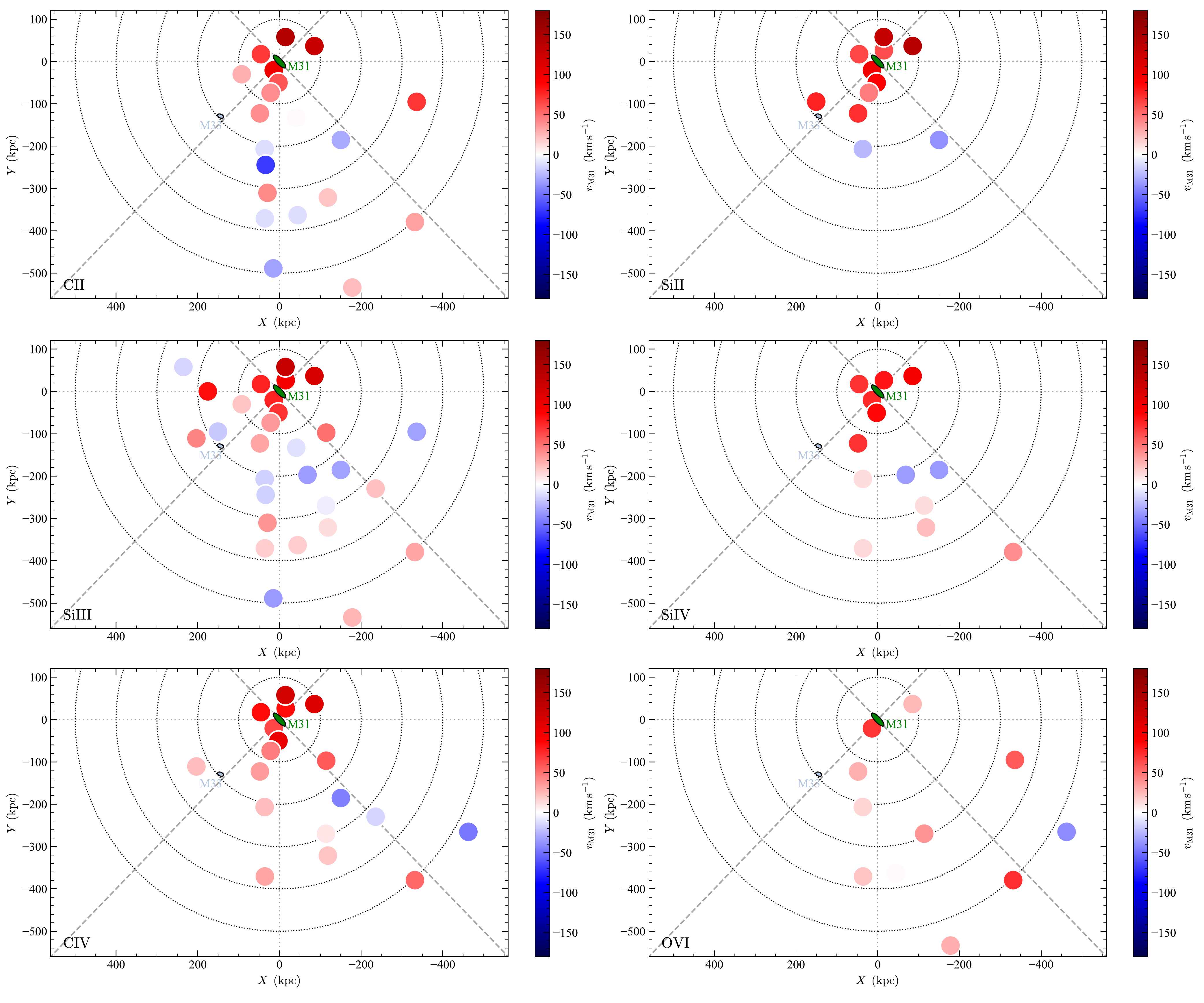}
\caption{Same as Fig.~\ref{f-velmap}, but for the average velocities.
\label{f-velavgmap}}
\end{figure*}

How the velocity field of the gas is distributed in $R$ and $\Phi$ beyond 25--50 kpc is a key diagnostic of accretion and feedback. However, a statistical survey using one sightline per galaxy (such as COS-Halos) cannot address this problem because it observes many galaxies in an essential random mix of orientations and inclinations, which necessarily washes out any coherent velocity structures. An experiment like Project AMIGA is needed to access information about large-scale flows in a sizable sample of lines of sight for a single galaxy. The velocity information remains limited because we have only the (projected) radial velocity along pencil beams piercing the CGM at various $R$ and $\Phi$. Nevertheless as we show below some trends are apparent thanks to the large size of the sample. We use here the $v_{\rm M31}$ peculiar velocities as defined by Eqn.~\ref{e-vm31}. By definition, in the M31 velocity frame, an absorber with no peculiar velocity relative to M31's bulk motion has $v_{\rm M31}=0$ \km. In \S\ref{s-dwarfs-vel}, we show that the M31 peculiar velocities of the absorbers seen toward the QSOs and the velocities of the M31 dwarf satellites largely overlap. We now review how the velocities of the absorbers are distributed in the CGM of M31 over the entire surveyed range of $R$.

In Figs.~\ref{f-velmap} and \ref{f-velavgmap}, we show the distribution of the M31 peculiar velocities of the individual components identified for each ion and column-density-weighted average velocities of each ion, respectively. Circles with several colors indicate that the observed absorption appears in more than one component. Both Figs.~\ref{f-velmap} and \ref{f-velavgmap} demonstrate that in many cases there is some overlap in the velocities between low ions (\siii, \cii, \siiii) and higher ions (\siiv, \civ, \ovi). This strongly implies that the CGM of M31 has multiple gas-phases with overlapping kinematics when they are observed in projection (a property also readily observed from the normalized profiles shown in Fig.~\ref{f-example-spectrum} and as supplemental material in Appendix~\ref{a-supp-fig}). There are also some rarer cases where there is no velocity correspondence in the velocities between \siiii\ and higher ions (see, e.g., near $X\simeq -335$ $Y\simeq -95$ kpc), indicating that the observed absorption in each ion is dominated by a single phase--that is, the components are likely to be distinct single-phase objects.

The full range of velocities associated with the CGM of M31 are between $-249 \le v_{\rm M31}\le +175$ \km\ for \siiii, but for all the other ions it is $-53 \la v_{\rm M31}\le +175$ \km. Furthermore there is only one absorber/component of \siiii\ that has $ v_{\rm M31}=-249$ \km. We emphasize that the rarity of velocity $v_{\rm M31}<-249$ \km\ (corresponding to  $\vlsr <-510$ \km\ in the direction of this sightline) is not an artifact since velocities below these values are not contaminated by any foreground gaseous features. 

We show in \S\ref{s-dwarfs-vel} that the velocity dispersion of the M31 dwarf satellites have a velocity dispersion that is larger (110 \km\ for the dwarfs vs. 68 \km\ for the \siiii\ absorbers) and the M31 dwarfs have some velocities in the velocity range contaminated by the MW and MS. While the CGM gas velocity field distribution may not follow that of the dwarf satellites, it remains plausible that some of the absorption from the extended region of the M31 CGM could be lost owing to contamination from the MW or MS. Therefore we may not be fully probing the entire velocity distribution of the M31 CGM.  However, as discussed in \S\ref{s-ms}, there is no evidence that the velocity distributions of the \siiii\ component in and outside the MS contamination zone are different (see also Figs.~\ref{f-map} and \ref{f-velmap}), and hence it is quite possible that at least the MS contamination does not affect much the velocity distribution of the M31 CGM. With these caveats, we now proceed describing the apparent trends of the velocity distribution in the CGM of M31.

From Fig.~\ref{f-velmap}, the first apparent property was already noted in the previous section:  the velocity complexity (and hence full-width) of the absorption profiles increases with decreasing $R$ (see \S\ref{s-map-metal}). Within $R\la 100$ kpc or $\la 200$ kpc, about 75\% of the \siiii\ absorbers have at least two components (at the COS G130M-G160M resolution). This drops to about 33\% at $200<R\la 569$ kpc.

The second property evident from either Fig.~\ref{f-velmap} or Fig.~\ref{f-velavgmap} is that the M31 peculiar velocities are larger at $R\la 100$ kpc than at higher $R$. Table~\ref{t-vel} lists the average M31 velocities, their standard deviations, and their interquartile ranges (IQRs) for the individual components and averaged components in three samples; the full AMIGA set, the subset with $R\le 100$ kpc, and the subset with $R>100$ kpc. From this table and for all the ions besides \ovi, $\langle v_{\rm M31} \rangle=90$ \km\ at $R\le 100$ kpc, while at $R>100$ kpc,  $\langle v_{\rm M31} \rangle=20$ \km, a factor 4.5 times smaller. There are only two data points for \ovi, at $R\le 100$ kpc, but the average at $R>100$ kpc is  $\langle v_{\rm M31} \rangle=22$ \km, following a similar pattern as observed for the other ions. For all the ions but \cii, the velocity dispersions or IQRs are smaller at $R\le 100$ kpc than at $R>100$ kpc.

The third property observed in Fig.~\ref{f-velmap} or Fig.~\ref{f-velavgmap} is that at $R\le 100$ kpc, there is no evidence for negative M31 velocities, while at $R>100$ kpc, about 40\% of the \siiii\ sample has blueshifted $v_{\rm M31}$ velocities. This partially explains the previous result, but even if we consider the absolute velocities, $\langle|v_{\rm M31}|\rangle=40$ \km\ at $R>100$ kpc, implying $\langle |v_{\rm M31}(R>100)|\rangle = 0.44\langle |v_{\rm M31}|(R\le 100) \rangle$, i.e., in absolute terms or not, $v_{\rm M31}$ is smaller at $R>100$ kpc than at $R\le 100$ kpc. Therefore at $R>100$ kpc, not only are the peculiar velocities of the CGM gas less extreme, but they are also more uniformly distributed around the bulk motion of M31.  At $R<100$ kpc, the peculiar velocities of the CGM gas are more extreme and systematically redshifted relative to the bulk motion of M31.

The fourth property appears in Fig.~\ref{f-velmap-dwarfs} where we compare $v_{\rm M31}$ velocities of the M31 dwarfs and \siiii\ absorbers, which shows that overall the velocities of the satellites and the CGM gas do not follow each other. As noted in \S\ref{s-dwarfs-cgm} (see also Table~\ref{t-xmatch-dwarf}), some velocity components seen in absorption toward the QSOs are found with $\Delta_{\rm sep}<1$ and have $\delta v<v_{\rm esc,dwarf}$. However, the last two trends found for the CGM gas are not observed for the dwarfs. Fig.~\ref{f-velmap-dwarfs} shows that both blue- and redshifted $v_{\rm M31}$ velocities are observed at any $R$ and $v_{\rm M31}$ above and below 50 \km\ are also observed at any $R$. More quantitatively, at $R>100$ kpc or $R\le 100$ kpc, $\langle v_{\rm M31,dwarf} \rangle \simeq 34$ \km ($\langle |v_{\rm M31,dwarf}|\rangle \simeq 102$ \km), remarkably contrasting with the properties of the CGM gas described in the previous two paragraphs. These findings strongly suggests that the velocity fields of the dwarfs and CGM gas are decoupled. We infer from this decoupling that (1) gas bound to satellites does not make a significant contribution to CGM gas observed in this way, and (2) the velocities of gas removed from satellites via tidal or ram-pressure interactions, if it is present, becomes decoupled from the dwarf that brought it in (as one might expect from its definition as unbound to the satellites). 

The fifth property is more readily apparent considering the average velocities shown in Fig.~\ref{f-velavgmap} where considering the CGM gas in different annuli, there is an apparent change in the sign of the average $v_{\rm M31}$ velocities with on average a positive velocity in at $R<200$ kpc, negative velocity in $200<R<300$ kpc, and again positive velocity in $300<R<400$ kpc. This is more evident with \siiii\ where the sample of absorbers is larger, but taking the average velocities in the different annuli, the same pattern is observed for \cii, \siiii, \siiv, and \civ. Beyond $\ga 1.1 \rvir$ (330 kpc), there is the region of gas that we have identified in \S\ref{s-map-metal} that is observed between near the southern projected major axis and about $52\degr$ east off near the $X=0$ kpc axis. In that region $v_{\rm M31}$ is predominantly positive.

We emphasize again that the MS contamination does not really alter these properties and neither is  the source of these properties. As shown in Fig.~\ref{f-map} (see also \S\ref{s-ms}), the MS contamination dominantly occurs in the region $X<0$ for any $Y$. There is no evidence that these properties change with $\Phi$ and in particular between the quadrants $X<0$ and $X>0$ (see \S\ref{s-ms}). Absorption occurring in the velocity range $-50 \la v_{\rm M31}\la +150$ \km\ is also not contaminated by the MS.

\section{Discussion}\label{s-disc}

The major goal of Project AMIGA is to determine the global distribution of the gas phases and metals through the entire CGM of a representative galaxy. With a large sample of QSOs accumulated over many surveys, and newly observed by \hst/COS, we are able to probe multiple sightlines that pierce M31 at different radii and azimuthal angles. Undertaking this study in the UV has been critical since only in this wavelength band there are the diagnostics and spectral resolution to constrain the physical properties of the multiple gas-phases existing in the CGM over $10^{4-5.5}$ K (for $ z = 0$, the hottest phase can only be probed with X-ray observations). With 25 sightlines within about $1.1\rvir$ and 43 within 569 kpc ($\la 1.9\rvir$) of M31, the size of the sample and the information as a function of $R$ and $\Phi$ are unparalleled. We will now consider the broad patterns and conclusions we can draw from this unique dataset. 

\subsection{Pervasive Metals in the CGM of M31}\label{s-disc-metal-mass}

A key finding of Project AMIGA is the ubiquitous presence of metals in the CGM of M31. While the search for \hi\ with $\mlnhi \ga 17.5$ in the CGM of M31 toward pointed radio observations has been unsuccessful in the current sample (\citetalias{howk17} and see Fig.~\ref{f-map}), the covering factor of \siiii\ (29 sightlines) is essentially 100\% out to $1.2 \rvir$, while \ovi\ associated with M31 is detected toward all 11 sightlines with FUSE data, all the way out to $1.9 \rvir$, the maximum radius of our survey (see \S\S\ref{s-n-vs-r}, \ref{s-fc}). From the ionization range probed by Project AMIGA, we further show that \siiii\ and \ovi\ are key probes of the diffuse gas (see \S\ref{s-ratio-vs-r}). With information from \siii, \siiii, and \siiv, we demonstrate that \siiii\ is the dominant ion in the ionizing energy range 8--45 eV (see \S\ref{s-si2-si3-si4}). The fact that \siiii\ and \ovi\ have such high covering factors suggests that these ions are not produced in small clumps within a hotter medium; instead it must be more pervasively distributed. 

The finding of pervasive metals in the CGM of M31 is a strong indication of ongoing and past gas outflows that ejected metals well beyond their formation site. Based on a specific star-formation rate of ${\rm SFR/M_\star} = (5\pm 1) \times 10^{-12}$ yr$^{-1}$ \citep[using the stellar mass M$_\star$ and SFR from][]{geehan06,kang09}, M31 is not currently in an active star-forming episode. In fact, \citet{williams17} show that the bulk of star formation occurred in the first $\sim$6 billion years and the last strong episode happened over $\sim$2 billion years ago (see also Fig.~6 in \citealt{telford19} for a metal production model of M31). Hence most of the metals seen in the CGM of M31 have most likely been ejected by previous star-forming episodes and/or stripped from its dwarfs and more massive companions. However, the fact that metals are detected beyond \rvir, and, that beyond \rvir\ they are found predominantly in a certain direction, also suggests that some of the metals may be coming from the Local group medium, possibly recycling metals from the MW or M31 (see \S\ref{s-map-metal}), or from an IGM filament in that particular direction. 

In \S\ref{s-mass}, we estimate that the mass of metals $M^{\rm cool}_{\rm Z} = (2.0 \pm 0.5)  \times 10^7$ M$_\sun$ within \rvir\ for the predominantly photoionized gas probed by \siii, \siiii, and \siiv. For the gas probed by \ovi, we find that $M^{\rm warm}_{\rm Z} > 4.4 M^{\rm cool}_{\rm Z} \ga 9\times 10^7$ M$_\sun$ at \rvir\ (this is a lower limit because the fractional amount of \ovi\ is an upper limit, see \S\ref{s-mass}). The sum of these two phases yields a lower limit to the CGM metal mass because the hotter phase probed by the X-ray and metals bound in dust are not included. If the hot baryon mass of M31 is not too different from that estimated for the MW (see \S\ref{s-mass}), then we expect $M^{\rm hot}_{\rm Z} \approx M^{\rm warm}_{\rm Z}$. The CLUES simulation of the Local group estimates that the mass of the hot ($>10^5$ K) gas is a factor 3 larger than the cooler ($<10^5$ K) gas \citep{nuza14}.  The dust CGM mass remains quite uncertain, but could be at the level of $5\times 10^7$ M$_\sun$ according to estimates around $0.1$--$1L^*$ galaxies \citep{menard10,peeples14,peek15}. Hence the total metal mass of the CGM of M31 out to \rvir\ could be as large as $M^{\rm CGM}_{\rm Z}\ga 2.5\times 10^8$ M$_\sun$.

The stellar mass of M31 is $(1.5\pm 0.2) \times 10^{11}$ M$_\sun$ \citep[e.g.,][]{williams17}.  Using this result, \citet{telford19} estimated that the current metal mass in stars is $3.9\times 10^8$ M$_\sun$, i.e., about the same amount that is found in the entire CGM of M31 up to \rvir.  \citet{telford19} also estimated the metal mass of the gas in the disk of M31 to be around $(0.8$--$3.2)\times 10^7$ M$_\sun$, while \citet{draine14} estimated the dust mass in the disk to be around $5.4\times 10^7$ M$_\sun$, yielding a total metal mass in the disk of M31 of about  $M^{\rm disk}_{\rm Z}\simeq 5\times 10^8$ M$_\sun$. Therefore M31 has in its CGM within \rvir\ at least 50\% of the present-day metal mass in its disk. As we show in \S\ref{s-n-vs-r} and \S\ref{s-mass} and discuss above, metals are also found beyond \rvir, especially in the more highly ionized phase traced by \ovi\ (and even higher ions). These metals could come from M31 or being recycled in the Local group from the MW or dwarf galaxies. 

\subsection{Comparison with COS-Halos Galaxies}\label{s-coshalos}
The Project AMIGA experiment is quite different from most of the surveys of the CGM of galaxies done so far. Outside the local universe, surveys of the CGM of galaxies involve assembling samples of CGM gas in aggregate by using one sightline per galaxy (see \S\ref{s-intro}), and in some nearby cases up to 3--4 sightlines (e.g., \citealt{bowen16,keeney17}). By assembling a sizable sample of absorbers associated with galaxies in a particular sub-population (e.g. $L^*$, sub-$L^*$, passive or star-forming galaxies), one can then assess how the column densities change with radii around that kind of galaxy, and from this estimate average surface densities, mass budgets, etc. can then be evaluated. By contrast, Project AMIGA has assembled almost as many sightlines surrounding M31 as COS-Halos had for its full sample of 44 galaxies. We can now make a direct comparison between these two types of experiments. For this comparison, we use the COS-Halos survey of $0.3<L/L^*<2$ galaxies at $z\simeq 0.2$, which selected galaxies within about 160 kpc from the sightline  \citep{tumlinson11a,tumlinson13,werk13,werk14}. The full mass range of the COS-Halos galaxies is quite large, $11.5 \la \log M_{200} \la 13.7$, but most of the star-forming galaxies are in the range  $11.5 \la \log M_{200} \la 12.5$ and most of the passive quiescent galaxies have  $13.0 \la \log M_{200} \la 13.7$. As a reminder, M31 has $\log M_{200} =12.1$ (see \S\ref{s-intro}).

\begin{figure}[tbp]
\epsscale{1.2}
\plotone{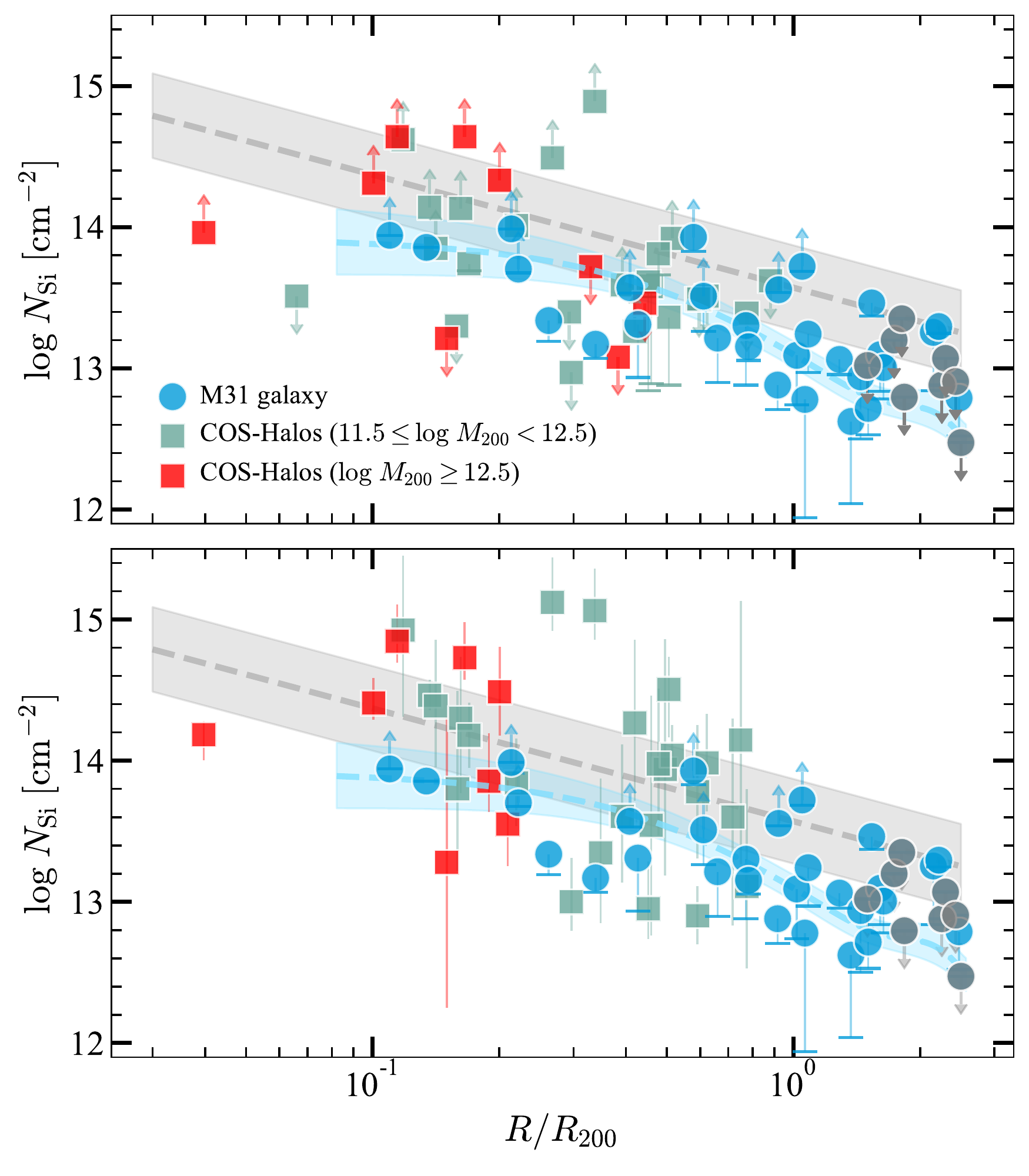}
\caption{Comparison of the total column densities of Si from M31 and the COS-Halos galaxies as a function of $R/R_{200}$.  {\it Top panel}: $N_{\rm Si}$ are directly constrained by the estimates on $N_{\rm Si\,II}$, $N_{\rm Si\,III}$, and $N_{\rm Si\,IV}$ from the observations for both Project AMIGA and COS-Halos \citep{werk13}. The error bars are less than the size of the square and the vertical bars include the range of possible values if the non-detections are either near their $3\sigma$ upper limits or so low as to be negligible. {\it Bottom panel:} same as top panel but $N_{\rm Si}$ for the COS-Halos data is derived from Cloudy photoionization models \citep{werk14}. The gray dashed line and shaded region represent the best fit between $N_{\rm Si}$ and $R/R_{200}$ and its dispersion using COS-Halos modeled data. The blue area shows the full range of the GP model of the Project AMIGA data (see \S\ref{s-nsi-vs-r}).
\label{f-cos-halos}}
\end{figure}

\subsubsection{Column Densities of Si vs. $R$}\label{s-cos-halos-col}
In Fig.~\ref{f-cos-halos}, we show the total column densities of Si as a function of $R/R_{200}$ for the COS-Halos galaxies and M31. For the COS-Halos survey, each data point corresponds to an absorber at some impact parameter from a galaxy, while for M31, each data point is an absorber probing the CGM at a different impact parameter from the same galaxy. For COS-Halos, we consider two cases: 1) the column densities of Si estimated in a similar fashion as those for M31; and 2) the column densities of Si estimated from photoionization modeling. For case 2) we use the results from \citet{werk14} (see also \citealt{prochaska17}), which were used to determine the metal mass of the CGM of the COS-Halos galaxies in \citet{peeples14}. For case 1), we use the results from \citet{werk13} and follow the procedure in \S\ref{s-nsi-vs-r} to estimate $N_{\rm Si}$ from the column densities of \siii, \siiii, and \siiv. We require that all three Si ions are available, except in the cases where there are only lower limits for two ions (typically \siii\ and \siiii) since in that case the resulting column density is a lower limit that encompasses any missing column density from the remaining Si ion (typically \siiv). The sample size in case 1) is 35, while it is 33 in case 2) with some overlap between the two subsamples. In Fig.~\ref{f-cos-halos}, we also show the modeled column density of Si as a function of $R$, in gray for COS-Halos and blue for M31 (we show the adopted GP model, see \S\ref{s-nsi-vs-r}).

A striking difference between the COS-Halos and M31 data immediately apparent from Fig.~\ref{f-cos-halos} is that, at $R/R_{200}\la 0.3$, there is a large amount of very high Si column densities in COS-Halos that are absent in Project AMIGA. The reason that the COS-Halos Si column densities have higher lower limits than those of M31 is because the weak transitions of \siii\ are saturated in COS-Halos, a situation not observed in M31 toward any of the sightlines---the lower limits of Si arise only because \siiii\ is saturated. These high Si column densities also correspond to the very strong \nhi\ ($\mlnhi \ga 18$) absorbers observed in COS-Halos, but again not in M31 \citepalias{howk17}. However, while for \hi, the beam dilution could have affected somewhat the interpretation of the difference between COS-Halos (\hi\ absorption)  and M31 (\hi\ emission), for the metal ions this is not an issue. Therefore the higher frequency of saturated weak \siii\ transitions in COS-Halos compared to M31 is a real effect, not an artifact. 

Besides this difference, the estimated Si column densities from the observations in the COS-Halos and Project AMIGA surveys are distributed with a similar scatter at larger impact parameters ($R/R_{200}\ga 0.4$) where the gas is more ionized (see top panel in Fig.~\ref{f-cos-halos}). The photoionization-modeled COS-Halos Si displayed on the bottom panel have some higher values than observed in the top panel, but in the impact parameter region $0.4\la R/R_{200}\la 0.8$ where they are observed, there are also several lower limits. Beyond $R>0.9R_{200}$, there is no COS-Halos observation (owing to the design of the survey). The extrapolated model to the COS-Halos observations shown in gray in Fig.~\ref{f-cos-halos} is a factor 2--4 higher than the models of the Project AMIGA data shown in blue depending on $R/R_{200}$. 

A likely explanation for the higher column density absorbers is that some of these COS-Halos absorbers could be fully or partly associated with a closer galaxy than the initially targeted COS-Halos galaxies where the gas can contain more neutral and weakly ionized gas. Indeed, while the COS-Halos galaxies were selected to have no bright companion, that selection did not preclude fainter nearby companions such as dwarf satellites (see \citealt{tumlinson13}). Galaxy observation follow-up by \citet{werk12} found several $L > 0.1 L^*$ galaxies within 160 kpc of the targeted COS-Halos galaxy. Comparing the results from other surveys of galaxies/absorbers at low redshift \citep{stocke13,bowen02}, \citet{bregman18} also noted a higher preponderance of high \hi\ column density absorbers in the COS-Halos survey. However, the higher COS-Halos column densities at large radii could also be an effect of evolution in the typical CGM, as COS-Halos probed a slightly higher cosmological redshift. It is also possible that the M31 CGM is less rich in ionized gas at these radii than the typical $L^*$ galaxy at $z \sim 0.2$, because of its star formation history or environment. 

\subsubsection{CGM Mass Comparison}\label{s-cos-halos-mass}
Among a key physical parameter of the CGM is its mass, which is obtained from the column density distribution of the gas and assuming a certain geometry of the gas. For M31, we cannot derive the baryonic mass of CGM gas without assuming a metallicity since the \hi\ column density remains unknown toward all the targets in our sample (but see \S\ref{s-metallicity}, \ref{s-mass}). However, the metal mass of the cool gas probed by \siii, \siiii, and \siiv\ can be straightforwardly estimated directly from the observations without any ionization modeling (see top panel of Fig.~\ref{f-cos-halos}).  

Even though both \citet{peeples14} and \citet{werk14} use the Si column densities derived from phoionization models, as illustrated in Fig.~\ref{f-cos-halos}, this would not change the outcome that the metal mass of the cool CGM gas derived from the COS-Halos survey is about a factor 2--3 higher than the metal mass derived in Project AMIGA. This is because there are 7 COS-Halos Si column densities at $R/R_{200}<0.3$ that are much higher owing to saturation in the weak \siii\ transitions (see above), driving the overall model of $N_{\rm Si}(R)$ substantially higher. The fact that these high $N_{\rm Si}$ are not found in the CGM of M31 or lower redshift galaxies at similar impact parameters \citep[e.g.,][]{bowen02,stocke13} suggests a source of high-column \hi\ and \siii\  absorbers in the COS-Halos sample that could be recent outflows, strong accretion/recycling, or gas associated with closer satellites to the sightline. With only 5 targets within $R/R_{200}<0.3$ and none below $R/R_{200}<0.1$ for M31, it would be quite useful to target more QSOs in the inner region of the CGM of M31 to better determine how  $N_{\rm Si}(R)$ varies with $R$ at small impact parameters.

For the warm-hot gas probed by \ovi, the COS-Halos star-forming galaxies have $\langle N_{\rm O\,VI}\rangle = 10^{14.5}$ \cmm, a detection rate close to 100\%, and no large variation of \novi\ with $R$  \citep{tumlinson11a}. For M31, we have a similar average \ovi\ column density, hit rate, and little evidence for any large variation of \novi\ with $R$ (see \S\ref{s-mass}). This implies that the masses of the warm-hot CGM of M31 and COS-Halos star-forming galaxies are similar.  M31 has a specific SFR that is a factor $\ga 10$ lower than the COS-Halos star-forming galaxies, but its halo mass is on the higher side of the COS-Halos star-forming galaxies (but lower than the COS-Halos quiescent galaxies). As discussed in \S\ref{s-disc-pers-comp} in more detail, M31 and the COS-Halos star-forming galaxies have halo masses in the range $M_{200} \simeq 10^{11.7}$--$10^{12.3}$ M$_\sun$, corresponding to a virial temperature range that overlaps with the temperature at which the ionization fraction of \ovi\ peaks, which may naturally explaining some of the properties of the \ovi\ in the CGM of ``$L^*$" galaxies \citep{oppenheimer18}. It is also possible that some \ovi\ arises in photoionized gas or combinations of different phases (see \S\ref{s-disc-pers-comp}). 

Based on the comparison above, we find that the \ovi\ is less subject to the uncertainty in the association of the absorber to the correct galaxy owing to its column density being less dependent on $R$ (see also \S\ref{s-disc-change}).  Therefore this leads to similar metal masses of the CGM of the $z\sim 0.2$ COS-Halos galaxies and M31 for the \ovi\ gas-phase. For the lower ions, their column densities are more dependent on $R$ (see also \S\ref{s-disc-change}). Therefore the association of the absorber to the correct galaxy is more critical to derive an accurate column density profile with $R$ and hence derive an accurate CGM metal mass. However, we note that despite these uncertainties the metal mass of the cool CGM of the COS-Halos galaxies is only a factor 2--3 higher than that derived for M31.

\subsection{A Changing CGM with Radius}\label{s-disc-change}
A key discovery from Project AMIGA is that the properties of the CGM of M31 change with $R$. This is reminiscent of our earlier survey \citepalias{lehner15}, but the increase in the size sample has transformed some of the tentative results of our earlier survey into robust findings. In particular the radius around $R\sim 100$--150 kpc appears critical in view of several properties changing near this threshold radius:
\begin{enumerate}[wide, labelwidth=!, labelindent=0pt]
    \item For any ions, the frequency of strong absorption is larger at $R\la 100$--150 kpc than at larger $R$.
    \item The column densities of Si and C ions change by a factor $>5$--$10$ between about 25 kpc and 150 kpc, while they change only by a factor $\la 2$ between 150 kpc and 300 kpc.
    \item The detection rate of singly ionized species (\cii, \siii) is close to 100\% at $R<150$ kpc, but sharply decreases beyond (see Fig.~\ref{f-col-vs-rho}), and therefore the gas has a more complex gas-phase structure at $R<150$ kpc.
    \item The peculiar velocities of the CGM gas are more extreme and systematically redshifted relative to the bulk motion of M31 at $R\la 100$ kpc, while at $R\ga 100$ kpc, the peculiar velocities of the CGM gas are less extreme and more uniformly distributed around the bulk motion of M31.
\end{enumerate}
There are also two other significant regions: 1) beyond $R_{200}\simeq 230$ kpc the gas is becoming more ionized and more highly ionized than at lower $R$ (e.g., there is a near total absence of \siii\ absorption beyond $R_{200}$---see Fig.~\ref{f-col-vs-rho}, or, a higher \cii/\civ\ ratio on average at $R\ga R_{200}$ than at lower $R$---see \S\ref{s-c2-c4}); and 2) beyond $1.1\rvir$ the gas is not detected in all the directions away from M31, as it is at smaller radii, but only in a cone near the southern projected major axis and about $52\degr$ east off the $X=0$ kpc axis (see \S\ref{s-map-metal}). 

The overall picture that can be drawn out from these properties is that the inner regions of the CGM of M31 are more dynamic and complex, while the more diffuse regions at $R\ga 0.5\rvir$ are more static and simpler. Zoom-in cosmological simulations capture in more detail and more accurately the structures of the CGM than large-scale cosmological simulations thanks to their higher mass and spatial resolution. Below we use several results from zoom simulations to gain some insights on these observed changes with $R$. However, the results laid out in \S\ref{s-properties} also now provide a new testbed for zoom simulations, so that not only qualitative but also quantitative comparison can be undertaken. We note that most of the zoom simulations discussed here have only a single massive halo. However, according to the ELVIS simulations of Local group analogs \citep{garrison-kimmel14}, there should be no major difference at least within about \rvir\ for the distribution of the gas between isolated and paired galaxies.

\subsubsection{Visualization and Origins of the CGM Variation}\label{s-disc-qual-comp}
To help visualize the properties described above and gain some insights into the possible origins of these trends, we begin by qualitatively examining two zoom simulations. First, we consider the Local group zoom simulations from the CLUES project \citep{nuza14} where the gas distribution around MW and M31-like galaxies is studied. This paper does not show the distribution of the individual ions, but examines the two main gas-phases above and below $10^5$ K in an environment that is a constrained analog to the Local group. Interestingly, considering Fig.~3 (simulated M31) or Fig.~6 (simulated MW) in \citeauthor{nuza14}, the region within 100--150 kpc appears more complex, with a large covering factor for both cool and hot gas phases and higher velocities than at larger radii. In these simulations, this is a result of the combined effects of cooling and supernova heating affecting the closer regions of the CGM of M31. This simulation also provides an explanation for the gas observed beyond $1.1\rvir$ that is preferentially observed in a limited region of the CGM of M31 (see Fig.~\ref{f-colmap} and see middle right panel of their Fig.~3) whereby the $\la 10^5$ K gas might be accreting onto the CGM of M31. We also note that \citet{nuza14} find a mass for the $\la 10^5$ K CGM gas of $1.7\times 10^{10}$ M$_\sun$, broadly consistent with our findings (see \S\ref{s-mass}). More quantitative comparisons between the CLUES (or Local group analog simulations like ELVIS-FIRE simulations, \citealt{garrison-kimmel14,garrison-kimmel19}) and Project AMIGA results are beyond the scope of this paper, but they would be valuable to undertake in the future. 

Second, we consider the zoom Eris2 simulation of a massive, star-forming galaxy at $z = 2.8$ presented in  \citet{shen13}. The Eris2 galaxy being $z = 2.8$ and with a star-formation rate of 20 M$_\sun$\,yr$^{−1}$ is nothing like M31, but this paper shows the distribution of the gas around the central galaxy using some of the same ions that are studied in Project AMIGA, specifically \siii, \siiv, \cii, \civ, and \ovi\ (see their Figs.~3a and 4a, b). Because Eris2 is so different from M31, we would naively expect their CGM properties to be different, and yet:  1) Eris2 is surrounded by a large diffuse \ovi\ halo with a near unity covering factor all the way out to about $3\rvir$; 2) the covering factor of absorbing material in the CGM of Eris2 declines less rapidly with impact parameter for \civ\ or \ovi\ compared to \cii, \siii, or \siiv; 3) beyond \rvir, the covering factor of \siii\ drops more sharply than \cii. There are also key differences, like the strongest absorption in any of these ions being observed in the bipolar outflows perpendicular to the plane of the disk, which is unsurprisingly not observed in M31 since it currently has a low star-formation rate \citep[e.g.,][]{williams17}. However, the broad picture of the CGM of M31 and the simulated Eris2 galaxy are remarkably similar. This implies that some of the properties of the CGM may depend more on the micro-physics producing the various gas-phases than the large-scale physical processes (outflow, accretion) that vary substantially over time. In fact, the Eris2 simulation shows that inflows and outflows coexist and are both traced by diffuse \ovi; In Eris2, a high covering factor of strong \ovi\ absorbers seems to be the least unambiguous tracer of large-scale outflows.

\subsubsection{Quantitative Comparison in the CGM Variation between Observations and Simulations}\label{s-disc-quant-comp}

Two simulations of M31-like galaxies in different environments at widely separated epochs show some similarity with some of the observed trends in the CGM of M31. We now take one step further by quantitatively comparing the column density variation of the different ions as a function of $R$ in three different zoom-in cosmological simulations, two being led by members of the Project AMIGA team (FIRE and FOGGIE collaborations), and a zoom-in simulation from the Evolution and Assembly of GaLaxies and their Environments (EAGLE) simulation project \citep{oppenheimer18a,schaye15,crain15}. 

\noindent
{\it $\bullet$ Comparison with FIRE-2 Zoom Simulations}

We first compare our observations with column densities modeled using cosmological zoom-in simulations from the FIRE project\footnote{FIRE project website: \url{http://fire.northwestern.edu}}. Details of the simulation setup and CGM modeling methods are presented in \citet{ji19}. Briefly, the outputs analyzed  here are FIRE-2 simulations evolved with the {\small GIZMO} code using the meshless finite mass (MFM) solver \citep{hopkins15}. The simulations include a detailed model for stellar feedback including core-collapse and Type Ia SNe, stellar winds from OB and AGB stars, photoionization, and radiation pressure \citep[for details, see][]{hopkins18}. We focus on the ``m12i'' FIRE halo, which has a mass $M_{\rm vir} \approx 1.2 M_{200} \approx 1.2\times10^{12}\,M_\odot$ at $z=0$, which is comparable to the halo mass of M31. However, neither the SFR history nor the present-day SFR are similar. The ``m12i'' FIRE halo has a factor 10--12 higher SFR (see Fig.~3 in \citealt{hopkins19}) than the present-day SFR of M31 of 0.5\,M$_\sun$\,yr$^{-1}$ \citep[e.g.][]{kang09}. We compare Project AMIGA to FIRE-2 simulations with two different sets of physical ingredients. The ``MHD'' run includes magnetic fields, anisotropic thermal conduction and viscosity, and the ``CR'' run includes all these processes plus the ``full physics'' treatment of stellar cosmic rays. The CR simulation assumes a diffusion coefficient $\kappa_{||}=3\times10^{29}$ \cmm\,s$^{-1}$, which was calibrated to be consistent with observational constraints from $\gamma-$ray emission of the MW and some other nearby galaxies \citep{hopkins19,chan19}.  \citet{ji19} showed cosmic rays can potentially provide a large or even dominant non-thermal fraction of the total pressure support in the CGM of low-redshift $\sim L^*$ galaxies. As a result, in the fiducial CR run analyzed here, the volume-filling CGM is much cooler ($\sim 10^{4}-10^{5}$ K) and is thus photoionized in regions where in the run without CRs prefers by hot gas that is more collisionally ionized.

The column densities are generated as discussed in \citet{ji19}.  For the ionization modeling, a hybrid treatment combining the FG09 \citep{faucher-giguere09} and HM12 \citep{haardt12} UV background models is used.\footnote{We use this mixture because, based on the recent UV background analysis of \citet{faucher-giguere19}, the FG09 model is in better agreement with the most up-to-date low-redshift empirical constraints at energies relevant for low and intermediate ions (\cii, \siii, \siiii, \siiv, and \civ).  However, the HM12 model is likely more accurate for high ions such as \ovi\ because the FG09 model used a crude AGN spectral model which under-predicted the higher-energy part of the UV/X-ray background. \citet{ji19} shows how some ion columns depend on the assumed UV background model.}

\begin{figure}[tbp]
\epsscale{1.2}
\plotone{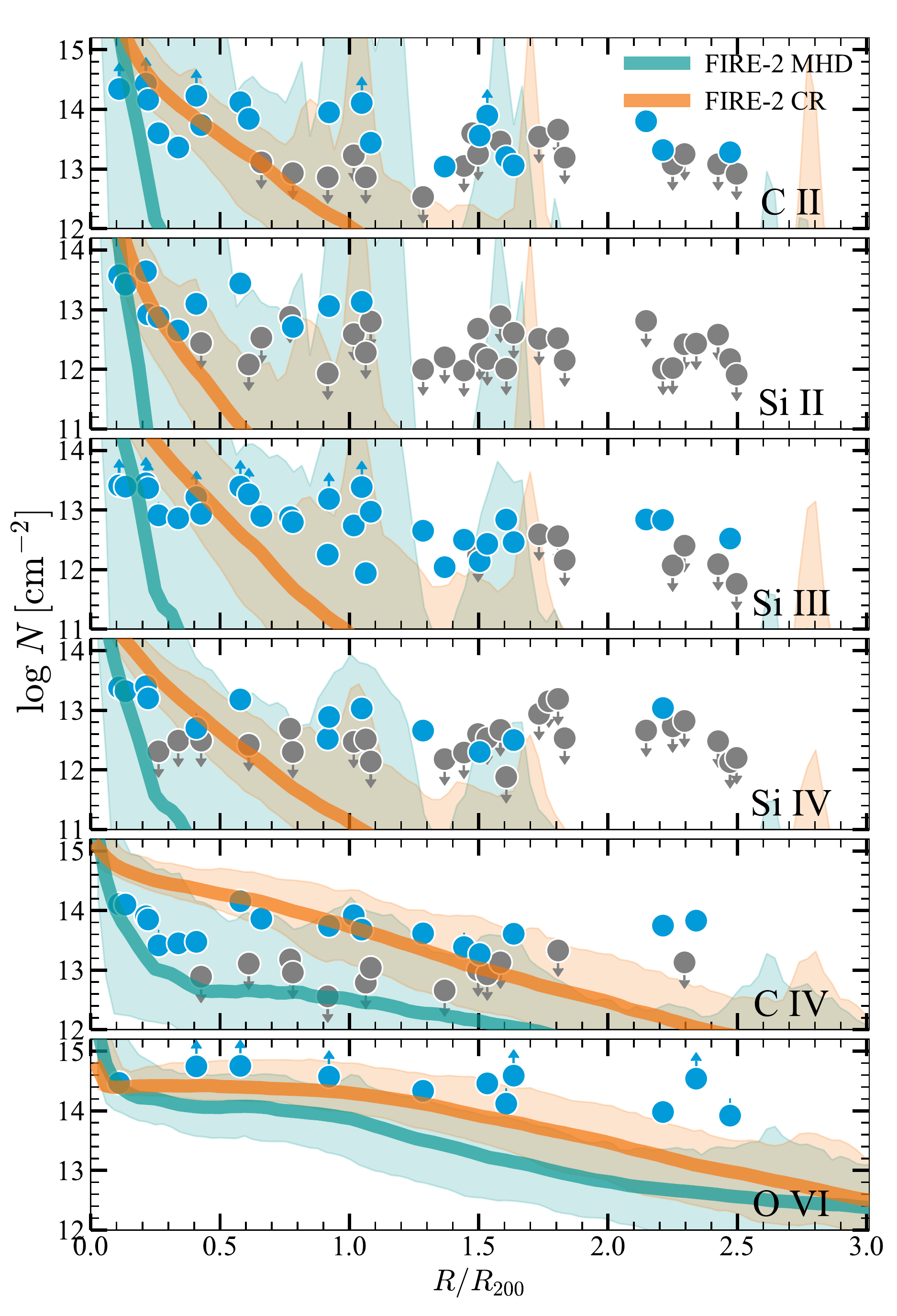}
\caption{Comparison of ion column density profiles between Project AMIGA (total column densities) and FIRE-2 simulations, where ``MHD'' and ``CR'' runs. Thick curves show median values of an ensemble of sightlines produced from simulations, and shaded regions show the full range across all model sightlines. 
\label{f-col-vs-rho-fire}}
\end{figure}

In Fig.~\ref{f-col-vs-rho-fire}, we compare the ion column densities from FIRE-2 simulations with observationally-derived total column densities around M31 as a function of $R/R_{200}$. The green and orange curves show the median simulated column densities for the MHD and CR runs, respectively, while the shaded regions show the full range of columns for all sightlines at a given impact parameter (the lowest values are truncated to match the scales that are adequate for the observations, see \citealt{ji19} for the full range of values). The CR run produces higher column densities and better agreement with observations than the MHD run for all ions presented. The much higher column densities of low/intermediate ions (\cii, \siii, \siiii, and \siiv) in the CR run owing to the more volume-filling and uniform cool phase, which produces higher median values of ion column densities and smaller variations across different sightlines. In contrast, in the MHD run the cool phase is pressure confined by the hot phase to compact and dense regions, leading to smaller median columns but larger scatter for the low and intermediate ions. We note, however, that even in the CR runs the predicted column densities are lower than observations at the larger impact parameters $R\ga 0.5 R_{200}$. This might be due to insufficient resolution to resolve fine-scale structure in outer halos, or it may indicate that feedback effects are more important at large radii than in the present simulations. This difference is quite notable owing to the fact that the star formation of the ``m12i" galaxy has been continuous with a SFR in the range 5--20 M$_\sun$\,yr$^{-1}$ \citep{hopkins19} over the last $\sim$8 billion years while M31 had only a continuous SFR around 6--8 M$_\sun$\,yr$^{-1}$ over its first 5 billion years while over the last 8 billion years it had only two short bursts of star formation about 4 and 2 billion years ago \citep{williams17}. While there are some discrepancies, the simulations also follow some similar trends: 1) the simulated column densities of the low ions decrease more rapidly with $R$ than the high ions, 2) \ovi\ is observed beyond $1.7R_{200}$ where there is no substantial amount of low/intermediate ions, and 3) a larger scatter is observed in the column densities of the low and intermediate ions than \ovi. 

In the FIRE-2 simulations, both collisional ionization and photoionization can contribute significantly to the simulated \ovi\ columns, typically with an increasing contribution from photoionization with increasing impact parameter, driven by decreasing gas densities. In the MHD run, most of the \ovi\ in the inner halo ($R\la 0.5 R_{200}$) is produced by collisional ionization, but photoionization can dominate at larger impact parameters. In the CR run, collisional ionization and photionization contribute comparably to the \ovi\ mass at radii $50 < R < 200$ kpc \citep{ji19}. The actual origins of the CGM in terms of gas flows in FIRE-2 simulations without magnetic fields or cosmic rays were analyzed in \cite{hafen19a}, although the results are expected to be similar for simulations with MHD only. In these simulations, \ovi\ exists as part of a well-mixed hot halo, with contributions from all the primary channels of CGM mass growth: IGM accretion, wind, and contributions from satellite halos (reminiscent of the Eris2 simulations, see above and \citealt{shen13}). The metals responsible for \ovi\ absorption originate primarily in winds, but IGM accretion may contribute a large fraction of total gas mass traced by \ovi\ since the halo is well-mixed and IGM accretion contributes $\ga 60\%$ of the total CGM mass. In the simulations, the hot halo gas persists in the CGM for billions of years, and the gas that leaves the CGM does so primarily by accreting onto the central galaxy \citep{hafen19b}.

\noindent
{\it $\bullet$ Comparison with FOGGIE Simulations}

We also compare the observed total column densities to the Milky-Way like-mass ``Tempest'' ($M_{\rm 200} \approx 4.2 \times 10^{11}$ M$_{\odot}$) halo from the FOGGIE simulations,\footnote{FOGGIE project website: \url{http://foggie.science}} which has a halo mass of $M_{\rm 200} \approx 4.2 \times 10^{11}$ M$_{\odot}$ \citep{peeples19}. We use the $z=0$ output (see \citealt{zheng20} for simulation details), but because of the size difference between M31 and the Tempest galaxy, we again scale all distances by $R_{200}$ ($R_{200} = 159$ kpc for the simulated halo compared to 230 kpc for M31). The only ``feedback'' included in this FOGGIE run is thermal explosion-driven SNe outflows. While this feedback is limited in scope compared to FIRE, FOGGIE achieves higher mass resolution than FIRE-2 by using a ``forced refinement'' scheme that applies a fixed computational cell size of $\sim 381 h^{-1}$ pc within a moving cube centered on the galaxy that is $\sim 200 h^{-1}$ ckpc on a side. This refinement scheme enforces constant {\it spatial resolution} on the CGM, resulting in a variable and very small mass resolution in the low density gas, with typical cell masses of ($\la 1$--100 M$_\sun$). The individual small-scale structures that contribute to the observed absorption profiles can therefore be resolved. These small-scale structures that become only apparent in high-resolution simulations are hosts to a significant amount of cool gas, enhancing the column densities in especially the low ionization state of the gas (\citealt{peeples19,corlies19}, and see also \citealt{vandevoort18,hummels19,rhodin19}).  

\begin{figure}[tbp]
\epsscale{1.2}
\plotone{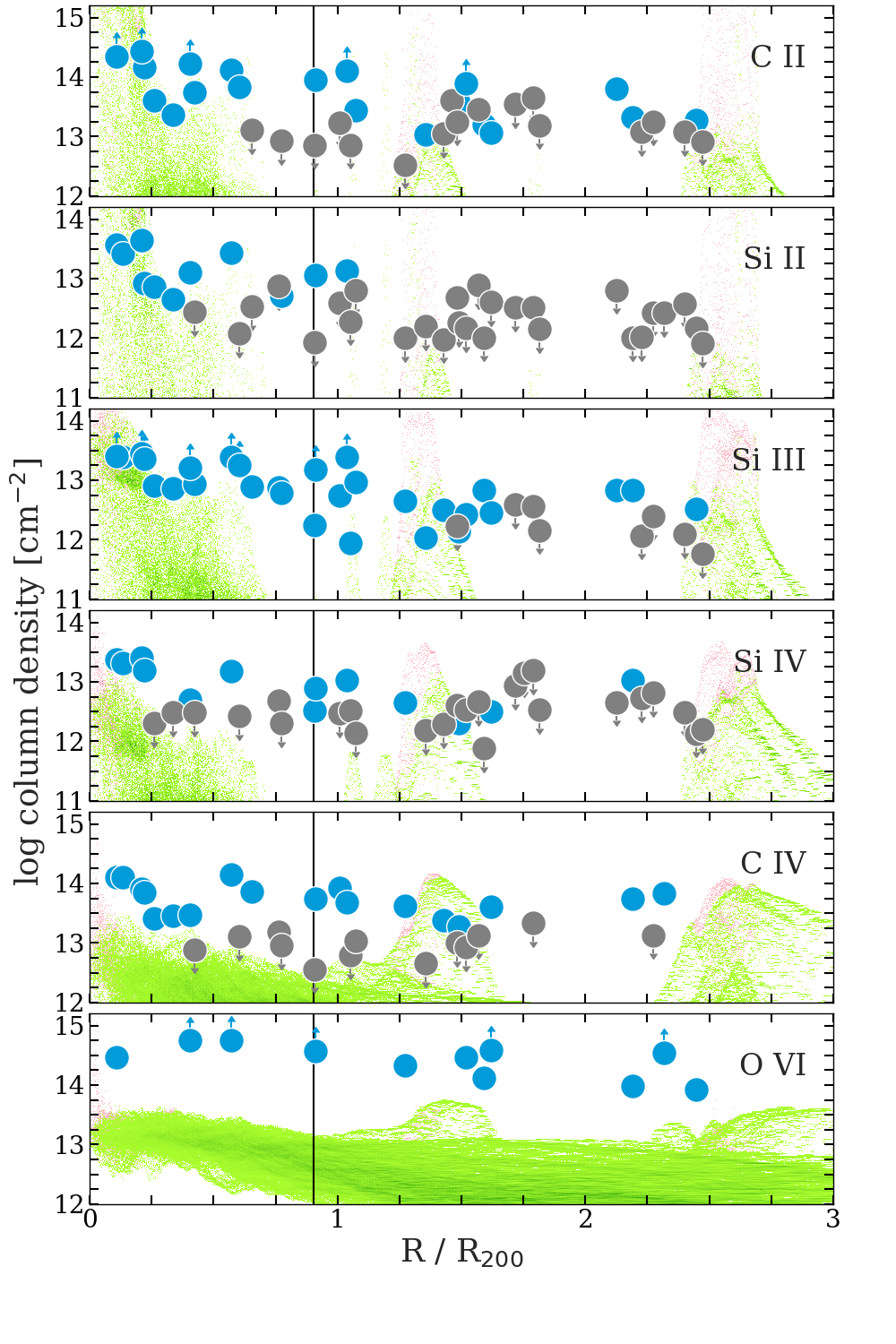}
\caption{Comparison of ion column density profiles between Project AMIGA (total column densities) and the ``Tempest" halo from the FOGGIE simulations. The pink and green shaded areas are projected total column densities from the simulated halo with and without galaxy/satellite contributions, respectively, while the rest of the figure is analogous to Fig.~\ref{f-coltot-vs-rho}. The vertical line shows the extent of the forced resolution cube in the FOGGIE simulation. 
\label{f-col-vs-rho-foggie}}
\end{figure}

As for the FIRE-2 simulations, we compare the total Project AMIGA column densities to FOGGIE because in the simulation we do not (yet) separate individual components, but look at the projected column densities through the halo. We note that the CGM is not necessarily self-similar, so some differences between the simulation predictions and M31 observations at rescaled impact parameter could be due to the halo mass difference. This is especially so since the halo mass range $M_{\rm h}\approx 3\times10^{11}$--$10^{12}$ M$_{\odot}$ corresponds to the expected transition between cold and hot accretion \citep[e.g.,][]{birnboim03,keres03,faucher-giguere11,stern19}.

In Fig.~\ref{f-col-vs-rho-foggie}, we compare the simulated and observed column densities for each ion probed by our survey. The pink and green shaded areas are the data points from the simulation (with and without satellite contribution, respectively) and show the total column density in projection through the halo. The scatter in the simulated data points comes from variation in the structures along the mock sightline and most of the scatter is in fact below $10^{11}$ cm$^{-2}$. The peaks in the column densities are due to small satellites in the halo, which enhance primarily the low-ion column densities. We show the green points to highlight the difference between the mock column densities with and without satellites. For the high ionization lines the difference is negligible, while the difference in the low ions is significant. 

Overall, the metal line column densities are systematically lower than in the observations at any $R$. Only at $R\la 0.3R_{200}$, there is some overlap for the singly ionized species between the FOGGIE simulation and observations. However, the discrepancy is particularly striking for \siiii\ and the high ions. This can be understood by the current feedback implementation in FOGGIE, which does not expel enough metals from the stellar disk into the CGM \citep{hamilton20} to be consistent with known galactic metal budgets\citep{peeples14}. This effect is expected to be stronger for the high ions than the low ions, due to the additional heating and ionization of the CGM that would be expected from stronger feedback, and indeed the discrepancy between the simulation and observations is larger for the high ions (and \siiii) than for the singly ionized species. However, while the absolute scale of the column densities is off, there are also some similarities between the simulation and observations in the behavior of the relative scale of the column density profiles with $R$: 1) the column densities of the low ions drop more rapidly with $R$ than the high ions; 2) despite the inadequate feedback in the current simulations, the \ovi-bearing gas (and \civ\ to a lesser extent) is observed well-beyond $R_{200}$; 3) a large scatter is observed in the column densities of the low and intermediate ions than \ovi. It is striking that the overall slope of the \ovi\ profile resembles the observations but at significantly lower absolute column density. In the FOGGIE simulation, the low ions tracing mainly dense, cool gas are preferentially found in the disk or satellites, while the hotter gas traced by the higher ions is more homogeneously distributed in the halo.

\noindent
{\it $\bullet$ Comparison with EAGLE Simulations}

Finally, we compare our results with the EAGLE zoom-in simulations (EAGLE \emph{Recal-L025N0752} high-resolution volume) discussed in length in \citet{oppenheimer18a}. The EAGLE simulations have successfully reproduced a variety of galaxy observables \citep[e.g.,][]{crain15,schaye15} and achieved ``broad but imperfect" agreement with some of the extant CGM observations (e.g., \citealt{turner16,rahmati18,oppenheimer18a,lehner19,wotta19}). 

 \citet{oppenheimer18a} aimed to directly study the multiphase CGM traced by low metal ions and to compare with the COS-Halos survey (see \S\ref{s-coshalos}). As such, they explored the circumgalactic metal content traced by the same ions explored in Project AMIGA in the CGM galaxies with masses that comprise that of M31. Overall \citeauthor{oppenheimer18a} find agreement between the simulated and COS-Halos samples for \siii, \siiii, \siiv, and \cii\ within a factor two or so and larger disagreement with \ovi, where the column density is systematically lower. With Project AMIGA, we can directly compare the results with one of the EAGLE galaxies that has a mass very close to M31 and also compare the column densities beyond 160 kpc, the maximum radius of the COS-Halos survey \citep{tumlinson13,werk13}. We refer the reader to \citet{oppenheimer16}, \citet{rahmati18}, and \citet{oppenheimer18a} for more detail on the EAGLE zoom-in simulations. We also refer the reader to Fig.~1 in \citet{oppenheimer18a} where in the middle column they show the column density map for galaxy halo mass of  $\log M_{200} = 12.2$ at $z \simeq 0.2$, which qualitatively shows similar trends described in \S\ref{s-disc-qual-comp}. 

\begin{figure}[tbp]
\epsscale{1.2}
\plotone{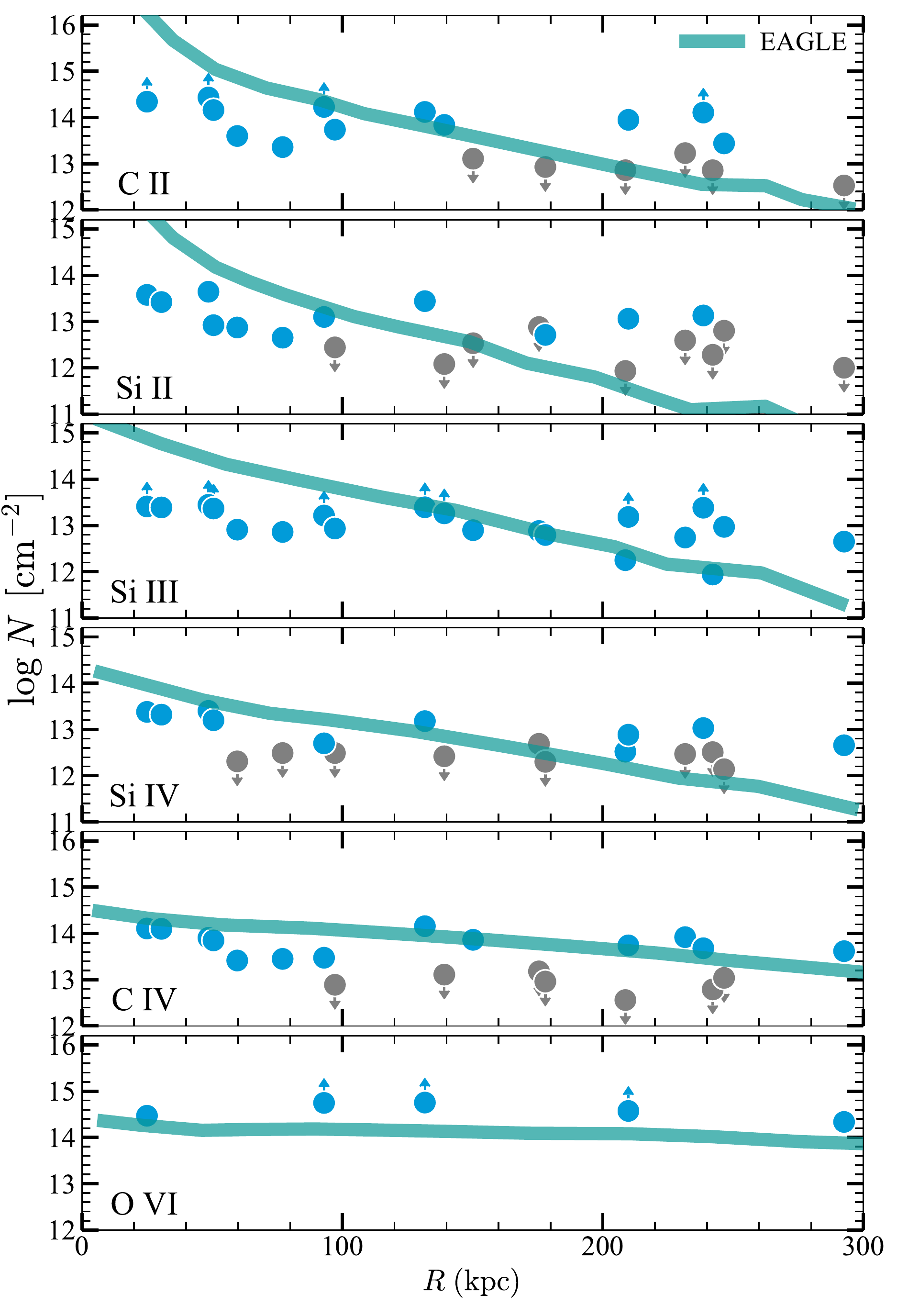}
\caption{Comparison of ion column density profiles between Project AMIGA and EAGLE zoom-in simulation of a galaxy with $\log M_{200}\simeq 12.1 $ at $z=0$ (from the models presented in \citealt{oppenheimer18a}). For the EAGLE simulation, the mean column densities are shown. Note that here we only plot the column density profiles out to about \rvir.
\label{f-col-vs-rho-eagle}}
\end{figure}

In  Fig.~\ref{f-col-vs-rho-eagle}, we compare the EAGLE and observed column densities as a function of the impact parameter out to \rvir. As in the previous two figures, the blue and gray circles are detections and non-detections in the halo of M31. The green curve in each panel represents the mean column density for each ion as a function of the impact parameter for the EAGLE galaxy with $\log M_{200} = 12.1$ at $z=0$. In contrast to FIRE-2 or FOGGIE simulations, the EAGLE simulations appear to produce a better agreement between $N$ and $R$ for low and intermediate ions (\siii, \siiii, \siiv), and \civ\ out to larger impact parameters. However, as already noted in \citet{oppenheimer18a}, this agreement is offset by producing too much column density for the low and intermediate ions at small impact parameters (see, e.g., \siii, which is not affected by lower limits, is clearly overproduced at $R\la 80$ kpc).  The flat profile of \ovi, with very little dependence on $R$, is similar to the observations and other models, but overall the EAGLE \ovi\ column densities are a factor 0.2--0.6 dex smaller than observed. \citet{oppenheimer18a} (and also \citealt{oppenheimer16}) already noted that issue from their comparison with the COS-Halos galaxies (see also \S\ref{s-coshalos}), requiring additional source(s) of ionization for the \ovi\ such as AGN flickering \citep{oppenheimer13a,oppenheimer18} or possibly CRs as shown for the FIRE-2 simulations (see \citealt{ji19} and above). While the results are shown only to \rvir, as in the other simulations and M31, \ovi\ is also observed well beyond \rvir\ in the EAGLE simulations (see Fig.~1 in \citealt{oppenheimer18a}). 

\subsubsection{Insights from the Observation/Simulation Comparison}\label{s-disc-pers-comp}

The comparison with the simulations shows that the CGM is changing in zoom-in simulations on length scales roughly similar to those observed in M31. The low ions and high ions follow substantially different profiles with radius, in both data and simulations. In the zoom-in simulations described above, the inner regions of the CGM of galaxies are more directly affected by large-scale feedback and recycling processes between the disk and CGM of galaxies. Therefore it is not surprising that the M31 CGM within 100--150 kpc shows a large variation in column density profiles with $R$, a more complex gas-phase structure, and larger peculiar velocities even though the current star formation rate is low. While both accretion and large-scale outflow coexist in the CGM and are responsible for the gas flow properties, stellar feedback is required to produce substantial amount of metals in the CGM at large impact parameters (see Figs.~\ref{f-col-vs-rho-fire}, \ref{f-col-vs-rho-foggie}, \ref{f-col-vs-rho-eagle}). M31 has currently a low SFR, but it had several episodes of bursting star formation in the past \citep[e.g.,][]{williams17}, likely ejecting a large portion of its metals in the CGM during these episodes.

Various models simulating different galaxy masses at different epochs, with distinct SFRs or feedback processes can reproduce at some level the diffuse \ovi\ observed beyond \rvir. All the simulations we have reviewed produce \ovi\ profiles that are flatter than the low ions and which extend to beyond \rvir\ with significant column density. While the galaxy halo masses are different, they are all roughly in the range of about  $10^{11.5}$--$10^{12.3}$ M$_\sun$, which is a mass range where their virial temperatures overlap with the temperature at which the ionization fraction of \ovi\ peaks \citep{oppenheimer16}. Using the EAGLE simulations, \citet{oppenheimer16} show that the virial temperature of the galaxy halos can explain the presence of strong \ovi\ in the CGM of star-forming galaxies with $M_{200} \simeq 10^{11.5}$--$10^{12.3}$ M$_\sun$ and the absence of strong \ovi\ in the CGM of quiescent galaxies that have overall higher halo masses ($M_{200} 10^{12.5}$--$10^{13.5}$ M$_\sun$) and hence higher virial temperatures, i.e., halo mass, not SFR largely drives the presence of strong \ovi\ in the CGM of galaxies (cf. \S\ref{s-coshalos}). Production the \ovi\ in volume-filling virialized gas could explain why \ovi\ is widely spread in the CGM of simulated galaxies and the real M31. Additional ionization mechanisms from cosmic rays (\citealt{ji19} and see Fig.~\ref{f-col-vs-rho-fire}) or fluctuating AGNs \citep{oppenheimer13a,oppenheimer18} can further boost the \ovi\ production, but halo masses with their virial temperatures close to the temperature at which the ionization fraction of \ovi\ peaks appear to provide a natural source for the diffuse, extended \ovi\ in the CGM of $L^*$ galaxies. Conversely, a number of studies have shown that significant \ovi\ can arise in active outflows, with the outflow column densities varying strongly with the degree of feedback \citep{hummels13, hafen19a}. Right now, no clear observational test can distinguish \ovi\ in warm virialized gas and direct outflows. However, any model that attempts to distinguish them will be constrained by the flat profile and low scatter seen by Project AMIGA. 

On the other hand, the cooler, diffuse ionized gas probed predominantly by \siiii, and also low ions (\cii, \siii) at smaller impact parameters, is not well-reproduced in the simulations. In the FIRE-2 and FOGGIE simulations, the column densities of \siiii\ and low ions within $\la 0.3 R_{200}$ are reasonably matched, but their covering fractions drop sharply and much more rapidly than observed for M31 when $ R > 0.3 R_{200}$. Only near satellite galaxies within $0.3 R_{200}$ do the column densities of these ions increases. This is, however, not a fair comparison as M31 lacks gas-rich satellites within this radius. Furthermore the near unity covering factor of \siiii\ out to $1.65 R_{200}$ in the CGM of M31 could not be explained by dwarf satellites anyway. For the EAGLE simulation, this problem is not as extreme as in the other simulations, but EAGLE does overproduce low and intermediate ions in the inner regions ($\la 0.3 R_{200}$) of the CGM. Possibly maintaining a high resolution out to \rvir\ would be needed to accurately model the small-scale structures of the cool gas content and preserve it over longer periods of times \citep{hummels19,peeples19,vandevoort18}. 

While the observations of M31 and simulations discussed above show some discrepancy, there is an overall trend that is universally observed: when the ionization energies increase from the singly-ionized species (\siii, \cii) to intermediate ions (\siiii, \siiv) to \civ\ to \ovi, the column density dispersions and dependence on $R$ decrease. While the larger scatter in the low and intermediate column densities compared to \ovi\ was observed previously \cite[e.g.,][]{werk13,liang16}, that trend with $R$ was not as obvious owing to a larger scatter at any $R$, in part caused by neighboring galaxies or different galaxy masses \citep{oppenheimer18a}. This general trend is the primary point of agreement between the observations and simulations, especially considering that the simulations were not tuned to match the CGM properties. This trend most likely arises from the physical conditions of the gas: in the inner regions of the CGM the gas takes on a density that favors the production of the low and intermediate ions. At these densities \ovi\ would need to be collisionally ionized or distributed in pockets of low-density photoionized gas. In the outer regions of the CGM, the overall gas must have a much lower density where \ovi\ and weak \siiii\ and nearly no singly ionized species can be produced predominantly by photoionization processes. This basic structure of the CGM appears in broad agreement between Project AMIGA, statistical sampling of many galaxies like COS-Halos, and three different suites of simulations. 

\subsection{Implications for the MW CGM}

Based the findings from Project AMIGA, it is likely that the MW has not only an extended hot CGM \citep{gupta14,gupta17}, but also an extended CGM of cool (\siii, \siiii, \siiv) and warm-hot (\civ, \ovi) gas that extends all the way to about 300 kpc (\rvir), and even farther away for the \ovi. In fact, the MW and M31 \ovi\ CGMs most likely already overlap as it can be seen, e.g., in the CLUES simulations of the Local group \citep{nuza14} since the distance between M31 and MW is only 752 kpc. 

A large covering factor of the CGM of M31 is not detected at high peculiar velocities (see Fig.~\ref{f-velavgmap}), and in fact beyond 100 kpc, the velocities $v_{\rm M31}$ are scattered around 20 \km. Even within 100 kpc, the average velocity is about 90 \km, which would barely constitute a HVC studied in the MW. In the MW most of the absorption within $\pm 90$ \km\ relative to the systemic velocity of the MW in a given direction is dominated by the disk, i.e., material within a few hundreds of pc from the galactic plane. Because the HVC velocities are high enough to separate them from the disk absorption, HVCs in the MW have been studied for many years to determine the ``halo" properties of the MW \citep[e.g.][]{wakker97,putman12,richter17}. However, we know now that the distances of these HVCs, including the predominantly ionized HVCs, are not at 100s of kpc from the MW, but most of them are within 15--20 kpc from the sun \citep[e.g.][]{wakker01,wakker08,thom08,lehner11a,lehner12}, i.e., in a radius not even explored by Project AMIGA and many other surveys of the galaxy CGM at higher redshifts \citep[e.g.,][]{werk13,liang14,borthakur16,burchett16}. Only the MS allows us to probe the interaction between the MW and the Magellanic clouds in the CGM of the MW out to about 50--100 kpc \citep[e.g.,][]{donghia16}. The results from Project AMIGA strongly suggest that the CGM of the MW is hidden in the low velocity absorption arising from its disk (see also \citealt{zheng15}. To complicate the matter, the column densities of the low, intermediate ions, and \civ\ drop substantially beyond 100-150 kpc (see, e.g., Figs.~\ref{f-coltot-vs-rho}, \ref{f-coltotsi-vs-rho}). Owing to its strength and little dependence on $R$, \ovi\ is among the best ultraviolet diagnostic of the extended CGM (see also the recent FOGGIE simulation results by \citealt{zheng20}). 

\section{Summary}\label{s-sum}
With Project AMIGA, we have surveyed the CGM of a galaxy with an unprecedented number of background targets (43) piercing it at various azimuths and impact parameters, 25 from $0.08 \rvir$ to about $1.1\rvir$ and the additional 18 between $1.1<R/\rvir \la 1.9$. The 43 QSOs were all observed with COS G130M/G160M or G130M (providing in particular \oi, \cii, \civ, \siii, \siiii, \siiv) and 11 were also observed with \fuse\ (providing \ovi). The resolution of the COS G130M/G160M and the SNRs have been key for the success of this program.  All the data were uniformly reduced and analyzed. For the 43 QSOs in our sample, we have identified all the absorption features in their spectra to determine if any transitions used to probe the CGM of M31 could be contaminated. We provide the entire line identification in Appendix~\ref{a-lineid}.  While we survey only a single galaxy, M31, the uniqueness of our experiment has allowed us to gain a wealth of new information that can be summarized as follows.

\begin{enumerate}[wide, labelwidth=!, labelindent=0pt]
\item Ionized gas traced by \siiii\ and \ovi\ have near unity covering factor all the way out to  $1.2\rvir$ and $1.9\rvir$, respectively. All the other ions have their covering factors monotonically decreasing as $R$ increases.
\item We do not find that the properties of the CGM of M31 strongly depend  on the azimuth with respect to the major and minor axes, but several properties of the CGM depend on the projected distance. 
\item The gas has a more complex gas-phase structure at $R\la 0.5\rvir$ with high covering factors of all the ions. At larger $R$, the gas becomes more highly ionized, with a paucity of singly ionized species. Stronger absorbers are also observed closer to M31 with the column densities of all the ions but \ovi\ decreasing sharply as $R$ increases up to $R\la 0.5\rvir$; beyond $R\ga 0.5\rvir$ the column densities decrease much more mildly with increasing $R$.
\item The velocity structure of the absorption profiles is  more complex with  $R\la 0.5\rvir$ where frequently more than one velocity component is observed, while at larger $R$, the absorption profiles predominantly show only one velocity component (at the COS resolution).  The peculiar velocities of the CGM gas are also more extreme and systematically redshifted by about $+90$ \km\ relative to the bulk motion of M31 at $R\la 0.5\rvir$. On the other hand, at $R\ga 0.5\rvir$, the peculiar velocities are both blue- and redshifted relative to the bulk motion of M31 and only by 10--20 \km\ on average.
\item Cosmological zoom-in simulations $\sim L^*$ galaxies (individual galaxies or galaxies in Local group analogs) show that \ovi\ does extend well beyond \rvir\ as observed for M31. On the other hand, cosmological zoom-in simulations do not reproduce well the column density profiles of the low ions (\siii, \cii) or intermediate ions (\siiii, \siiv). All the zoom in simulations explored in this work show some common traits with the observations of the CGM of M31: 1) the column densities of \ovi\ do not vary much with $R$ while those of the lower ions have a strong dependence with $R$; 2) the scatter in the column densities at $R$ is smaller in \ovi\ than any lower ions; 3) \ovi\ is observed at $R\gg \rvir$. In other words, the dispersion and the dependence of the column densities on the impact parameter decline going from singly through doubly to highly ionized species.
\item We estimate that the mass of the cool metal mass probed by \siii, \siiii, and \siiv\  of the CGM within \rvir\ is $2\times 10^7$ M$_\sun$ and by \ovi\ is $>8 \times 10^7$ M$_\sun$. The total metal mass could be as large as $\ga 2.5 \times 10^8$ M$_\sun$ if the dust and hot X-ray gas are accounted for. Since the total metal mass in the disk of M31 is about $M^{\rm disk}_{\rm Z}\simeq 5\times 10^8$ M$_\sun$, the CGM of M31 has at least 50\% of the present-day metal mass of its disk and possibly much more. 
\item We estimate the baryon mass of the $\sim 10^4$--$10^{5.5}$ K gas is $\ga 3.7 \times 10^{10}\,(Z/0.3\,Z_\sun)^{-1}$  M$_\sun$ at \rvir. The dependence on the largely unknown metallicity of the CGM makes the baryon mass estimate uncertain, but it is broadly comparable to other recent observational results or estimates in zoom-in simulations. 
\item We study if any of the M31 dwarf satellites could give rise to some of the observed absorption associated with the CGM of M31. We find it is plausible that few absorbers within close spatial and velocity proximity of the dwarfs  could be associated with the CGM of dwarfs if they have a gaseous CGM. However, these are Sph galaxies, which have had their gas stripped via ram-pressure and unlikely to have much gas left in their CGM. And, indeed, none of the properties of the absorbers in close proximity to these dwarf galaxies show any peculiarity that would associate them to the CGM of the satellites rather than the CGM of M31.
\end{enumerate}

\section*{Acknowledgements}
We thank David Nidever for sharing his original fits of the MS \hi\ emission and Ben Oppenheimer for sharing the EAGLE simulations shown in Fig.~\ref{f-col-vs-rho-eagle}. Support for this research was provided by NASA through grant HST-GO-14268 from the Space Telescope Science Institute, which is operated by the Association of Universities for Research in Astronomy, Incorporated, under NASA contract NAS5-26555. CAFG and ZH were also supported by NSF through grants AST-1517491, AST-1715216, and CAREER award AST-1652522; by NASA through grants NNX15AB22G and 17-ATP17-0067; by STScI through grants HST-GO-14681.011 and HST-AR-14293.001-A; and by a Cottrell Scholar Award from the Research Corporation for Science Advancement.  Based on observations made with the NASA-CNES-CSA Far Ultraviolet Spectroscopic Explorer, which was operated for NASA by the Johns Hopkins University under NASA contract NAS5-32985.

\software{Astropy \citep{price-whelan18}, emcee \citep{foreman-mackey13}, Matplotlib \citep{hunter07}, PyIGM \citep{prochaska17a}}

\facilities{HST(COS); HST(STIS); FUSE}

\bibliographystyle{aasjournal}

\clearpage
\begin{deluxetable*}{lccccccrrclc}
\tabletypesize{\scriptsize}
\tablecaption{Sample Summary\label{t-sum}}
\tablehead{\colhead{Target} & \colhead{$z_{\rm em}$} & \colhead{RA} & \colhead{Dec} & \colhead{$l_{\rm MS}$} & \colhead{$b_{\rm MS}$} & \colhead{$R$} & \colhead{$X$} & \colhead{$Y$} & \colhead{PID} & \colhead{COS Grating} & \colhead{SNR}\\ 
\colhead{} & \colhead{} & \colhead{($\mathrm{{}^{\circ}}$)} & \colhead{($\mathrm{{}^{\circ}}$)} & \colhead{($\mathrm{{}^{\circ}}$)} & \colhead{($\mathrm{{}^{\circ}}$)} & \colhead{ ($\mathrm{kpc}$)} & \colhead{($\mathrm{kpc}$)} & \colhead{($\mathrm{kpc}$)} & \colhead{ }& \colhead{ } & \colhead{ }}
\startdata
RX\_J0048.3+3941 & $0.134$ & $12.079$ & $39.687$ & $-125.0$ & $24.8$ & $25.0$ & $14.1$ & $-20.7$ & $11632$ & G130M-G160M & 31.5 \\
HS0033+4300 & $0.120$ & $9.096$ & $43.278$ & $-129.1$ & $22.7$ & $30.5$ & $-15.2$ & $26.5$ & $11632$ & G130M-G160M & 5.9 \\
HS0058+4213 & $0.190$ & $15.380$ & $42.493$ & $-128.1$ & $27.3$ & $48.6$ & $45.4$ & $17.3$ & $11632$ & G130M-G160M & 8.1 \\
RX\_J0043.6+3725 & $0.080$ & $10.927$ & $37.422$ & $-122.6$ & $23.7$ & $50.5$ & $ 2.5$ & $-50.5$ & $14268$ & G130M-G160M & 17.9 \\
Zw535.012 & $0.048$ & $9.087$ & $45.665$ & $-131.7$ & $22.9$ & $59.7$ & $-14.7$ & $57.8$ & $14268$ & G130M-G160M & 17.8 \\
RX\_J0050.8+3536 & $0.058$ & $12.711$ & $35.612$ & $-120.5$ & $25.0$ & $77.1$ & $21.6$ & $-74.0$ & $14268$ & G130M-G160M & 18.7 \\
IRAS\_F00040+4325 & $0.163$ & $1.652$ & $43.708$ & $-130.1$ & $17.4$ & $93.0$ & $-85.5$ & $36.5$ & $14268$ & G130M-G160M & 24.9 \\
RXS\_J0118.8+3836 & $0.216$ & $19.706$ & $38.606$ & $-123.7$ & $30.7$ & $97.2$ & $92.3$ & $-30.2$ & $14268$ & G130M-G160M & 14.4 \\
\enddata
\tablecomments{The 18 sightlines from our large \hst\ program have the PID 14268; for the supplemental archival 25 other targets, the \hst\ PID is listed. All the projected distances are computed using the three dimensional separation (coordinates of the target and distance of M31 assumed to be 752 kpc). The coordinates $l_{\rm MS}$ and $b_{\rm MS}$ are the MS longitudes and latitudes as defined by \citet{nidever08}. The SNR is given per COS resolution element (assuming $R\sim 17,000$) and estimated in the continuum near  \siiii\ $\lambda$1206.
}
\end{deluxetable*}

\startlongtable
\begin{deluxetable*}{llcccccccrr}
\tabletypesize{\scriptsize}
\tablecaption{Summary of the Results\label{t-results}}
\tablehead{\colhead{Target} & \colhead{Ion} & \colhead{$v_1$} & \colhead{$v_2$} & \colhead{$v$} & \colhead{$\sigma_v$} & \colhead{$\log N$} & \colhead{$\sigma_{ \log N}^1$} & \colhead{$\sigma_{ \log N}^2$} & \colhead{$f_{\rm N}$} & \colhead{$f_{\rm MS}$}\\ 
\colhead{} & \colhead{} & \colhead{(\km)} & \colhead{(\km)} & \colhead{(\km)} & \colhead{(\km)} & \colhead{[cm$^{-2}$]} & \colhead{} & \colhead{} & \colhead{}& \colhead{}}
\startdata
RX\_J0048.3+3941 & \ion{Al}{2} & $-480.0$ & $-320.0$ & $-397.2$ & 19.6 & 11.92 & 0.15 & 0.23 & $ 0$ & $-1$ \\
RX\_J0048.3+3941 & \ion{C}{2} & $-480.0$ & $-320.0$ & $-381.4$ & 1.2 & 14.15 & 0.01 & 0.01 & $-2$ & $-1$ \\
RX\_J0048.3+3941 & \ion{C}{4} & $-480.0$ & $-320.0$ & $-390.8$ & 5.3 & 13.25 & 0.05 & 0.05 & $ 0$ & $-1$ \\
RX\_J0048.3+3941 & \ion{Fe}{2} & $-480.0$ & $-320.0$ & $-432.9$ & 25.2 & 13.40 & 0.15 & 0.23 & $ 0$ & $-1$ \\
RX\_J0048.3+3941 & \ion{N}{1} & $-480.0$ & $-320.0$ & \nodata   & \nodata   & 12.88 & 0.18 & 0.30 & $-1$ & $-1$ \\
RX\_J0048.3+3941 & \ion{N}{5} & $-480.0$ & $-320.0$ & \nodata   & \nodata   & 12.77 & 0.18 & 0.30 & $-1$ & $-1$ \\
RX\_J0048.3+3941 & \ion{O}{1} & $-480.0$ & $-320.0$ & \nodata   & \nodata   & 13.47 & 0.18 & 0.30 & $-1$ & $-1$ \\
RX\_J0048.3+3941 & \ion{O}{6} & $-480.0$ & $-320.0$ & $-375.5$ & 6.2 & 13.85 & 0.05 & 0.06 & $ 0$ & $-1$ \\
\enddata
\tablecomments{The velocities $v_1$ and $v_2$ correspond to the integration of the absorption component. The $1\sigma$ errors $\sigma_{ \log N}^1$ and $\sigma_{ \log N}^2$ are the positive and negative errors, respectively, on the logarithm of the column density. The flag $f_{\rm N}$ has the following definition: $0=$ detection (not saturated or contaminated); $-1 =$ upper limit; $-2 =$ lower limit (due to saturation of the line). The flag $f_{\rm MS}$ has the following definition: $0=$ not contaminated by the MS; $-1 =$ contaminated by the MS (see \S\ref{s-ms}).
}
\end{deluxetable*}

\begin{deluxetable*}{llccccccccccccc}
\tablecaption{Summary of the M31 Dwarf Galaxies\label{t-dwarf}}
\tablehead{\colhead{Name} & \colhead{Type} & \colhead{$l_{\rm MS}$} & \colhead{$b_{\rm MS}$} & \colhead{\vlsr} & \colhead{$D_{\rm dwarf}$} & \colhead{$R$} & \colhead{$X$} & \colhead{$Y$} & \colhead{$M_*$} & \colhead{$M_{\rm 200}$} & \colhead{$R_{200}$} & \colhead{$v_{200}$}\\ 
\colhead{}  & \colhead{}& \colhead{($\mathrm{{}^{\circ}}$)} & \colhead{($\mathrm{{}^{\circ}}$)} & \colhead{(\km)} & \colhead{(kpc)} & \colhead{(kpc)} & \colhead{(kpc)} & \colhead{(kpc)} & \colhead{($10^5$\,M$_\sun$)} & \colhead{($10^8$\,M$_\sun$)} & \colhead{(kpc)} & \colhead{(\km)}}
\startdata
M32 			& cE 		& $-126.5$ & $23.8$ & $-194.6$ & $805.0$ & $ 5.3$ & $-0.1$ & $-5.3$ & $3200.0$ & $852.2$ & $92.8$ & $88.9$ \\
NGC 205 		& dE/dSph 	& $-127.5$ & $23.4$ & $-241.3$ & $824.0$ & $ 8.0$ & $-5.8$ & $ 5.5$ & $3300.0$ & $864.8$ & $93.2$ & $89.3$ \\
And IX 			& dSph		& $-129.0$ & $25.7$ & $-203.5$ & $766.0$ & $35.3$ & $24.3$ & $25.6$ & $  1.5$ & $21.5$ & $27.2$ & $26.1$ \\
And XVII 		& dSph	 	& $-130.3$ & $22.9$ & $-246.3$ & $794.0$ & $42.3$ & $-13.2$ & $40.2$ & $  2.6$ & $28.0$ & $29.7$ & $28.5$ \\
And I 			& dSph 		& $-123.4$ & $24.2$ & $-372.0$ & $745.0$ & $43.0$ & $ 7.6$ & $-42.3$ & $ 39.0$ & $102.7$ & $45.8$ & $43.9$ \\
And XXVII 		& dSph 		& $-131.5$ & $23.1$ & $-534.2$ & $828.0$ & $55.5$ & $-12.2$ & $54.1$ & $  1.2$ & $19.3$ & $26.3$ & $25.2$ \\
And III 		& dSph 		& $-121.9$ & $22.0$ & $-341.5$ & $748.0$ & $65.2$ & $-18.9$ & $-62.4$ & $  8.3$ & $48.9$ & $35.8$ & $34.3$ \\
And X 			& dSph 		& $-130.8$ & $28.2$ & $-160.0$ & $701.0$ & $73.5$ & $55.4$ & $48.3$ & $  1.0$ & $17.4$ & $25.3$ & $24.3$ \\
\enddata
\tablecomments{Galaxy parameters are from \citet{mcconnachie12}, \citet{martin14,martin16} and references therein.  }
\end{deluxetable*}

\begin{deluxetable*}{llcccc}
\tablecaption{QSO Absorbers within $R_{200}$ of M31 Dwarf Galaxies \label{t-xmatch-dwarf}}
\tablehead{\colhead{Dwarf} & \colhead{QSO} & \colhead{$v_{\rm LSR, Si\,III}$} & \colhead{$\Delta_{\rm sep}$} & \colhead{$v_{\rm esc}$} & \colhead{$\delta v$}\\ 
\colhead{ } & \colhead{} & \colhead{(\km)} & \colhead{} & \colhead{(\km)} & \colhead{(\km)}}
\startdata
M32 & RX\_J0048.3+3941 & $-550.4$ & $ 0.23$ & $187.3$ & $ 55.1$ \\
M32 & RX\_J0048.3+3941 & $-485.4$ & $ 0.23$ & $187.3$ & $  9.9$ \\
M32 & HS0033+4300 & $-513.2$ & $ 0.38$ & $144.3$ & $  7.9$ \\
M32 & HS0058+4213 & $-535.6$ & $ 0.55$ & $120.1$ & $ 42.9$ \\
M32 & HS0058+4213 & $-498.7$ & $ 0.55$ & $120.1$ & $  6.0$ \\
M32 & RX\_J0043.6+3725 & $-595.9$ & $ 0.49$ & $127.3$ & $101.9$ \\
M32 & RX\_J0043.6+3725 & $-489.1$ & $ 0.49$ & $127.3$ & $  4.9$ \\
M32 & Zw535.012 & $-535.4$ & $ 0.70$ & $106.4$ & $ 27.8$ \\
\enddata
\tablecomments{The match between dwarfs and QSO absorbers was made so that the projected separation between the dwarf and QSO is within $R_{200}$ of the dwarf, i.e., $\Delta_{\rm sep} = r/R^{\rm dwarf}_{200}\le 1$ (where $r$ is the projected distance between the QSO and dwarf). Here $v_{\rm esc}$ is the escape velocity of the dwarf at the projected distance of the QSO assuming a point-like mass with halo mass listed in Table~\ref{t-dwarf}. The velocity separation between the QSO absorber and the  dwarf is $\delta v = |v_{\rm M31, Si\,III} - v_{\rm M31, dwarf}|$.
}
\end{deluxetable*}

\begin{deluxetable}{lcc}
\tablecaption{Strength and Abundance of Key Ions\label{t-strength}}
\tablehead{\colhead{Ion} & \colhead{$f \lambda$} & \colhead{$[{\rm X/H}]_{\sun}$}}
\startdata
\cii\ $\lambda$1334     &171 & $     -3.57$ \\
\civ\ $\lambda$1550     &147 & $     -3.31$ \\
\civ\ $\lambda$1548     &294 & $     -3.57$ \\
\siii\ $\lambda$1526    &203 & $     -4.49$ \\
\siii\ $\lambda$1190    &348 & $     -4.49$ \\
\siii\ $\lambda$1193    &695 & $     -4.49$ \\
\siii\ $\lambda$1260    &1487 & $     -4.49$ \\
\siiii\ $\lambda$1206   &1967 & $     -4.49$ \\
\siiv\ $\lambda$1402    &358 & $     -4.49$ \\
\siiv\ $\lambda$1393    &715 & $     -4.49$ \\
\ovi\ $\lambda$1031     &137 & $     -3.31$
\enddata
\tablecomments{$[{\rm X/H}]_{\sun}$ is the solar abundance of the element X, i.e., C, O, or Si (solar abundances from \citealt{asplund09}).}
\end{deluxetable}

\begin{deluxetable}{lccc}
\tablecaption{Metal mass of the Cool CGM of M31\label{t-mass}}
\tablehead{\colhead{Range} & \colhead{$M_{\rm Z}$} & \colhead{$M_{\rm Z}$} & \colhead{$M_{\rm Z}$}\\
\colhead{} & \colhead{H Model} & \colhead{SPL Model}& \colhead{GP Model}\\
\colhead{(kpc)} & \colhead{($10^7$ M$_\sun$)} & \colhead{($10^7$ M$_\sun$)}& \colhead{($10^7$ M$_\sun$)}
}
\startdata
5--230      &   1.5--2.1       & 1.4--1.7 & 1.2--2.1  \\
5--300      &   2.0--2.6       & 1.8--2.0 & 1.5--2.5   \\
5--150      &   1.0--1.5       & 1.0--1.2 & 0.73--1.4  \\
150--300    &   1.0--1.1       &0.77-0.78 & 0.72--1.1  \\
5--360      &   2.5--3.0       & 2.1--2.3 & 1.7--2.9   \\
\enddata
\tablecomments{$M_{\rm Z}$ corresponds to the metal mass traced by \siii, \siiii, and \siiv. The H Model corresponds to the low and high hyperbola models,  the SPL model represents the high and low the single-power law models, and the GP model encompasses the range allowed by the errors from the low and high model (see \S\ref{s-nsi-vs-r} and Appendix~\ref{a-model}).}
\end{deluxetable}

\begin{deluxetable}{llcc}
\tablecaption{Summmary of the Velocities of the CGM of M31\label{t-vel}}
\tablehead{\colhead{Ion} &\colhead{Region} & \colhead{$\langle v_{\rm M31} \rangle$} & \colhead{IQR} \\
\colhead{}  &\colhead{}  & \colhead{(\km)} &  \colhead{(\km)}
}
\startdata
\multicolumn{4}{c}{Individual Components} \\
\hline
\cii\  &  All $R$ & $ 31.3  \pm  52.3 $ & $ [ -6.5 , 68.0 ]$ \\ 
\cii\  &  $R \le 100$ kpc & $ 80.6  \pm  39.9 $ & $ [ 37.9 , 101.1 ]$ \\ 
\cii\  &  $R > 100$ kpc & $ 4.7  \pm  36.6 $ & $ [ -21.2 , 24.9 ]$ \\ 
\siii\  &  All $R$ & $ 65.7  \pm  53.1 $ & $ [ 53.8 , 95.2 ]$ \\ 
\siii\  &  $R \le 100$ kpc & $ 90.6  \pm  32.3 $ & $ [ 67.2 , 117.8 ]$ \\ 
\siii\  &  $R > 100$ kpc & $ 22.3  \pm  54.5 $ & $ [ -27.0 , 76.8 ]$ \\ 
\siiii\  &  All $R$ & $ 26.8  \pm  45.5 $ & $ [ -10.1 , 54.7 ]$ \\
\enddata
\tablecomments{Average, standard deviation, and IQR velocity range are listed for the individual components and  averaged M31 velocity components.}
\end{deluxetable}

\appendix

\clearpage
\makeatletter
\renewcommand{\thefigure}{A\@arabic\c@figure}
\setcounter{figure}{0}
\renewcommand{\thetable}{A\@arabic\c@table}
\setcounter{table}{0}

\section{Line Identification}\label{a-lineid}
In Table~\ref{t-linelist}, we provide the line identification for each absorption feature detected at about the 2$\sigma$ level (the line list is complete at this level, but can include also less significant absorption). The table is ordered by alphabetical order of the QSO name and for each QSO in order of increasing observed wavelength (second column). In this table, we define the various types of absorption features as follows (third column): ``ISMLG" is any ISM/CGM/IGM absorption from the Local group environment (mostly the MW and M31); ``IGMABS" is any intervening IGM/CGM absorber at $\Delta v > 3000 $ \km\ from the QSO redshift; ``PROXIMATE" is a  proximate/associated absorber at $500< \Delta v < 3000 $ \km\ from the QSO redshift; ``INTRINSIC" is an absorber  at $\Delta v < 500 $ \km\ from the QSO redshift. Any ``UNIDENTIFIED" feature at the $>2\sigma $ level is marked with that denomination. Finally, ``OTHER" includes known fixed-pattern noise feature (``FPN"),  special case of fixed-pattern noise feature occurring near the edge of the COS detector (``EDGE"), or the 1043 \AA\ detector flaw in the \fuse\ data that causes a fake line (``FLAW"). FPN, EDGE, and FLAW all appear in the fourth column, which is otherwise used to list the atom or ion detected. The fifth column gives the rest wavelength of the atom/ion. The sixth column provides the information regarding frame into which the velocity (sixth column) and redshift (seventh column) are defined (``L": LSR frame--- any absorption at $|\vlsr|\le 700$ \km, and otherwise ``H": heliocentric frame). Finally the last two columns give the approximate equivalent widths ($W_{\lambda}$) and errors that are only provided as guidelines, i.e., these should not be used for quantitative scientific purposes since the continuum placement is only approximate. We finally note that the H$_2$ lines are not individually measured, but are based on a model of the H$_2$ absorption (see \citealt{wakker06}), which is the reason for not providing an error on $W_{\lambda}$.

The process for identifying the absorption lines in the COS spectra is reviewed in \S\ref{s-lineid}. As discussed in this section, some of the QSOs do not have the full FUV wavelength coverage or their redshifts put \lya\ beyond the observed wavelength. There are seven such cases that are reviewed below:

\begin{itemize}[wide, labelwidth=!, labelindent=0pt]

\item 3C454.3: With $\zem =0.859$, the highest redshift \lya\ absorber would be at 2259 \AA.  However, there are also COS G225M and FOS G190H/G270H data that help disentangle any possible \lyb\ from \lya. In the velocity range \dvrange, there is detection of absorption in several ions, with all the velocity profiles being consistent, suggesting no contamination in the surveyed velocity range \dvrange.

\item PG0044+030: With $\zem =0.859$,  the highest redshift \lya\ absorber would be at 1973 \AA.  The FOS G190H/G270H data helps securely identifying \lyb\ above 1347 \AA\ in the G130M spectrum. Many lines between 1215  and 1347 \AA\ are clearly identified as higher redshift Lyman series and metal lines, leaving just 5 absorption features identified as \lya, which is about the expected number of absorbers given the SNR in the COS spectrum of this target. In the velocity range \dvrange, there is detection of absorption in several ions, with all the velocity profiles being consistent, suggesting no contamination in the surveyed velocity range.

\item PHL1226: This target has only COS G130M observations, and since $\zem =0.404$,   the highest redshift \lya\ absorber would be at 1705 \AA. There are 11 lines that are listed as \lya\ in Table~\ref{t-linelist}, which might be \lyb. However, some are very unlikely \lyb\ as they are so strong that \lyg\ and/or metal lines would be expected to be detected and are not. In all but \siiii\ there is no detection in the velocity range \dvrange. The absorption at  $-332$ \km\ from \siiii\ $\lambda$1206 is, however, clearly identified as \oi\ $\lambda$1039 associated with the super Lyman limit system at $z=0.15974$.

\item RX\_J0023.5+1547:  This target has also only COS G130M observations, and since $\zem =0.412$,   the highest redshift \lya\ absorber would be at 1716 \AA.  There are 7 lines that are listed as \lya\ in Table~\ref{t-linelist}, which could be \lyb. Given the SNR of the COS spectrum, the expected number of \lya\ lines between 1215 and 1460 \AA\ is 15,  while we identify 13. On the other hand, the expected number of \lyb\ is 4, and only 1 is identified. This is well within the possible cosmic variance, but it is possible that 2--3 lines that are identified as \lya\ could actually be \lyb. In the velocity range \dvrange, only \siiii\ is detected, but there is no likely contamination of this aborption feature.

\item RX\_J0028.1+3103:   This target has also COS G130M and G160M observations, but since $\zem =0.500$,   the highest redshift \lya\ absorber would be at 1823 \AA. Only one possible \lyb\ identified as \lya\ is above the spectral limit of 1792 A.  In the velocity range \dvrange, two components are detected, with one observed in  \cii, \siiii, \siiv, and a weaker one only in \siiii. Based on our line identification, there is no likely contamination of the weaker \siiii\ absorption.

\item SDSSJ015952.95+134554.3: This target has also only COS G130M observations, and since $\zem =0.504$,   the highest redshift \lya\ absorber would be at 1828 \AA. Given the SNR of the COS spectrum, we expect about 21 \lya\ and 7 \lyb\, while 19 and 2 are listed in Table~\ref{t-linelist}, respectively. So it is possible that  2--3 lines identified as \lya\ could actually be \lyb. There is no detection of \siiii\ (or other ions) in the velocity range \dvrange, and hence no contamination issue.

\item SDSSJ225738.20+134045.0:  This target has also COS G130M and G160M observations, but since $\zem =0.595$,   the highest redshift \lya\ absorber would be at 1938 \AA. There are 3 \lya\ with no corresponding \lyb \ that possibly could be \lyb, but it is very unlikely to be the case for the 3 identified \lya.  In the velocity range \dvrange, there is detection of absorption in several ions, with all the velocity profiles being consistent, suggesting no contamination in the surveyed velocity range.

\end{itemize}

\section{Visually Identifying Contamination}\label{a-conflicplot}

\begin{figure*}[tbp]
\epsscale{1.1}
\plotone{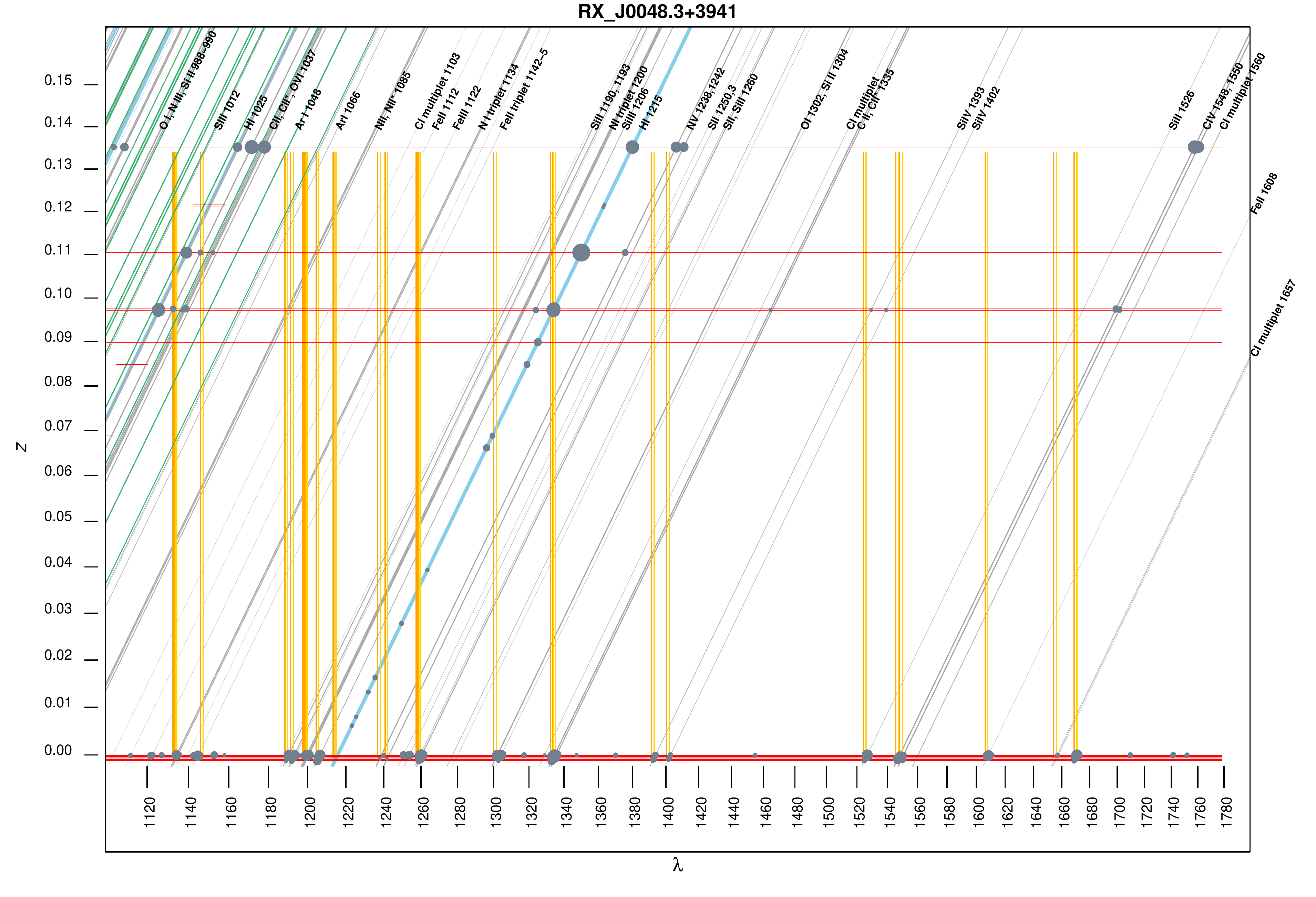}
\caption{Example of conflict plot used to determine potential contamination in the velocity range $-700\la \vlsr \la -150 $ \km. Locations on the $x$-axis are scaled according to $\log \lambda_{\rm obs}$, but the markers show the wavelength. The $y$-axis scale is proportional to $\log (1 + z)$, but the markers show $z$. The diagonal lines represent loci of different transitions and how their observed wavelengths change with $z$. The gray lines correspond to metal transitions, while the green ones indicate H$_2$ transitions. Lyman series lines of \hi\ are shown in blue. Some of the transitions are identified near the top and on the sides of the plot. Absorption features identified in the UV spectrum are shown with gray dots with varying sizes proportional to $\sqrt{W_\lambda}$. The red horizontal lines represent redshift systems with any line (usually \lya) that has $W_\lambda \ge 100$ m\AA; systems with \lya\ having strengths below this threshold do not  have lines that stretch across the whole plot, but do have short line segments that cover just \lyb\ and \lyg. Most of the lines of interests appear along the yellow and orange lines that represent a blue shift of  $-500$ and $-150$ \km, respectively. Potential conflicts arise at  the intersection of the horizontal red line, a gray/green/blue sloping line, and the vertical orange/yellow line.
\label{f-example-conflict}}
\end{figure*}

In order to visually assess possible line contamination as well as conflicts in independent line identifications, we have designed the ``conflict plot" that is shown in Fig.~\ref{f-example-conflict}. In this plot, the locations on the $x$-axis are scaled according to $\log \lambda_{\rm obs}$ and the $y$-axis scale is proportional to $\log (1 + z)$. The diagonal lines represent loci of different transitions and how their observed wavelengths change with $z$. The diagonal lines are color coded depending on whether they trace metal transitions (gray), H$_2$ (green), or \hi\ (blue). We identify some of the transitions at the top and on the sides of the plot. Any absorption feature that is identified in the COS spectrum is identified by a gray filled circle. Although not shown here, an independent line identification would be represented with a different symbol, immediately identifying the similarities and differences between two independent line identifications. The red horizontal lines represent redshift systems with any line (usually \lya) that has $W_\lambda \ge 100$ m\AA; systems with \lya\ having strengths below this threshold do not  have lines that stretch across the whole plot, but can have short line segments that cover just \lyb\ and \lyg. Some of the common interstellar lines appear along the yellow and orange lines that represent a blue shift of  $-500$ and $-150$ \km), respectively (i.e., within the velocity range where we observe M31 CGM gas). For Project AMIGA, we mostly concentrate on transitions at $\lambda>1145$ \AA\ that are available in the COS G130M and G160M wavelength bandpass. However, if there are \fuse\ data, which provide wavelength coverage $\lambda <1145$ \AA, we also use those.

In this figure, potential conflicts are readily detected at  the intersection of the horizontal red line, a gray/green/blue sloping line, and the vertical orange/yellow line. If for one of the line of interest, there is a gray circle at $z>0.01$ at the one of the intersection, there is a potential contamination in the velocity range \dvrange\ that needs to be checked. For each sightline, in our sample, we have produced these conflict plots to easily identify any potential contamination of the absorption in the velocity range \dvrange.

For the specific example shown in Fig.~\ref{f-example-conflict}, there are two potential conflicts for the components identified in the velocity \dvrange. One is near the \nni\ $\lambda$1134.1 line, where an \ovi\ absorber at $z=0.09753$ is present at 1132.58 (shifted by $-400$ \km\ relative to \nni, see Table~\ref{t-linelist}). However, there is no contamination near \nni\ $\lambda$1199.1 in the velocity range \dvrange, so we know there is no \nni\ absorption. The second conflict is for \cii\ $\lambda$1334 where \lya\ at $z=0.09723$ contaminates the $-140$ \km\ \cii\ component. Although there are several weaker Lyman series transitions that could have been used to correct for that contamination (which we did in some other cases), in this case we have also \fuse\ observations that provide \cii\ $\lambda$1036 where the absorption is not contaminated in this component.

\section{Comparison between COS G130M/G160M and STIS E140M Spectra}\label{s-comp-fit-aod}
For three targets (MRK335, UGC12163, and NGC7469) in our sample we have higher resolution STIS E140M spectra. Using the same integration velocity ranges that we used for COS (which is justified since the spectra were initially all aligned), we estimate the velocities and column densities in the STIS E140M spectra. The results are summarized in Table~\ref{t-comp}. In the footnote of this table we also list the SNRs  in the continuum near \cii\ and \siiii\ since not only resolution but also SNRs can explain some of the differences. The STIS data have systematically lower SNRs than the COS spectra. \citet{fox05a} show that in low SNR STIS E140M spectra (4--9 per resolution element), the AOD method can overestimate the apparent column densities by a factor 0.1--0.4 dex, especially when the absorption is weak. As for the COS spectra, we use the original binning sampling of the data to estimate the column densities. Binning by 2 or 3 pixels the STIS spectra did not change the results in contrast to the study of \citet{fox05a}, but in this study, the simulated spectra were affected only by Poisson noise while the STIS E140M are affected by both Poisson and fixed-pattern noises.

 As illustrated in Fig.~\ref{f-example-fit} (and see also normalized profiles in the supplemental figures) with the spectra of MRK335 where we show for \cii\ and \siiii\ the COS and STIS spectra, more components can be revealed in the higher resolution spectrum and the components appear sharper and deeper in the higher resolution spectrum. However, the STIS spectrum is also much noisier. For MRK335, the absorption in all the components are weak with a peak absorption depth at the 20\% level in \cii\ and 30--40\%  in \siiii. Within about the 1--$2\sigma$ errors the column densities between COS and STIS are in agreement even though additional components are revealed in the STIS spectrum. The column densities derived from the STIS spectrum are systematically higher, but this effect is consistent with the lower SNRs in the STIS data that was observed by \citet{fox05a}. For UGC12163, with a peak absorption depth at  60\% in \cii\ and 90\%  in \siiii\ in the component at $-425$ \km\ in the COS spectrum (see supplemental figures), the absorption in these two transitions is marked as saturated. The STIS \siiii\ reaches zero-flux level, confirming complete saturation in \siiii. The STIS apparent column density of \cii\ is 0.09 dex higher than the estimated lower limit from the COS spectrum, but in agreement within the $1\sigma$ error, implying that the adopted peak optical depth of $\tau_a >0.9$ is about right for saturation in the COS spectra (see \S\ref{s-aod}). The other components toward UGC12163 are weak and the STIS upper limits are in agreement with the COS detection. Toward NGC7469, the most negative absorption is again the strongest component and both COS \cii\ and \siiii\ were correctly identified as saturated (even though again these do not reach zero-flux levels, while they do in the STIS spectrum). The other two components in the spectra of NGC7469 are very weak.  For \siiii, the SNR effect is observed with the STIS spectrum having 0.1--0.2 dex higher than the column densities derived in the very high SNR COS spectrum. On the other hand, the SNR near \cii\ is higher in the STIS spectrum and for the component at $-251$ \km, the column densities derived from the COS and STIS spectra are in excellent agreement (see Table~\ref{t-comp}).

While the sample with both STIS and COS spectra is small, the comparison shows that there is overall a good agreement in the column densities derived from the STIS and COS data and our conservative choice of  $\tau_a \sim 0.9$ as the threshold for saturation in the COS data is adequate.

\section{Confronting the AOD Results with a Line-Profile Fitting Analysis}\label{s-pf}

For the most blended profiles (6 targets in our sample), we also use a Voigt profile fit (PF) analysis  to separate the absorption profiles into individual components with the goal to assess differences in column density estimates between the PF and AOD methods. With the PF method, we model the absorption profile as individual components using a modified version of the software described in \citet{fitzpatrick97} (and see also \citealt{lehner11b} for the updates).  The best-fit values describing the gas are determined by comparing the model profiles convolved with the COS G130M or G160M (and STIS E140M when available) instrumental line-spread function (LSF) of the data.  The COS and STIS LSFs are not purely Gaussian and we adopt the tabulated COS LSFs from the COS and STIS instrument handbooks \citep{riley19,fischer19}. As the COS LSFs vary with the FUV lifetime positions, we use the COS LSFs at the appropriate lifetime positions.\footnote{For targets obtained at different lifetime positions, we adopt the one with the longest exposure time, but we note that the results would not change quantitatively if we adopted another lifetime position (less than 0.02--0.04 dex on the column densities).} Three parameters $N_i$, $b_i$, and $v_i$ for each component, $i$, are input as initial guesses and were subsequently varied to minimize $\chi^2$.  The fitting process enables us to find the best fit of the component structure using the data from one or more transitions of the same ionic species simultaneously. However, all the ions are fitted independently  (i.e., we did not assume a common component structure for all the ions a priori). When STIS E140M data are available we also fit the COS and STIS independently to assess how different these are (see also Appendix \ref{s-comp-fit-aod}). We apply this method  to \cii, \civ, \siii, \siiii, \siiv, and \ovi\ if it is available.

\begin{figure*}[tbp]
\epsscale{1}
\plotone{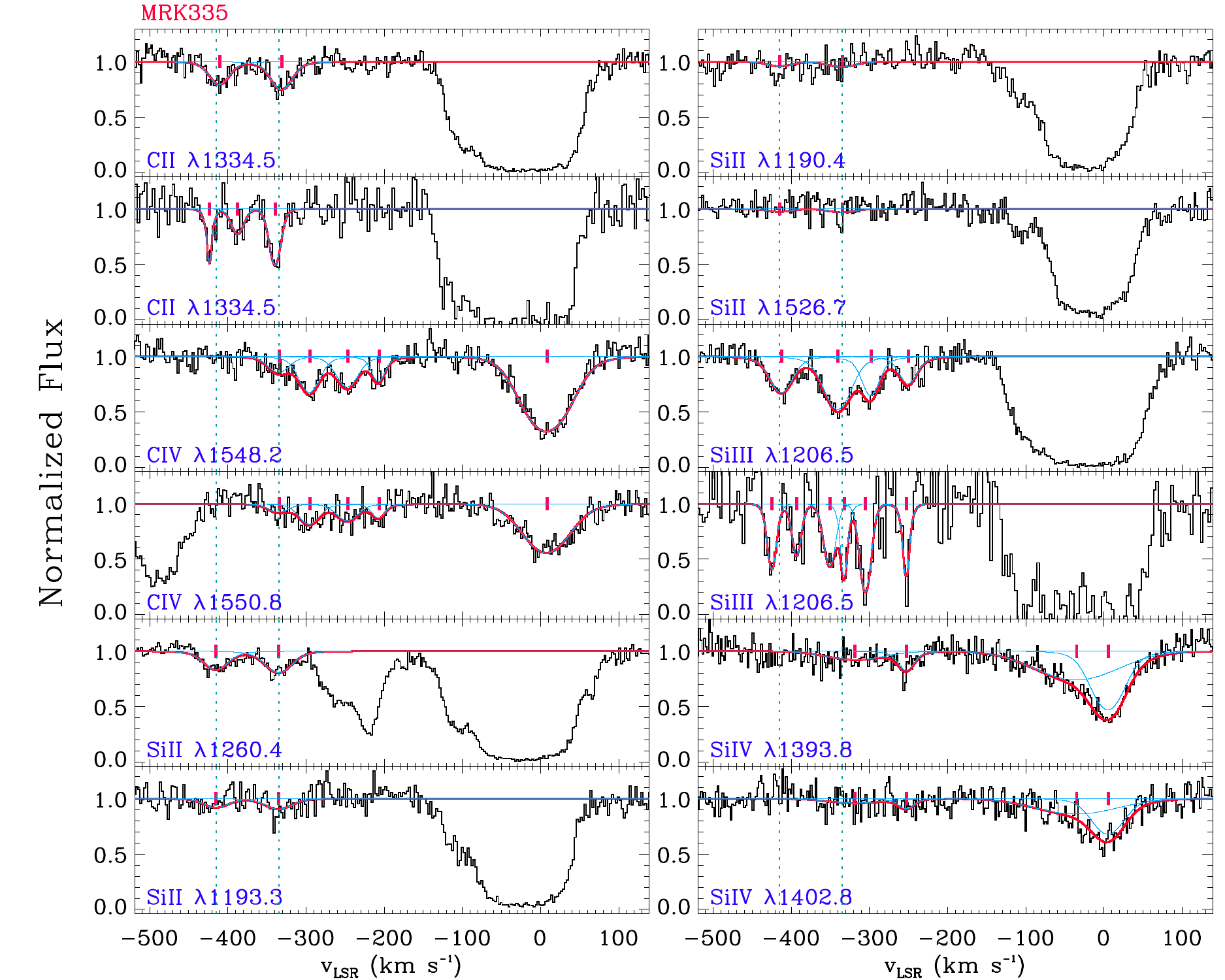}
\caption{Example of normalized absorption lines as a function of the LSR velocity toward MRK335 with a Voigt PF model to the data. The red tick-marks show the velocity centroids. In each panel, the red line shows the resulting PF while the blue line show the individual components. The green vertical dotted lines show the velocity centroids of \siii. When the same ions appear twice (here \cii\ and \siiii), the top and bottom panel shows the COS and STIS data, respectively. Note that the higher resolution of STIS E140M shows additional components in \cii\ and \siiii, but the low SNR near \siiii\ makes the results very uncertain.
\label{f-example-fit}}
\end{figure*}

 We always start each fit with the smallest number of components that reasonably model the profiles, and add more components as needed. We do not fix any of the input parameters, i.e., each input parameter is allowed to freely vary.  This procedure is repeated for each profile until the best fit is achieved. We finally bear in mind that even though the $\chi^2$ goodness of fit may be good, the PF method may still not assess correctly the saturation level especially since the COS resolution is relatively crude for PF. In low SNR or complicated profile, a broad component may also be fitted principally to reduce the $\chi^2$  while several narrower components could be more adequate and physically more appropriate (see also \citealt{lehner11b}). This is a limitation of the profile fitting especially when the spectral resolution is only $\sim$17 \km\ and/or the SNR is low. For \civ\ and \siiv\ we fit all the components including the MW low velocity components since those are not saturated, while for the other ions, we only fit the high-velocity components since the low velocity components are saturated.

 The results from component fitting are provided in  Table~\ref{t-fit}. In Fig.~\ref{f-example-fit}, we show an example of PF where both COS and STIS observations are available (in the supplemental material, we show the PFs for the 6 targets). Considering first the COS data only, although \cii\ and \siii\ are fitted independently, the velocity centroids of \cii\ and \siii\ match each other. The two components seen in \cii\ and \siii\ are also observed in \siiii\ at the same velocities. However, only the component at $-330$ \km\ is observed in \civ\ (and possibly \siiv). Two additional components are observed in \siiii\ but not in \cii\ or \siii. These additional components seen in \siiii\ are also observed in \civ, but shifted in velocity; an additional component is also observed at $-227$ \km\ only \civ. This demonstrates a clear change in the gas properties with velocity. The STIS E140M observations show additional components, but the results are far more uncertain (especially in the narrow components) owing to the lower SNR of the STIS data (for STIS column densities with less than 0.15 dex, those are quite consistent with the COS derived column densities).

 Much of these conclusions regarding the velocity structures and how the ionization properties change in the different components can be drawn from the AOD analysis by comparing the velocity profiles and derived averaged velocities in each determined components. The six targets that we model with Voigt PFs have the most complex velocity structure in our sample and therefore a comparison with the AOD results provides a way to assess how similar or dissimilar are the results between the AOD and PF analyses in the worst case scenarios. For \cii, \civ, \siii, \siiii, and \siiv, we use the information in Tables~\ref{t-results} and \ref{t-fit} to match the component. For  78\% (69/88) of the components, we can match them directly. For another 8, we coadd two components in the PF or AOD to match the AOD or PF results, and therefore in total we have 88\% matched AOD and PF components. For 3\% (3/88) and 9\% (8/88) of the components, the PF fits yield extremely narrow ($b<4$ \km) or broad ($b\ga 45$ \km) components, respectively; in these cases, the results are deemed uncertain because they appear in low SNR spectra and/or complicated profiles (see above). In particular, the majority of broad components (7/8) appear in \siiii\ and \siiv\ with no counterpart broad components in \civ.
\begin{figure}[tbp]
\epsscale{0.6}
\plotone{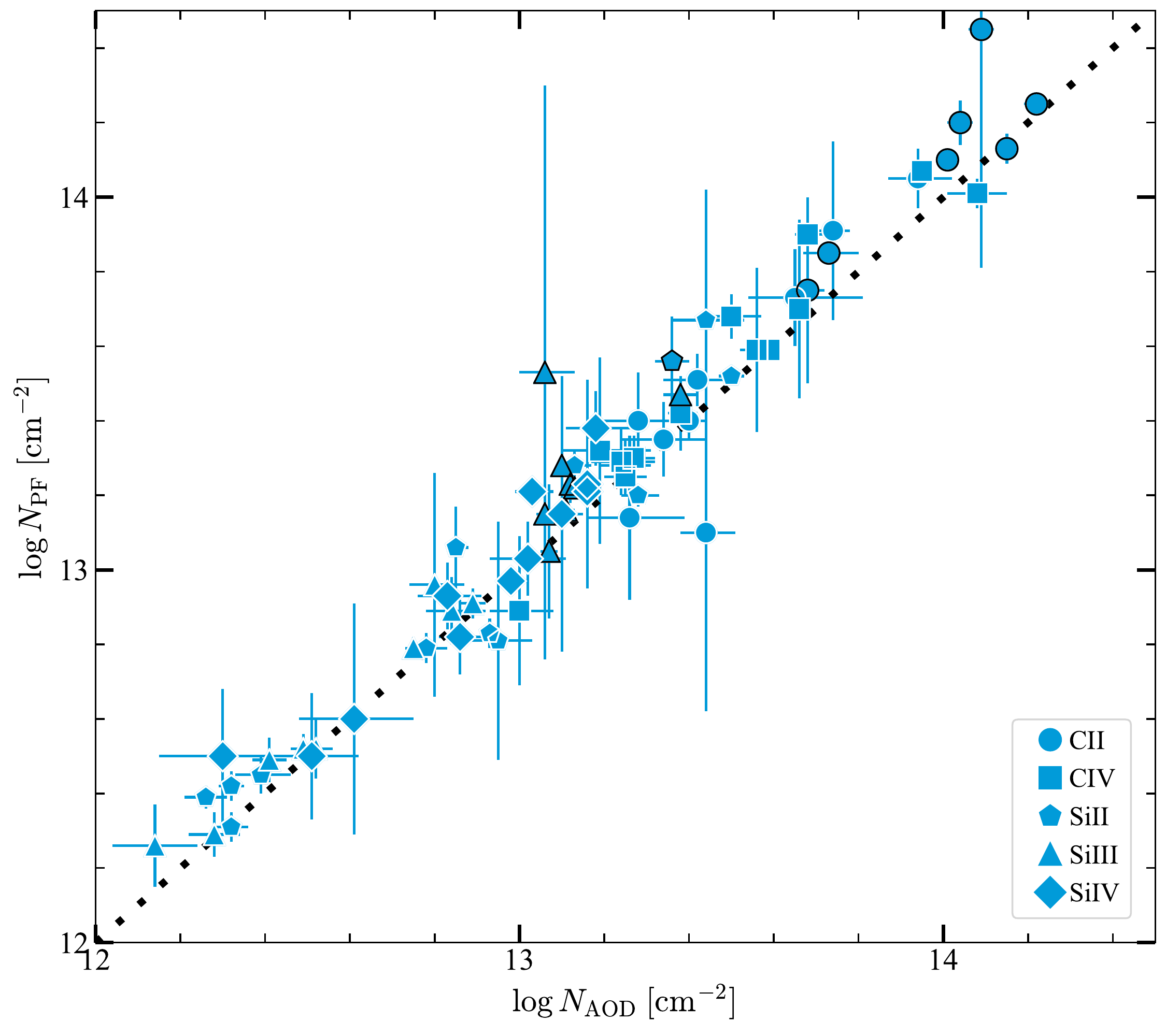}
\caption{Comparison of the column densities derived using the AOD and PF methods for the matched components. Black outlined circles indicate that the absorption is flagged as saturated using the AOD method.
\label{f-comp-fit-aod}}
\end{figure}

 For the matched components, the PF and AOD velocity centroids are in good agreement with a difference on average of 0.4--2.8 \km\ (and dispersion around the mean of 3--6 \km) depending on the ions. In the Fig.~\ref{f-comp-fit-aod}, we show the comparison of the column densities for the individual components derived from the PF ($y$-axis) and AOD ($x$-axis) methods. Within $1\sigma$ error, the majority of the data are within the 1:1 relationship. There is a slight systematic since $\langle \log N_{\rm PF} - \log N_{\rm AOD} \rangle \simeq +0.06 \pm 0.10$ for \cii, \siii, \civ, \siv; for \siiii\ that difference is somewhat larger with $+0.09 \pm 0.10$ dex, but within $1\sigma$ dispersion in agreement with the 1:1 relationship. This systematic can be understood as follows: 1) in blended absorption some extra absorption can be present in the wings of the profiles that is taken into account in the PF but not necessarily in the AOD method (this effect is more important for weak absorption features); 2) the width in the AOD integration is fixed, while in the PF it is a free parameter, which can increase the column if the width is larger than used in the AOD. Another systematic observed from Fig.~\ref{f-comp-fit-aod} is that the errors in the PF method are on average about $+0.07 \pm 0.15$ dex larger than those of the AOD. Several effects can explain this systematic: 1) broad shallow components can arise in PF but not in AOD (e.g., for \siiv, removing any components with $b>40$ \km\ from the sample would change the error difference from 0.08 to 0.04 dex); 2) PF of saturated components (removing \siiii\ saturated components from the sample would change the systematic from 0.08 to 0.04 dex---as noted above, with the AOD we, however, consider saturated components as lower limits only, which is likely to be the case also for the PF results); 3) in the PF, all the components influence the error in each component (typically relatively well-separated and not too shallow absorbing components lead to similar errors as, e.g., \cii, \siii, and \siiii\ shown in Fig.~\ref{f-example-fit}, see Table~\ref{t-fit}).

 In conclusion, while there are some systematic differences between the AOD and PF derived column densities, those are on average small (less than 15\%). Furthermore and importantly the targets considered in this section have the most complicated blending of components in our sample. Since a great part of this small systematic arises owing to the profiles being heavily blended, a  majority of our sample is not affected by those.

\section{M33 in the CGM of M31}\label{a-m33}
M33 is separated from M31 by about 190 kpc (see Fig.~\ref{f-map}) and is the third most massive galaxy in the Local group, but has still a mass about 20 times lower than M31 \citep{corbelli03}. It is considered a dwarf spiral galaxy, but its stellar mass of $(3$--$6) \times 10^9$ M$_\sun$ \citep{corbelli03} is at least 10 times larger than the next two most massive satellites (M32 and NGC205) of M31 (see Table~\ref{t-dwarf}), making M33 quite unique. \citet{kam17} show that the halo mass could be as large as $5.2\times 10^{11}$ M$_\sun$ within a virial radius of 168 kpc, but this would imply a very low baryonic mass fraction, suggesting a more plausible M33 virial radius (and hence halo mass) that is much smaller.

\begin{figure}[tbp]
\epsscale{0.6}
\plotone{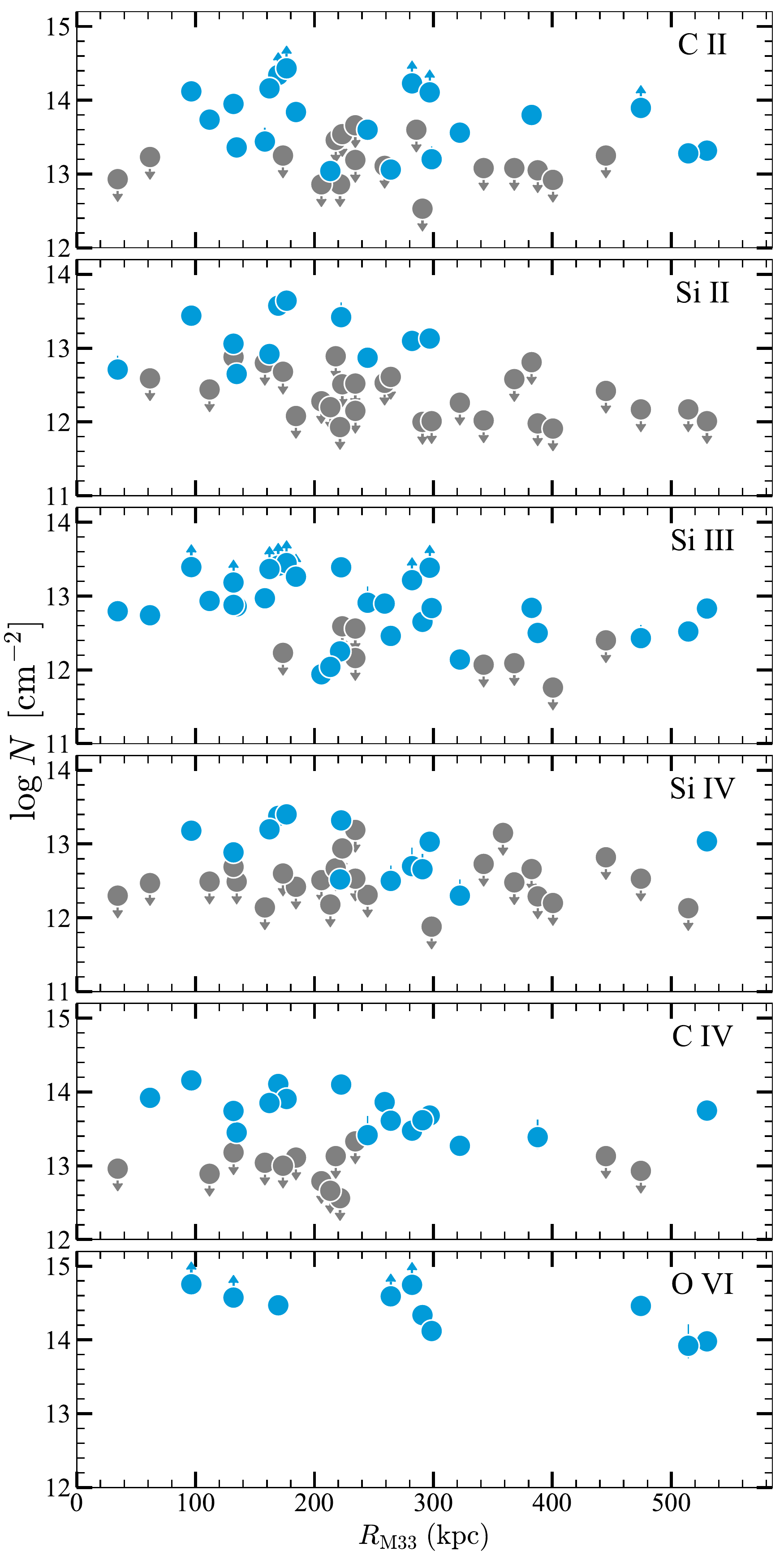}
\caption{Logarithm of the total column densities of the detected ions as a function of $R$. Blue circles are detections, while gray circles with downward arrows are non-detections. A blue circle with an upward arrow denotes that the absorption is saturated, resulting into a lower limit. 
\label{f-coltot-vs-rho-m33}}
\end{figure}

While M33 appears quite unique as a dwarf spiral galaxy, there are two main reasons that the CGM of M33 is unlikely to affect much the observed absorption observed toward the QSOs in our sample. First, the systemic velocity of M33 is $-180$ \km\ and the rotation velocity range is from $-300$ to $-75$ \km. Therefore, a large fraction of the M33 CGM absorption may actually be lost in the MW HVC and disk absorption ($\vlsr > -150$ \km). Second, checking the column density maps shown in Fig.~\ref{f-colmap}, there is no apparent trend between $N$ and the projected distance from M33, and, furthermore the two closest sightlines to M33 show a lack of strong absorption from singly ionized species. To show explicitly this lack of trend, we plot in Fig.~\ref{f-coltot-vs-rho-m33} the column densities of the various ions in our sample as a function of the projected distance from M33 ($R_{\rm M33}$). Contrary to Fig.~\ref{f-coltot-vs-rho}, there is no trend between $N$ and $R_{\rm M33}$. Furthermore, non-detections and detections are found at any projected distances from M33. All these strongly suggest that the CGM of M33 does not contribute significantly to the observed absorption associated with the CGM of M31. 

\section{Modeling the Column Density of Si as a Function of $R$}\label{a-model}
To model the functional form of $N_{\rm Si}$ with $R$ (see \S\ref{s-nsi-vs-r}), we consider three models, a hyperbolic (H) model, single-power law (SPL) model, and a Gaussian Process (GP) model.  Since our sample contains both lower and upper limits, we first need to determine how to treat censored points in the fit. Beyond 360 kpc, 8/10 data are upper limits but there is only 1 upper limit at $R< 360$ kpc. Upper limits correspond to non-detection of any Si ions, and hence the value of $N_{\rm Si}$ can only decrease. We have assessed the effect of these upper limits on the fit by refitting with these values decreased by a factor 2, 4, and 10; the overall effect on the fits is small. We adopt a factor 10 decrease for all the models. The lower limits are mostly observed at $R\la 140$ kpc (only 2 are observed at $R>140$ kpc, see Fig.~\ref{f-coltotsi-vs-rho}). As discussed in \S\ref{s-aod}, none of the absorption of \siiii\ associated with the CGM of M31 reaches the zero-flux level. We estimate that the lower limits at $R>50$ kpc should be increased by 0.1 dex while for the two inner lower limits, the increase is larger, and likely as large as 0.3 to 0.6 dex; we consider these two extrema in the fit. Finally, for data with vertical-ticked bars in Fig.~\ref{f-coltotsi-vs-rho}, we take the mean value between the high and low values.

For the H model, we first use the smaller correction on the inner region lower limits and consider the two populations of data below and above $R_{\rm th} =130$ kpc and fit each population with a linear fit. We note $R_{\rm th}$ can vary between 90 and 130 kpc without varying the fit results. We find $N_{\rm Si}/(10^{14} {\rm cm}^{-2})   = -1.76\times10^{-2}\, R  + 1.80$ at $R_{\rm th}<130$ kpc and $N_{\rm Si}/(10^{14} {\rm cm}^{-2})   =  -3.25\times10^{-4}\, R  + 0.204$ at $R_{\rm th}\ge 130$ kpc. The two lines intersect at $R_{0} = 92$ kpc and $N^0_{\rm Si} = 2.0\times 10^{13}$ \cmm. If stronger correction on the lower limits is applied, then we find $N_{\rm Si}/(10^{14} {\rm cm}^{-2})   = -3.43\times10^{-2}\, R  + 3.26$ at $R_{\rm th}<130$ kpc and $N_{\rm Si}/(10^{14} {\rm cm}^{-2})   =  -3.24\times10^{-4}\, R  + 0.204$ at $R_{\rm th}\ge 130$ kpc. The two lines intersect at $R_{0} = 90$ kpc and $N^0_{\rm Si} = 1.7\times 10^{13}$ \cmm. As expected, the effect of the different corrections on the lower limits is much stronger on the fit at $R<130$ kpc. The two regimes can be then modeled with a single hyperbola (e.g., \citealt{watts74}):
\begin{equation}\label{e-colsi-vs-r}
N_{\rm Si}/(10^{14}\,{\rm cm}^{-2}) = N^0_{\rm Si}/(10^{14}\,{\rm cm}^{-2}) \,+\, \beta_1 (R - R_0) \,+ \,\beta_2 [(R - R_0)^2\, + \,\delta^2/4]^{0.5}\,,
\end{equation}
where $\beta_1 = (\alpha_1 + \alpha_2)/2$ ($\alpha_{1,2}$ being the slopes of each straight-line),  $\beta_2 = (\alpha_2 - \alpha_1)/2$, and  the radius of curvature at $R = R_0$ being proportional to $\delta $ (an adjustable parameter that allows one to exactly follow the lines right to the intersection point--$\delta =0$ or smoothly merge the two asymptotes--$\delta>0$). We adopt $\delta = 30$. The dotted and dashed orange curves in Fig.~\ref{f-coltotsi-vs-rho} show the models where the lower limits at $R<50$ kpc are increased by a factor 0.3 or 0.6 dex, respectively.

The second model is the SPL, which can be written as:
\begin{equation}\label{e-colsi-vs-r1}
N_{\rm Si} = N^0_{\rm Si} \, R^\beta \; ({\rm cm^{-2}}),
\end{equation}
where here  $N^0_{\rm Si} =10^{15.91} $ \cmm\ and $\beta = -1.23$ and $N^0_{\rm Si} =10^{16.16} $ \cmm\ and $\beta = -1.33$ corresponding to lower limits at $R<50$ kpc being increased by a factor 0.3 or 0.6 dex, respectively. These two fits are shown in Fig.~\ref{f-coltotsi-vs-rho} with the dotted and dashed green lines, respectively.

Finally, we use the GP model, which is a generic supervised learning method designed to solve a regression, here between $\log N$ and $R$. The major advantages of this method are that the prediction interpolates the observations in a non-parametric way and is probabilistic so that empirical confidence intervals can be computed. We use the Python {\sc scikit-learn Gaussian Process Regression} \citep{pedregosa11,buitinck13} to model the data with a squared-exponential kernel (with a length-scale of 0.1 and lower and upper bounds on length-scale from 0.01 to 130). Changing the length-scale has little effect on the model, but changing the bounds changes the smoothness of the model (a small length-scale value means that function values can change quickly while large values characterize functions that change only slowly). The use of a more complex kernel like a Matern kernel would yield similar results when using similar bounds.  We treat all the data with the same weight using an error of 0.3 dex on the column density. This is larger than the typical measurements (0.05--0.15 dex) in our sample except for some of the data with vertical-ticked bars in Fig.~\ref{f-coltotsi-vs-rho}. This error can be understood as a prior factor to smooth out the scatter of the data; empirically, 0.3 dex is the minimal value to make the model converge.\footnote{Including another term in the kernel (e.g., adding another squared-exponential kernel to the original kernel) would produce a similar model with a larger standard deviation; in that case the error on each data point would have actually no effect on the model.}. The effect of an increase on the error on each data point would flatten the relationship with a somewhat larger deviation. The same weight is justified in order to not favor detections versus upper or lower limits. With these assumptions, we model the data and show the mean values of the predictive distribution from the GP models in  Fig.~\ref{f-coltotsi-vs-rho} with the dotted and dashed blue curves corresponding again to the cases where the lower limits at $R<50$ kpc are increased by a factor 0.3 or 0.6 dex, respectively. The blue area around each curve show the standard deviation determined by the GP model.

\section{Supplemental Figures}\label{a-supp-fig}

As part of the supplemental figures associated with this work, we provide for each absorber a figure  as shown in Fig.~\ref{f-example-spectrum} where we plot the normalized profiles of metal-line transitions for which we estimated the column densities. Each color corresponds to a component and the  velocity range of the absorption over which the velocity profile is integrated to derive the column densities and kinematics.

In a separate file, we also provide a figure as shown in Fig.~\ref{f-example-fit}, which shows the normalized profiles and the Voigt profile fitting. The profile fitting is done for 6 QSOs where the absorption components in the velocity range \dvrange\ are more severely blended.

\clearpage

\startlongtable
\begin{deluxetable*}{lcllccrrrr}
\tablecaption{Line List Identification\label{t-linelist}}
\tablehead{\colhead{Target} & \colhead{$\lambda_{\rm obs}$} & \colhead{Type} & \colhead{Ion} & \colhead{$\lambda_{\rm rest}$} & \colhead{Frame} & \colhead{$cz$} & \colhead{$z$} & \colhead{$W_{\rm \lambda}$} & \colhead{$\sigma_{W_{\lambda}}$}\\ \colhead{ } & \colhead{(\AA)} & \colhead{} & \colhead{} & \colhead{(\AA)} & \colhead{} & \colhead{(\km)} & \colhead{} & \colhead{(m\AA)} & \colhead{(m\AA)}}
\startdata
3C454.3 & $   1144.95$ & ISMLG & \ion{Fe}{2} & $    1144.9$ & L & $     2.0$ & $  -0.00002$ & $  309.4$ & $  96.6$ \\
3C454.3 & $   1146.29$ & IGMABS & \ion{H}{1} & $    1025.7$ & H & $ 35240.0$ & $   0.11758$ & $  260.5$ & $  71.2$ \\
3C454.3 & $   1152.82$ & ISMLG & \ion{P}{2} & $    1152.8$ & L & $    -1.0$ & $  -0.00002$ & $  166.3$ & $  38.3$ \\
3C454.3 & $   1153.25$ & IGMABS & \ion{O}{6} & $    1031.9$ & H & $ 35246.0$ & $   0.11758$ & $  155.6$ & $  30.3$ \\
3C454.3 & $   1158.39$ & IGMABS & \ion{O}{3} & $     832.9$ & H & $117142.0$ & $   0.39074$ & $  142.0$ & $  28.6$ \\
3C454.3 & $   1159.58$ & IGMABS & \ion{O}{6} & $    1037.6$ & H & $ 35239.0$ & $   0.11758$ & $   93.0$ & $  44.3$ \\
3C454.3 & $   1182.72$ & IGMABS & \ion{H}{1} & $    1025.7$ & H & $ 45885.0$ & $   0.15307$ & $   63.0$ & $  27.3$ \\
3C454.3 & $   1183.43$ & IGMABS & \ion{H}{1} & $    1025.7$ & H & $ 46094.0$ & $   0.15376$ & $  548.3$ & $  24.0$ \\
\enddata
\tablecomments{Definition of the identified types of absorption features: ISMLG: ISM/CGM/IGM absorption from the Local group environment (mostly the MW and M31); IGMABS: intervening IGM/CGM absorber at $\Delta v > 3000 $ \km\ from the QSO redshift; PROXIMATE: proximate/associated absorber at $500< \Delta v < 3000 $ \km\ from the QSO redshift; INTRINSIC: intrinsic absorber  at $\Delta v < 500 $ \km\ from the QSO redshift; UNIDENTIFIED: unknown origin of the absorption; OTHER: FPN: fixed-pattern noise feature; OTHER: EDGE: special case of fixed-pattern noise feature occurring near the edge of the COS detector; OTHER: FLAW: refers to the 1043 \AA\ detector flaw in the \fuse\ data that causes a fake line. In the frame column, ``L" stands for LSR frame (any absorption at $|\vlsr\|\le 500$),`H" for heliocentric frame (any absorption at $|\vlsr\|> 500$). Note that the equivalent widths ($W_{\lambda}$) and errors are provided for guidelines and should not be used for quantitative scientific purposes. The H$_2$ lines are not individually measured, but are based on a model of the H$_2$ absorption, which is the reason for not providing an error on $W_{\lambda}$ (i.e., in the table it is set to ``0.0").
}
\end{deluxetable*}
\begin{deluxetable*}{lccc}
\tabletypesize{\scriptsize}
\tablecaption{Comparison between COS and STIS \label{t-comp}}
\tablehead{\colhead{Ion/Instrument} & \colhead{$[v_1,v_2]$} & \colhead{$v$} & \colhead{$\log N$} \\ 
\colhead{} & \colhead{(\km)} & \colhead{(\km)} & \colhead{[cm$^{-2}$]}}
\startdata
\multicolumn{4}{c}{MRK335} \\
\hline
       \cii\ $\lambda$1334 COS  & $  -450, -372 $ & $  -406.9  \pm  3.7 $ & $13.22  \pm 0.08  $ \\
       \cii\ $\lambda$1334 STIS & $  -450, -372 $ & $  -403.2  \pm  4.7 $ & $13.34  \pm 0.11  $ \\
       \cii\ $\lambda$1334 COS  & $  -372, -310 $ & $  -334.6  \pm  2.1 $ & $13.34  \pm 0.06  $ \\
       \cii\ $\lambda$1334 STIS & $  -372, -310 $ & $  -338.8  \pm  2.0 $ & $13.52  \pm 0.06  $ \\
     \siiii\ $\lambda$1206 COS  & $  -450, -372 $ & $  -409.7  \pm  1.8 $ & $12.49  \pm 0.04  $ \\
     \siiii\ $\lambda$1206 STIS & $  -450, -372 $ & $  -406.5  \pm  5.3 $ & $12.51  \pm 0.13  $ \\
     \siiii\ $\lambda$1206 COS  & $  -372, -310 $ & $  -339.9  \pm  0.8 $ & $12.75  \pm 0.02  $ \\
     \siiii\ $\lambda$1206 STIS & $  -372, -310 $ & $  -337.8  \pm  2.2 $ & $12.82  \pm 0.07  $ \\
     \siiii\ $\lambda$1206 COS  & $  -310, -273 $ & $  -296.6  \pm  0.8 $ & $12.40  \pm 0.04  $ \\
     \siiii\ $\lambda$1206 STIS & $  -310, -273 $ & $  -301.8  \pm  2.2 $ & $12.64  \pm 0.17  $ \\
     \siiii\ $\lambda$1206 COS  & $  -273, -190 $ & $  -246.9  \pm  3.7 $ & $12.28  \pm 0.06  $ \\
     \siiii\ $\lambda$1206 STIS & $  -273, -190 $ &              \nodata  & $ <       12.2    $ \\
\hline
\multicolumn{4}{c}{UGC12163 } \\
\hline
       \cii\ $\lambda$1334 COS  & $  -475, -375 $ & $  -423.3  \pm  3.6 $ & $> 13.94 	   $ \\
       \cii\ $\lambda$1334 STIS & $  -475, -375 $ & $  -425.2  \pm  5.0 $ & $14.05  \pm 0.13  $ \\
       \cii\ $\lambda$1334 COS  & $  -375, -310 $ & $  -353.0  \pm  7.3 $ & $13.35  \pm 0.20  $ \\
       \cii\ $\lambda$1334 STIS & $  -375, -310 $ &              \nodata  & $ <       13.28   $ \\
       \cii\ $\lambda$1334 COS  & $  -220, -180 $ & $  -196.6  \pm  3.1 $ & $13.26  \pm 0.18  $ \\
       \cii\ $\lambda$1334 STIS & $  -220, -180 $ & $  -200.8  \pm  3.6 $ & $13.37\,^{+0.16}_{-0.26} $ \\
     \siiii\ $\lambda$1206 COS  & $  -475, -375 $ & $  -426.0  \pm  2.3 $ & $>13.30   	   $ \\
     \siiii\ $\lambda$1206 STIS & $  -475, -375 $ & $  -425.5  \pm 21.4 $ & $ >	   13.45   $ \\
     \siiii\ $\lambda$1206 COS  & $  -375, -310 $ & $  -349.8  \pm  6.2 $ & $12.43  \pm 0.18  $ \\
     \siiii\ $\lambda$1206 STIS & $  -375, -310 $ &	      \nodata  & $ <	   12.57   $ \\
\hline
\multicolumn{4}{c}{NGC7469} \\
\hline
       \cii\ $\lambda$1334 COS  & $  -400, -268 $ & $  -335.3  \pm  0.7 $ & $ >	   14.27     $ \\
       \cii\ $\lambda$1334 STIS & $  -400, -268 $ & $  -337.6  \pm  0.8 $ & $ >	   14.44     $ \\
       \cii\ $\lambda$1334 COS  & $  -268, -210 $ & $  -251.4  \pm  3.8 $ & $13.06  \pm 0.09    $ \\
       \cii\ $\lambda$1334 STIS & $  -268, -210 $ & $  -251.1  \pm  3.3 $ & $13.10  \pm 0.08  $ \\
       \cii\ $\lambda$1334 COS  & $  -202, -150 $ & $  -176.2  \pm  3.4 $ & $12.94  \pm 0.11  $ \\
       \cii\ $\lambda$1334 STIS & $  -202, -150 $ & $  -184.9  \pm  6.8 $ & $12.66\,^{+0.14}_{-0.20} $ \\
     \siiii\ $\lambda$1206 COS  & $  -400, -268 $ & $  -325.2  \pm  0.5 $ & $ >	   13.57     $ \\
     \siiii\ $\lambda$1206 STIS & $  -400, -268 $ & $  -332.7  \pm  5.9 $ & $ >	   13.76     $ \\
     \siiii\ $\lambda$1206 COS  & $  -268, -210 $ & $  -246.9  \pm  0.9 $ & $12.63  \pm 0.02  $ \\
     \siiii\ $\lambda$1206 STIS & $  -268, -210 $ & $  -250.1  \pm  1.7 $ & $12.76  \pm 0.16  $ \\
     \siiii\ $\lambda$1206 COS  & $  -202, -150 $ & $  -178.8  \pm  1.3 $ & $12.32  \pm 0.04  $ \\
     \siiii\ $\lambda$1206 STIS & $  -202, -150 $ & $  -174.8  \pm  1.2 $ & $12.58  \pm 0.04  $ 
\enddata
\tablecomments{COS stands here for COS G130M and STIS for STIS E140M. The SNRs (per resolution element) near \cii\ and \siiii\ are, respectively: MRK335: 36.6, 32.2 (COS), 9.5, 4.8 (STIS);  UGC12163: 10.7, 10.8 (COS), 5.1, 2.3 (STIS); NGC7469: 37.5, 35.4 (COS), 16.5, 9.0 (STIS).  }
\end{deluxetable*}

\clearpage
\startlongtable
\begin{deluxetable*}{llccccccc}
\tablecaption{Summary of the Profile Fit Results\label{t-fit}}
\tablehead{\colhead{Target} & \colhead{Ion} & \colhead{Comp.} & \colhead{$v$} & \colhead{$\sigma_v$} & \colhead{$b$} & \colhead{$\sigma_b$} & \colhead{$\log N$} & \colhead{$\sigma_{\log N}$}\\ \colhead{} & \colhead{} & \colhead{} & \colhead{(\km)} & \colhead{(\km)} & \colhead{(\km)} & \colhead{(\km)} & \colhead{[cm$^{-2}$]} & \colhead{}}
\startdata
RX\_J0048.3+3941 & \ion{C}{2} & $ 1$ & $-471.4$ & $ 4.3$ & $13.8$ & $ 7.7$ & 12.90 & 0.13 \\
RX\_J0048.3+3941 & \ion{C}{2} & $ 2$ & $-418.9$ & $ 4.5$ & $17.5$ & $ 7.6$ & 13.21 & 0.13 \\
RX\_J0048.3+3941 & \ion{C}{2} & $ 3$ & $-373.8$ & $ 0.9$ & $17.3$ & $ 1.7$ & 14.07 & 0.02 \\
RX\_J0048.3+3941 & \ion{C}{2} & $ 4$ & $-329.1$ & $ 3.2$ & $14.6$ & $ 5.1$ & 13.21 & 0.10 \\
RX\_J0048.3+3941 & \ion{C}{2} & $ 5$ & $-244.3$ & $ 0.8$ & $17.3$ & $ 1.3$ & 13.85 & 0.02 \\
RX\_J0048.3+3941 & \ion{C}{2} & $ 6$ & $-177.3$ & $ 1.0$ & $20.0$ & $ 1.3$ & 14.25 & 0.03 \\
RX\_J0048.3+3941 & \ion{C}{4} & $ 1$ & $-386.7$ & $ 4.2$ & $39.2$ & $ 6.3$ & 13.25 & 0.05 \\
RX\_J0048.3+3941 & \ion{C}{4} & $ 2$ & $-239.0$ & $ 0.6$ & $21.7$ & $ 0.9$ & 14.07 & 0.01 \\
RX\_J0048.3+3941 & \ion{C}{4} & $ 3$ & $-183.6$ & $ 0.8$ & $11.7$ & $ 1.3$ & 13.59 & 0.02 \\
RX\_J0048.3+3941 & \ion{C}{4} & $ 4$ & $-30.5$ & $ 4.6$ & $40.0$ & $ 3.7$ & 13.62 & 0.07 \\
RX\_J0048.3+3941 & \ion{C}{4} & $ 5$ & $-8.8$ & $ 2.6$ & $11.7$ & $ 6.0$ & 13.09 & 0.20 \\
RX\_J0048.3+3941 & \ion{Si}{2} & $ 1$ & $-374.7$ & $ 1.3$ & $15.4$ & $ 2.0$ & 12.83 & 0.04 \\
RX\_J0048.3+3941 & \ion{Si}{2} & $ 2$ & $-323.1$ & $ 5.2$ & $28.5$ & $10.3$ & 12.49 & 0.10 \\
RX\_J0048.3+3941 & \ion{Si}{2} & $ 3$ & $-251.8$ & $ 1.8$ & $13.1$ & $ 3.2$ & 12.79 & 0.04 \\
RX\_J0048.3+3941 & \ion{Si}{2} & $ 4$ & $-179.3$ & $ 0.5$ & $16.2$ & $ 0.8$ & 13.52 & 0.02 \\
RX\_J0048.3+3941 & \ion{Si}{3} & $ 1$ & $-374.6$ & $ 0.9$ & $10.2$ & $ 2.4$ & 12.97 & 0.08 \\
RX\_J0048.3+3941 & \ion{Si}{3} & $ 2$ & $-366.3$ & $ 6.0$ & $61.4$ & $13.0$ & 12.86 & 0.05 \\
RX\_J0048.3+3941 & \ion{Si}{3} & $ 3$ & $-278.2$ & $ 1.8$ & $ 1.6$ & $21.4$ & 12.89 & 24.70 \\
RX\_J0048.3+3941 & \ion{Si}{3} & $ 4$ & $-239.8$ & $ 0.7$ & $13.8$ & $ 1.5$ & 13.23 & 0.05 \\
RX\_J0048.3+3941 & \ion{Si}{3} & $ 5$ & $-180.6$ & $ 0.6$ & $16.9$ & $ 1.1$ & 13.28 & 0.04 \\
RX\_J0048.3+3941 & \ion{Si}{4} & $ 1$ & $-386.6$ & $ 3.8$ & $24.2$ & $ 6.1$ & 12.44 & 0.07 \\
RX\_J0048.3+3941 & \ion{Si}{4} & $ 2$ & $-238.0$ & $ 0.6$ & $16.9$ & $ 0.9$ & 13.21 & 0.01 \\
RX\_J0048.3+3941 & \ion{Si}{4} & $ 3$ & $-177.1$ & $ 0.7$ & $13.4$ & $ 1.2$ & 12.97 & 0.02 \\
RX\_J0048.3+3941 & \ion{Si}{4} & $ 4$ & $-49.8$ & $ 6.2$ & $64.4$ & $ 6.6$ & 13.08 & 0.05 \\
RX\_J0048.3+3941 & \ion{Si}{4} & $ 5$ & $-8.1$ & $ 0.7$ & $12.0$ & $ 1.5$ & 13.02 & 0.04 \\
RX\_J0048.3+3941 & \ion{O}{6} & $ 1$ & $-388.2$ & $ 6.3$ & $44.0$ & $ 8.8$ & 13.84 & 0.10 \\
RX\_J0048.3+3941 & \ion{O}{6} & $ 2$ & $-227.6$ & $ 2.4$ & $21.4$ & $ 4.5$ & 13.94 & 0.12 \\
RX\_J0048.3+3941 & \ion{O}{6} & $ 3$ & $-224.7$ & $ 7.1$ & $99.0$ & $16.6$ & 14.40 & 0.04 \\
RX\_J0048.3+3941 & \ion{O}{6} & $ 4$ & $-184.6$ & $ 1.3$ & $ 4.8$ & $ 3.4$ & 13.57 & 0.14 \\
RX\_J0048.3+3941 & \ion{O}{6} & $ 5$ & $-5.2$ & $ 2.7$ & $39.4$ & $ 3.7$ & 14.08 & 0.04 \\
MRK352 & \ion{C}{2} & $ 1$ & $-303.9$ & $ 9.3$ & $32.0$ & $16.5$ & 13.73 & 0.13 \\
MRK352 & \ion{C}{2} & $ 2$ & $-192.4$ & $ 4.8$ & $28.5$ & $ 8.0$ & 14.05 & 0.08 \\
MRK352 & \ion{C}{4} & $ 1$ & $-278.4$ & $ 2.8$ & $17.6$ & $ 4.8$ & 13.68 & 0.06 \\
MRK352 & \ion{C}{4} & $ 2$ & $-220.9$ & $ 1.8$ & $19.7$ & $ 2.9$ & 14.01 & 0.04 \\
MRK352 & \ion{C}{4} & $ 3$ & $-31.2$ & $ 1.9$ & $22.0$ & $ 3.0$ & 13.87 & 0.04 \\
MRK352 & \ion{C}{4} & $ 4$ & $54.4$ & $ 6.3$ & $23.8$ & $11.0$ & 13.34 & 0.11 \\
MRK352 & \ion{Si}{2} & $ 1$ & $-206.4$ & $ 2.0$ & $ 5.9$ & $ 2.4$ & 13.67 & 0.35 \\
MRK352 & \ion{Si}{3} & $ 1$ & $-300.0$ & $ 4.7$ & $20.2$ & $ 7.8$ & 12.89 & 0.09 \\
MRK352 & \ion{Si}{3} & $ 2$ & $-242.0$ & $ 3.3$ & $ 9.7$ & $12.3$ & 12.96 & 0.30 \\
MRK352 & \ion{Si}{3} & $ 3$ & $-199.0$ & $ 2.8$ & $10.0$ & $ 6.4$ & 13.53 & 0.77 \\
MRK352 & \ion{Si}{4} & $ 1$ & $-302.2$ & $23.7$ & $34.7$ & $71.6$ & 12.75 & 0.34 \\
MRK352 & \ion{Si}{4} & $ 2$ & $-199.8$ & $ 8.8$ & $47.9$ & $14.4$ & 13.38 & 0.10 \\
MRK352 & \ion{Si}{4} & $ 3$ & $-21.7$ & $ 3.9$ & $38.4$ & $ 5.6$ & 13.61 & 0.05 \\
RBS2055 & \ion{C}{2} & $ 1$ & $-407.1$ & $ 3.2$ & $15.6$ & $ 5.5$ & 13.51 & 0.07 \\
RBS2055 & \ion{C}{2} & $ 2$ & $-328.5$ & $ 1.2$ & $15.3$ & $ 2.1$ & 14.20 & 0.06 \\
RBS2055 & \ion{C}{2} & $ 3$ & $-266.2$ & $11.4$ & $23.8$ & $27.6$ & 13.14 & 0.22 \\
RBS2055 & \ion{C}{4} & $ 1$ & $-443.9$ & $14.4$ & $28.1$ & $36.4$ & 13.32 & 0.25 \\
RBS2055 & \ion{C}{4} & $ 2$ & $-322.1$ & $ 3.6$ & $49.9$ & $ 5.4$ & 13.90 & 0.04 \\
RBS2055 & \ion{C}{4} & $ 3$ & $-0.6$ & $ 1.4$ & $31.4$ & $ 2.2$ & 14.19 & 0.02 \\
RBS2055 & \ion{Si}{2} & $ 1$ & $-406.3$ & $ 1.0$ & $ 9.3$ & $ 2.2$ & 12.39 & 0.03 \\
RBS2055 & \ion{Si}{2} & $ 2$ & $-330.3$ & $ 1.3$ & $13.3$ & $ 2.2$ & 13.28 & 0.04 \\
RBS2055 & \ion{Si}{3} & $ 1$ & $-421.2$ & $ 1.4$ & $27.3$ & $ 2.3$ & 13.22 & 0.03 \\
RBS2055 & \ion{Si}{3} & $ 2$ & $-326.0$ & $ 1.1$ & $28.5$ & $ 1.9$ & 13.47 & 0.03 \\
RBS2055 & \ion{Si}{4} & $ 1$ & $-439.3$ & $ 7.6$ & $43.5$ & $13.3$ & 12.93 & 0.09 \\
RBS2055 & \ion{Si}{4} & $ 2$ & $-322.7$ & $ 3.0$ & $32.1$ & $ 4.6$ & 13.21 & 0.04 \\
RBS2055 & \ion{Si}{4} & $ 3$ & $-7.6$ & $ 1.0$ & $29.3$ & $ 1.6$ & 13.78 & 0.02 \\
MRK335 & \ion{C}{2} & $ 1$ & $-410.4$ & $ 2.5$ & $13.7$ & $ 4.4$ & 13.28 & 0.07 \\
MRK335 & \ion{C}{2} & $ 2$ & $-330.9$ & $ 2.1$ & $14.8$ & $ 3.5$ & 13.40 & 0.05 \\
MRK335 & \ion{C}{4} & $ 1$ & $-334.2$ & $ 7.0$ & $15.1$ & $14.4$ & 12.89 & 0.20 \\
MRK335 & \ion{C}{4} & $ 2$ & $-294.9$ & $ 2.6$ & $14.2$ & $ 5.4$ & 13.29 & 0.09 \\
MRK335 & \ion{C}{4} & $ 3$ & $-246.4$ & $ 2.7$ & $15.2$ & $ 5.7$ & 13.22 & 0.08 \\
MRK335 & \ion{C}{4} & $ 4$ & $-206.6$ & $ 2.5$ & $ 6.7$ & $ 6.7$ & 12.98 & 0.09 \\
MRK335 & \ion{C}{4} & $ 5$ & $ 8.2$ & $ 1.1$ & $35.3$ & $ 1.5$ & 14.05 & 0.01 \\
MRK335 & \ion{Si}{2} & $ 1$ & $-414.8$ & $ 2.0$ & $17.4$ & $ 3.4$ & 12.31 & 0.04 \\
MRK335 & \ion{Si}{2} & $ 2$ & $-334.8$ & $ 1.9$ & $19.4$ & $ 3.1$ & 12.42 & 0.04 \\
MRK335 & \ion{Si}{3} & $ 1$ & $-412.6$ & $ 1.7$ & $15.0$ & $ 2.9$ & 12.52 & 0.04 \\
MRK335 & \ion{Si}{3} & $ 2$ & $-340.3$ & $ 1.6$ & $18.0$ & $ 2.7$ & 12.79 & 0.03 \\
MRK335 & \ion{Si}{3} & $ 3$ & $-298.1$ & $ 1.8$ & $ 9.9$ & $ 3.8$ & 12.49 & 0.06 \\
MRK335 & \ion{Si}{3} & $ 4$ & $-250.3$ & $ 2.1$ & $10.4$ & $ 4.2$ & 12.29 & 0.06 \\
MRK335 & \ion{Si}{4} & $ 1$ & $-318.5$ & $17.9$ & $47.5$ & $48.3$ & 12.60 & 0.31 \\
MRK335 & \ion{Si}{4} & $ 2$ & $-252.7$ & $ 3.2$ & $ 9.9$ & $ 6.7$ & 12.50 & 0.17 \\
MRK335 & \ion{Si}{4} & $ 3$ & $-35.0$ & $10.0$ & $68.5$ & $ 7.2$ & 13.31 & 0.08 \\
MRK335 & \ion{Si}{4} & $ 4$ & $ 5.3$ & $ 1.5$ & $23.8$ & $ 3.1$ & 13.35 & 0.07 \\
PG0003+158 & \ion{C}{2} & $ 1$ & $-399.9$ & $ 1.7$ & $23.1$ & $ 2.7$ & 13.83 & 0.03 \\
PG0003+158 & \ion{C}{2} & $ 2$ & $-325.8$ & $ 1.0$ & $18.7$ & $ 1.8$ & 14.10 & 0.03 \\
PG0003+158 & \ion{C}{2} & $ 3$ & $-279.6$ & $ 9.7$ & $11.4$ & $999.0$ & 12.90 & 0.53 \\
PG0003+158 & \ion{C}{2} & $ 4$ & $-240.0$ & $ 3.9$ & $19.9$ & $ 8.2$ & 13.50 & 0.11 \\
PG0003+158 & \ion{C}{2} & $ 5$ & $-176.9$ & $ 0.9$ & $20.3$ & $ 1.5$ & 14.26 & 0.03 \\
PG0003+158 & \ion{C}{4} & $ 1$ & $-396.7$ & $20.1$ & $44.2$ & $21.2$ & 13.59 & 0.22 \\
PG0003+158 & \ion{C}{4} & $ 2$ & $-355.7$ & $ 2.9$ & $14.8$ & $ 6.7$ & 13.49 & 0.24 \\
PG0003+158 & \ion{C}{4} & $ 3$ & $-307.2$ & $14.2$ & $36.6$ & $20.5$ & 13.30 & 0.21 \\
PG0003+158 & \ion{C}{4} & $ 4$ & $-227.8$ & $ 1.4$ & $ 8.2$ & $ 2.8$ & 13.30 & 0.06 \\
PG0003+158 & \ion{C}{4} & $ 5$ & $-3.5$ & $ 3.4$ & $32.4$ & $ 4.2$ & 13.82 & 0.05 \\
PG0003+158 & \ion{C}{4} & $ 6$ & $57.2$ & $ 4.2$ & $24.2$ & $ 5.4$ & 13.52 & 0.09 \\
PG0003+158 & \ion{Si}{2} & $ 1$ & $-387.6$ & $ 1.8$ & $10.8$ & $ 3.5$ & 12.45 & 0.05 \\
PG0003+158 & \ion{Si}{2} & $ 2$ & $-323.7$ & $ 0.6$ & $13.6$ & $ 1.0$ & 13.20 & 0.03 \\
PG0003+158 & \ion{Si}{3} & $ 1$ & $-406.0$ & $ 2.8$ & $29.7$ & $ 4.5$ & 12.91 & 0.04 \\
PG0003+158 & \ion{Si}{3} & $ 2$ & $-335.6$ & $ 1.5$ & $22.0$ & $ 2.5$ & 13.16 & 0.03 \\
PG0003+158 & \ion{Si}{3} & $ 3$ & $-249.3$ & $ 4.9$ & $35.8$ & $ 8.3$ & 12.62 & 0.06 \\
PG0003+158 & \ion{Si}{4} & $ 1$ & $-431.5$ & $ 9.3$ & $16.9$ & $26.4$ & 12.32 & 0.27 \\
PG0003+158 & \ion{Si}{4} & $ 2$ & $-358.2$ & $ 6.1$ & $38.0$ & $ 9.6$ & 12.94 & 0.08 \\
PG0003+158 & \ion{Si}{4} & $ 3$ & $-232.1$ & $ 2.5$ & $ 3.6$ & $ 6.4$ & 12.50 & 0.18 \\
PG0003+158 & \ion{Si}{4} & $ 4$ & $-4.5$ & $ 1.6$ & $26.1$ & $ 2.4$ & 13.39 & 0.03 \\
PG0003+158 & \ion{Si}{4} & $ 5$ & $57.2$ & $ 2.0$ & $14.4$ & $ 3.3$ & 12.97 & 0.05 \\
PG2349$-$014 & \ion{C}{2} & $ 1$ & $-328.2$ & $ 5.5$ & $23.3$ & $ 8.9$ & 13.91 & 0.24 \\
PG2349$-$014 & \ion{C}{2} & $ 2$ & $-294.3$ & $ 1.9$ & $ 8.8$ & $16.5$ & 14.45 & 0.64 \\
PG2349$-$014 & \ion{C}{2} & $ 3$ & $-259.6$ & $ 3.2$ & $15.9$ & $ 7.3$ & 13.75 & 0.25 \\
PG2349$-$014 & \ion{C}{2} & $ 4$ & $-204.0$ & $ 7.4$ & $11.3$ & $999.0$ & 13.10 & 0.48 \\
PG2349$-$014 & \ion{C}{2} & $ 5$ & $-172.1$ & $ 2.4$ & $ 5.8$ & $ 8.7$ & 13.40 & 0.13 \\
PG2349$-$014 & \ion{Si}{2} & $ 1$ & $-327.3$ & $ 6.4$ & $25.2$ & $ 9.4$ & 13.06 & 0.11 \\
PG2349$-$014 & \ion{Si}{2} & $ 2$ & $-289.0$ & $ 1.2$ & $ 7.9$ & $ 2.2$ & 13.56 & 0.12 \\
PG2349$-$014 & \ion{Si}{2} & $ 3$ & $-259.0$ & $ 2.6$ & $ 3.0$ & $ 4.1$ & 12.81 & 0.32 \\
PG2349$-$014 & \ion{Si}{3} & $ 1$ & $-336.6$ & $ 4.0$ & $15.4$ & $ 5.8$ & 13.05 & 0.18 \\
PG2349$-$014 & \ion{Si}{3} & $ 2$ & $-301.0$ & $ 3.9$ & $ 6.5$ & $ 2.9$ & 14.70 & 1.44 \\
PG2349$-$014 & \ion{Si}{3} & $ 3$ & $-267.0$ & $10.9$ & $51.2$ & $ 8.6$ & 13.63 & 0.12 \\
PG2349$-$014 & \ion{Si}{3} & $ 4$ & $-166.5$ & $ 1.7$ & $ 7.9$ & $ 4.0$ & 12.52 & 0.08 \\
PG2349$-$014 & \ion{Si}{4} & $ 1$ & $-349.9$ & $ 1.8$ & $ 9.4$ & $ 4.0$ & 12.82 & 0.10 \\
PG2349$-$014 & \ion{Si}{4} & $ 2$ & $-296.9$ & $ 1.8$ & $18.0$ & $ 6.2$ & 13.23 & 0.28 \\
PG2349$-$014 & \ion{Si}{4} & $ 3$ & $-256.7$ & $35.8$ & $48.5$ & $33.4$ & 13.15 & 0.37 \\
PG2349$-$014 & \ion{Si}{4} & $ 4$ & $-8.6$ & $ 5.6$ & $26.3$ & $10.3$ & 13.04 & 0.52 \\
PG2349$-$014 & \ion{Si}{4} & $ 5$ & $26.6$ & $36.2$ & $46.1$ & $22.4$ & 13.12 & 0.43
\enddata
\tablecomments{Error $v$, $b$, and $N$ are $1\sigma$ errors.}
\end{deluxetable*}

\phantom{This is needed for the table to fully resolve.}

\end{document}